\RequirePackage{rotating}
\documentclass[fleqn,usenatbib]{mnras}

\usepackage[T1]{fontenc}
\usepackage{ae,aecompl}

\usepackage{CJK}
\usepackage{multirow}
\usepackage{xcolor}
\usepackage{amssymb}
\usepackage{amsmath,amstext}
\usepackage{multirow}
\usepackage{booktabs}
\usepackage{graphicx}
\usepackage{subfig}
\usepackage{array}
\usepackage{longtable,booktabs,threeparttablex}
\usepackage{url}
\usepackage{hyperref}
\usepackage{newtxtext}
\usepackage[varvw]{newtxmath}
\usepackage{rotating}
\usepackage{multirow}

\newcommand{\Ha}{\ensuremath{\rm H\alpha}}

\newcommand{\lya}{\ensuremath{\rm Ly\alpha}}

\newcommand{\kms}{\rm km~s\ensuremath{^{-1}\,}}

\newcommand{\nhi}{\ensuremath{N_{\rm HI}}}

\newcommand{\flya}{\ensuremath{F_{\mathrm{Ly}\alpha}}}
\newcommand{\fblue}{\ensuremath{F_{\mathrm{Ly}\alpha}(\mathrm{blue})}}
\newcommand{\fred}{\ensuremath{F_{\mathrm{Ly}\alpha}(\mathrm{red})}}
\newcommand{\wlya}{\ensuremath{W_{\lambda}(\lya)}}

\newcommand{\secpoint}{\mbox{$''\mskip-7.6mu.\,$}}

\newcommand{\pa}{\ensuremath{\rm PA}}
\def\ltsima{$\; \buildrel < \over \sim \;$}
\def\simlt{\lower.5ex\hbox{\ltsima}}
\def\gtsima{$\; \buildrel > \over \sim \;$}
\def\simgt{\lower.5ex\hbox{\gtsima}}
\def\arcs{$''~$}

\newcommand{\RNum}[1]{\uppercase\expandafter{\romannumeral #1\relax}}
\newcommand{\rnum}[1]{\lowercase\expandafter{\romannumeral #1\relax}}

\title[Ly$\alpha$ Emission \& Galaxy Azimuthal Angle]{ The KBSS-KCWI Survey: The connection between extended Ly$\alpha$ halos and galaxy azimuthal angle at $z\sim 2-3$ 
\thanks{Based on data obtained at the W. M. Keck Observatory, which is operated as a scientific partnership among the California Institute of Technology, the University of California and the National Aeronautics and Space Administration. The Observatory was made possible by the generous financial support of the W. M. Keck Foundation.}}

\begin{document}
\begin{CJK*}{UTF8}{gbsn}

\author[Y. Chen et al.]{Yuguang Chen (陈昱光)$^{1}$\thanks{Email: yuguangchen@astro.caltech.edu},
Charles C. Steidel$^{1}$, Dawn K. Erb$^2$, David R. Law$^3$, Ryan F. Trainor$^{4}$,
\newauthor Naveen A. Reddy$^{5}$, Alice E. Shapley$^{6}$, Anthony J. Pahl$^{6}$, Allison L. Strom$^{7}$, Noah R. Lamb$^{8}$, 
\newauthor Zhihui Li (李智慧)$^{1}$ and Gwen C. Rudie$^{9}$
\\
$^{1}$Cahill Center for Astronomy and Astrophysics, California Institute of Technology, 
1200 E California Blvd, MC249-17, 
Pasadena, CA 91125, USA\\
$^{2}$Center for Gravitation, Cosmology, and Astrophysics, Department of Physics, University of Wisconsin-Milwaukee, 3135 N. Maryland Avenue, \\Milwaukee, WI 53211, USA\\
$^{3}$ Space Telescope Science Institute, 3700 San Martin Drive, Baltimore, MD 21218, USA \\
$^{4}$Department of Physics and Astronomy, Franklin \& Marshall College, 
637 College Ave., 
Lancaster, PA 17603, USA\\
$^{5}$Department of Physics and Astronomy, University of California, Riverside, 900 University Avenue, Riverside, CA 92521, USA \\
$^{6}$Department of Physics and Astronomy, University of California, Los Angeles, 430 Portola Plaza, Los Angeles, CA 90095, USA\\
$^{7}$Department of Astrophysical Sciences, 4 Ivy Lane, Princeton University, Princeton, NJ 08544, USA \\
$^{8}$Department of Physics, Drexel University, 32 S. 32nd Street, Philadelphia, PA 19104, USA\\
$^{9}$The Observatories of the Carnegie Institution for Science, 813 Santa Barbara Street, Pasadena, CA 91101, USA \\ 
}
\date{Accepted XXX. Received YYY; in original form ZZZ}


\label{firstpage}
\pagerange{\pageref{firstpage}--\pageref{lastpage}}
\maketitle

\begin{abstract}
We present the first statistical analysis of kinematically-resolved, spatially-extended \lya\ emission around $z = 2-3$ galaxies in the Keck Baryonic Structure Survey (KBSS) using the Keck Cosmic Web Imager (KCWI). Our sample of 59 star-forming galaxies ($z_\mathrm{med} = 2.29$) comprises the subset with typical KCWI integration times of $\sim 5$ hours and with existing imaging data from the Hubble Space Telescope and/or adaptive optics-assisted integral field spectroscopy. 
The high resolution images were used to evaluate the azimuthal dependence of the diffuse \lya\ emission with respect to the stellar continuum within projected galactocentric distances of $\lesssim 30$ proper kpc. We introduce cylindrically-projected 2D spectra (CP2D) that map the averaged \lya\ spectral profile over a specified range of azimuthal angle, as a function of impact parameter around galaxies. The averaged CP2D spectrum of all galaxies shows clear signatures of \lya\ resonant scattering by outflowing gas. {We stacked the CP2D spectra of individual galaxies over ranges of azimuthal angle with respect to their major axes. The extended \lya\ emission along the galaxy principal axes are statistically indistinguishable, with residual asymmetry of $\le 2$\% ($\sim 2 \sigma$) of the integrated \lya\ emission.} The symmetry implies that the \lya\ scattering medium is dominated by outflows in all directions within 30 kpc.  Meanwhile, we find that the blueshifted component of \lya\ emission is marginally stronger along galaxy minor axes for galaxies with relatively weak \lya\ emission. We speculate that this weak directional dependence of \lya\ emission becomes discernible only when the \lya\ escape fraction is low. These discoveries highlight the need for similar analyses in simulations with \lya\ radiative transfer modeling.  
\end{abstract}




\section{Introduction}
\label{sec:intro}
Extended ``halos'' of diffuse {Lyman-alpha (Ly$\alpha$)} emission, extending to many times larger radii than starlight, are nearly ubiquitous around rapidly-star-forming galaxies at redshift $z > 2$ \citep[e.g.][]{steidel11, wisotzki16, leclercq17, wisotzki18}.
While there is not yet consensus on the dominant physical mechanism giving rise to diffuse \lya\ halos, there is general agreement on the list of potential sources, all of which depend on substantial cool hydrogen gas in the circumgalactic medium (CGM; e.g., \citealt{ouchi20})): (1) resonant scattering of \lya\ produced by recombination of gas photoionised by massive stars or {active galactic nuclei (AGN), i.e., a central source}, where \lya\ photons are subsequently scattered until they find optically-thin channels to escape (2) {\it in situ} photoionization of \ion{H}{I}  by the metagalactic {ultra-violet (UV)} ionising radiation field combined with local sources and followed by recombination (sometimes called ``fluorescence'') (3) accreting gas losing  energy via collisional excitation of \lya\ (sometimes referred to as ``gravitational cooling'' (4) emission from unresolved satellite galaxies in the halos of larger central galaxies.  In principle, the observed surface brightness, spatial distribution, and kinematics of \lya\ emission can discriminate between the various mechanisms and, perhaps more importantly, can provide direct information on the degree to which gas in the CGM is accreting, outflowing, or quiescent. \lya\ emission from the CGM, if interpreted correctly, can provide a detailed map of the cool component of the dominant baryon reservoir associated with forming galaxies, as well as constraints on large-scale gas flows that are an essential part of the current galaxy formation paradigm.

Because a \lya\ photon is produced by nearly every photoionisation of hydrogen, the intrinsic \lya\ luminosity of a rapidly star-forming galaxy can be very high, and thus easily detected \citep{partridge67}. However, due to its very high transition probability, \lya\ is resonantly scattered until the last scattering event gives it an emitted frequency and direction such that the optical depth remains low along a trajectory that allows it to escape from the host galaxy. When the \lya\ optical depth is high in all directions, the vastly increased effective path length -- due to large numbers of scattering events during the time the photon is radiatively trapped -- increases the probability that the photon is destroyed by dust or emitted via two-photon mechanism. But \lya\ photons that are not absorbed by dust grains or converted to two-photon radiation must eventually escape, with the final scattering resulting in the photon having a frequency and direction such that the photon can freely stream without further interaction with a hydrogen atom. The radiative transfer of \lya\ thus depends in a complex way on the distribution, clumpiness, and kinematics of \ion{H}{I} within the host galaxy, as well as on where and how the photon was produced initially. But the added complexity is counter-balanced by the availability of a great deal of information about the scattering medium itself that is otherwise difficult or impossible to observe directly: i.e., the neutral hydrogen distribution and kinematics in the CGM.

Since the commissioning of sensitive integral-field spectrometers on large ground-based telescopes  -- {the Multi Unit Spectroscopic Explorer (MUSE; \citealt{bacon10}) on the Very Large Telescopes (VLT) of the European Southern Observatory (ESO) and the Keck Cosmic Web Imager (KCWI; \citealt{morrissey18})} at the Keck Observatory -- it has become possible to routinely detect diffuse \lya\ emission halos around individual galaxies at high redshift {(e.g., \citealt{wisotzki16, leclercq17})}, and to simultaneously measure the spatially-resolved \lya\ kinematics {(e.g., \citealt{erb18, claeyssens19, leclercq20})}. Such observations can then be interpreted in terms of simple expectations based on \lya\ radiative transfer; for example, a generic expectation is that most \lya\ emission involving scattered photons (i.e., those that must pass through an \ion{H}{I} gas distribution before escaping their host) will exhibit a ``double-peaked'' spectral profile, where the relative strength of the blue-shifted and red-shifted peaks may be modulated by the net velocity field of the emitting gas.  In the idealised case of a spherical shell of outflowing (infalling) gas, one predicts that an external observer will measure a dominant red (blue) peak {\citep{verhamme06}}.  The fact that most ($\sim 90$\%) of star-forming galaxy ``down the barrel'' (DTB) spectra with \lya\ in emission in the central portions exhibit dominant red peaks (e.g., \citealt{pettini01, steidel10, kulas12, trainor15, verhamme18, matthee21}) has led to the conclusion that outflowing gas dominates \lya\ radiative transfer, at least at small galactocentric distances. 

Essentially every simulation of galaxy formation (\citealt{fg10}) predicts that gaseous accretion is also important -- particularly at high redshifts ($z \simgt 2$) -- and this has focused attention on systems in which a double \lya\ profile with a blue-dominant peak is observed, often cited as evidence for on-going accretion of cool gas (e.g., \citealt{vanzella17, martin14, martin16, ao20}).  Quantitative predictions of \lya\ emission from accreting baryons depend sensitively on the thermal state and the small-scale structure of the gas (e.g., \citealt{kollmeier10,fg10,goerdt10}), leading to large uncertainties in the predictions. The role played by ``local'' sources of ionising photons over and above that of the metagalactic ionising radiation field is likely to be substantial for regions near QSOs (\citealt{cantalupo14, borisova16, cai19, osullivan20}) but much more uncertain for star-forming galaxies, where the escape of scattered \lya\ photons is much more likely than that of ionising photons. 

Models of \lya\ radiative transfer have attempted to understand the dominant physics responsible for producing \lya\ halos around galaxies. Using photon-tracing algorithms with Monte-Carlo simulations, \citet{verhamme06, dijkstra14,gronke16a, gronke16b} have explored the effects of resonant scattering on the emergent central \lya\ line profile using various idealised \ion{H}{I} geometries and velocity fields;  {in most cases simple models can be made to fit the observed 1-D profiles \citep{gronke17, song20}. There have also been attempts to model or predict spatially-resolved \lya\ which almost certainly depends on a galaxy's immediate environment (\citealt{zheng11, kakiichi18}), including both outflows and accretion flows, as well as the radiative transfer of \lya\ photons from the site of initial production to escape \citep[e.g.][]{fg10, lake15, smith19, byrohl20}. However, the conclusions reached as to the dominant process responsible for the extended \lya\ emission have not converged, indicating that more realistic, high resolution, cosmological zoom-in simulations may be required to capture all of the physical processes.}  Despite the variety of \lya\ radiative transfer models to date, as far as we are aware, no specific effort has been made to statistically compare full 2-D model predictions (simultaneous spatial and kinematic) to observed \lya\ halos.

Some insight into the relationship between galaxy properties and the kinematics and spatial distribution of cool gas in the CGM has been provided by studies at lower redshifts, where galaxy morphology is more easily measured.  Statistical studies using absorption line probes have clearly shown that the strength of low-ionization metal lines such as \ion{Mg}{II}, \ion{Fe}{II} depends on where the line of sight passes through the galaxy CGM relative to the projected major axis of the galaxy -- the ``azimuthal angle'' -- \citep[e.g.][]{bordoloi11, bouche12, kacprzak12, nielsen15, lan18, martin19}, and the inclination of the galaxy disk relative to the line of sight (e.g., \citealt{steidel02, kacprzak11}). More recently, clear trends have also been observed for high ions (\ion{O}{VI}) \citep{kacprzak15}.
In general, these trends support a picture of star-forming galaxies in which high-velocity, collimated outflows perpendicular to the disk are responsible for the strongest absorption lines in both low and high ions, with low ions also being strong near the disk plane. Theoretically at least, accretion flows might also be quasi-collimated in the form of cold streams of gas that would tend to deposit cool gas near the disk plane (see, e.g., \citealt{tumlinson17}).  It is less clear how such a geometry for gas flows in the CGM would manifest as emission in a resonantly-scattered line like \lya. One might expect that \lya\ photons would escape most readily along the minor axis, since the large velocity gradients and lower \ion{H}{I} optical depths of outflowing material both favour \lya\ escape from the host galaxy. This picture is consistent with \citet{verhamme12}, who showed that \lya\ escape is enhanced when the simulated galaxies are viewed face-on. 

Observations of low-redshift, spatially resolved \lya\ emission have so far been limited to small samples -- e.g., in the local universe ($z<0.2$), using the Hubble Space Telescope (HST), the ``\lya\ reference sample'' (LARS; \citealt{ostlin14}) has obtained images probing \lya\ emission around galaxies in great spatial detail, reaffirming the complex nature of \lya\ radiative transfer and its relation to the host galaxies. In most cases LARS found evidence that extended \lya\ emission is most easily explained by photons produced by active star formation that then diffuse into the CGM before a last scattering event allows escape in the observer's direction. Although there are small-scale enhancements associated with outflows, even for galaxies observed edge-on the \lya\ emission is perhaps smoother than expected \citep{duval16}.

At $z > 2$, \lya\ emission is more readily observed but detailed analyses are challenged by the relatively small galaxy sizes (both physical and angular) and the need for high spatial resolution to determine the morphology of the stellar light. 
In this paper, we present a statistical sample of $z > 2$ galaxies drawn from a survey using KCWI of selected regions within the Keck Baryonic Structure Survey (KBSS; \citealt{rudie12a, steidel14, strom17}). 
Since the commissioning of KCWI in late 2017, we have obtained deep IFU data ($\sim 5$ hour integrations) for $> 100$ KBSS galaxies with $z = 2 - 3.5$, so that the \lya\ line is covered within the KCWI wavelength range; some initial results from the survey have been presented by \citet{erb18, law18}. The 59 galaxies included in our current analysis are those that, in addition to the KCWI data, have also been observed at high spatial resolution by either {HST} or adaptive-optics-assisted near-IR spectroscopy using Keck/OSIRIS. The overarching goal of the study is to evaluate the spatial and spectral distribution of \lya\ emission within $\simeq 5$ arcseconds as compared to the principle axes defined by the galaxy morphology on smaller angular scales by each galaxy's UV/optical continuum emission. In particular, we seek to use the observed kinematics and spatial distribution of \lya\ emission to evaluate whether the cool gas in the CGM of forming galaxies shows evidence for directional dependence -- e.g,, inflows or outflows along preferred directions -- with respect to the central galaxy.

This paper is organised as follows. In \S\ref{sec:sample}, we describe the KBSS-KCWI sample; \S\ref{sec:obs} introduces the high-resolution imaging and IFU dataset; \S\ref{sec:pa} covers the details on the measurement of the galaxy principle axes providing the definition of the galactic azimuthal angle; \S\ref{sec:analyses} presents results on the connection between Ly$\alpha$ halos and galactic azimuthal angle. \S\ref{sec:az_halo} looks into the connection between the \lya\ azimuthal asymmetry and the overall \lya\ emission properties. \S\ref{sec:three_bins} checks higher order azimuthal asymmetry of \lya\ emission by dividing the sample into finer azimuthal bins. \S\ref{sec:discussions} discusses the implications of the results, with a summary in \S\ref{sec:summary}. Throughout the paper, we assume a $\Lambda$CDM cosmology with $\Omega_m = 0.3$, $\Omega_\Lambda =0.7$, and $h=0.7$. Distances are given in proper units, i.e., physical kpc (pkpc).

\section{The KBSS-KCWI Galaxy Sample}

\label{sec:sample}

In late 2017, we began using the recently-commissioned Keck Cosmic Web Imager (KCWI; \citealt{morrissey18}) on the Keck \RNum{2} 10m telescope to target selected regions within the survey fields of the Keck Baryonic Structure Survey (KBSS; \citealt{rudie12a,steidel14,strom17}).  The main goal of the KCWI observations has been to detect diffuse emission from the CGM (within impact parameter, $D_{\rm tran} \simlt{} 100$ pkpc) of a substantial sample of rapidly star-forming galaxies and optically-faint AGN host galaxies, reaching surface brightness sensitivity of $\sim 5\times 10^{-20}$ ergs s$^{-1}$ cm$^{-2}$ arcsec$^{-2}$ ($1\sigma$) for unresolved emission lines.  Such limiting surface brightness would allow detection of the extended \lya\ halos of individual galaxies at redshifts $z \sim 2-3$ (e.g., \citealt{steidel11}), and would be capable of detecting extended diffuse UV metallic cooling line emission as predicted by simulations of galaxies with comparable mass and redshift (e.g., \citealt{sravan16}). 

KCWI offers three selectable image slicer scales, each of which can be used with three different regimes of spectral resolving power, $R \equiv \lambda/\Delta \lambda$. All of the observations used in the present study were obtained using the ``medium'' slicer scale and low resolution grating (BL), providing integral field spectra over a contiguous field of view (FoV) of 20\secpoint3$\times$16\secpoint5 covering the common wavelength range 3530-5530 \AA\ with resolving power $\langle R \rangle = 1800$.  

Given the relatively small solid angle of the KCWI FoV, and the total integration time desired for each pointing of $\sim 5$ hours, it was necessary to choose the KCWI pointings carefully. In general, we chose KCWI pointings to maximize the number of previously-identified KBSS galaxies with $2 \simlt z \simlt 3.4$ within the field of view, so that the KCWI spectra would include the \lya\ line as well as many other rest-frame far-UV transitions.  Most of the targeted galaxies within each pointing were observed as part of KBSS in both the optical (Keck/LRIS; \citealt{oke95,steidel04}) and the near-IR (Keck/MOSFIRE; \citealt{mclean12, steidel14}). The pointings were chosen so that the total sample of KBSS catalog galaxies observed would span the full range of galaxy properties represented in the KBSS survey in terms of stellar mass (M$_{\ast}$), star formation rate (SFR), \lya\ emission strength, and rest-optical nebular properties.  Most of the KCWI pointings were also directed within the regions of the KBSS survey fields that have been observed at high spatial resolution by {HST}.

As of this writing, the KBSS-KCWI survey comprises 39 pointings of KCWI, including observations of 101 KBSS galaxies, of which 91 have $2 \le z \le 3.4$ placing \lya\ within the KCWI wavelength range. {In this work, we focus only on galaxies without obvious spectroscopic or photometric evidence for the presence of an AGN, and have therefore excluded 14 objects with spectroscopic evidence for the presence of AGN, to be discussed elsewhere. The remaining objects show no sign of significant AGN activity -- e.g., they lack emission lines of high ionisation species in the rest-UV KCWI and existing LRIS spectra, their nebular line ratios in the rest-frame optical are consistent with stellar excitation based on spectra taken with Keck/MOSFIRE (see e.g. \citealt{steidel14}), they lack power-law SEDs in the NIR-MIR,  etc. Because measurements of galaxy morphology are important to the analysis, we considered only the subset of the star-forming (non-AGN) galaxies that have also been observed at high spatial resolution using HST imaging or Keck/OSIRIS IFU spectroscopcy behind adaptive optics (\S\ref{sec:image}).} 

In addition to the known KBSS targets, many of the KCWI pointings include continuum-detected serendipitous galaxies whose KCWI spectra constitute the first identification of their redshifts. A total of 50 galaxies with $z > 2$ (most of which are fainter in the optical continuum than the KBSS limit of {\cal R}$=25.5$) have been identified. We have included 10 such objects in our analysis sample, based on their having HST observations with sufficient S/N for determination of morphology. 

A minimum total integration of 2.5 hours (5 hours is more typical) at the position of a galaxy in the KCWI data cube was also imposed, in order to ensure the relative uniformity of the data set. Ultimately, after inspection of the high resolution images (\S\ref{sec:image}), 6 galaxies were removed because of source ambiguity or obvious contamination from nearby unrelated objects in the images, and two were excluded because they were not sufficiently resolved by HST to measure their position angle reliably (see \S\ref{sec:pa}). 

\begin{figure}
\includegraphics[width=8cm]{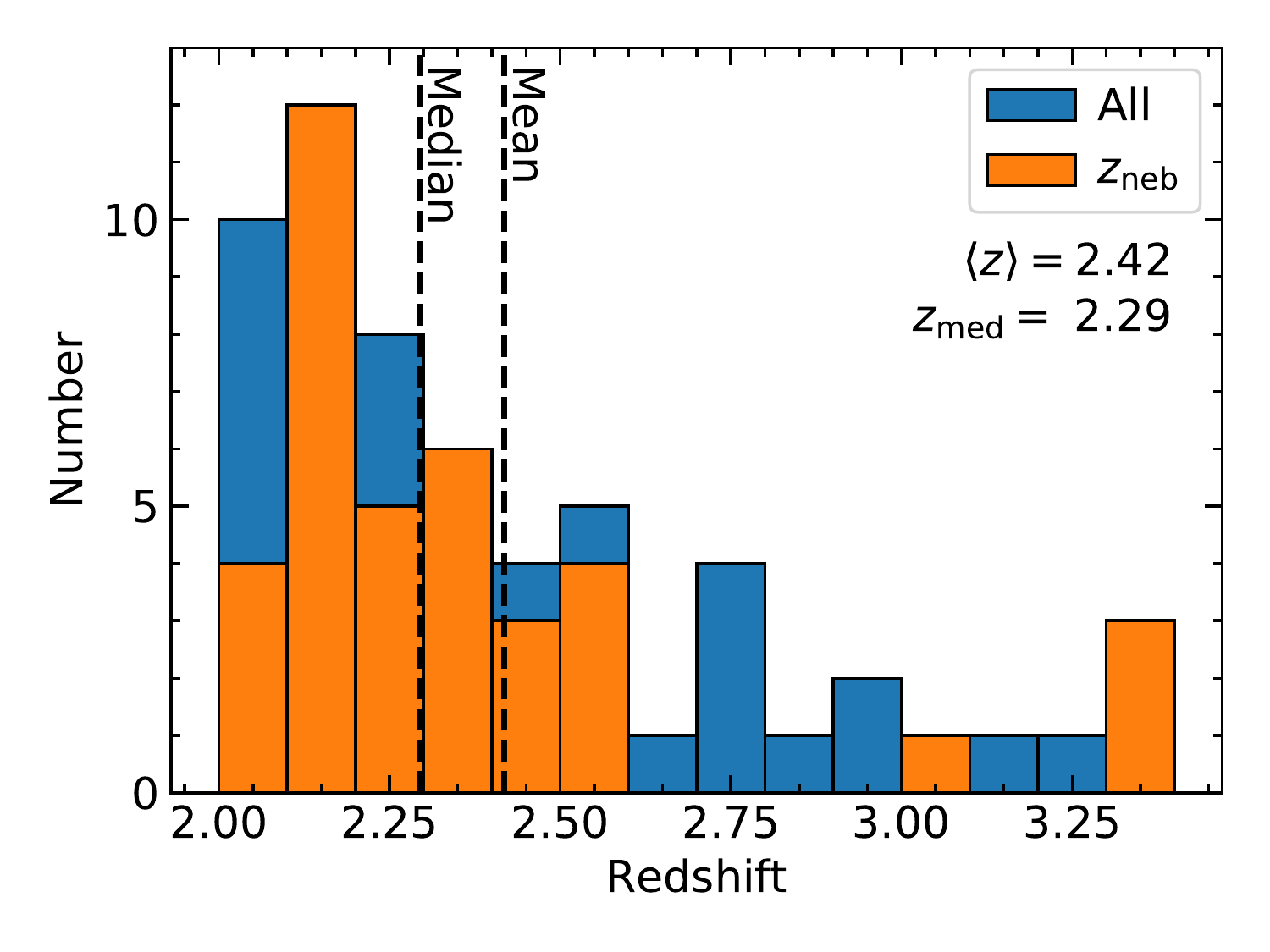}
\caption{   Redshift distribution of the galaxy sample. The blue histogram represents the full sample of 59 galaxies. The orange histogram shows the distribution for the 38 galaxies with nebular redshift measurements from MOSFIRE near-IR spectra, while the rest are calibrated based on \citet{chen20} using rest-UV absorption lines or \lya\ emission from Keck/LRIS and/or KCWI spectra. The mean (median) redshift of the full sample is 2.42 (2.29). } 
\label{fig:zdistribution}
\end{figure}

The final sample to be considered in this work contains 59 galaxies, listed in Table~\ref{tab:sample}. The redshifts given in Table~\ref{tab:sample} are based on MOSFIRE nebular spectra for 38 of the 59 galaxies, which have a precision of $\sim \pm 20$ \kms\ and should accurately reflect the galaxy systemic redshift. {In the remaining cases, features in the rest-frame UV spectra including \lya\ emission and strong interstellar absorption features (e.g., \ion{Si}{II}, \ion{Si}{IV}, \ion{C}{II}, \ion{C}{IV}, \ion{O}{I}) were used to estimate the systemic redshift of the galaxy, using the calibration described by \citet{chen20}. Briefly, the method uses the statistics -- based on several hundred KBSS galaxies with both nebular and UV observations -- of the velocity offsets between nebular redshifts and redshifts defined by UV spectral features for samples divided by their UV spectral morphology, i.e., \lya\ emission only, \lya\ emission and interstellar absorption, or interstellar absorption only. The mean offsets for the appropriate sub-sample were applied to the UV-based redshifts in cases where nebular redshifts are not available;  systemic redshifts obtained using such calibrations have an uncertainty of $\simeq 100$ \kms\  when the rest-UV spectra are of high quality (see, e.g., \citealt{steidel18}.)   Figure~\ref{fig:zdistribution} shows the redshift distribution of the KCWI sample, which has $z_\mathrm{med} =2.29 \pm 0.40$ (median and standard deviation), for which the conversion between angular and physical scales is 8.21 pkpc/" with our assumed cosmology. }

\begin{figure}
\includegraphics[width=8cm]{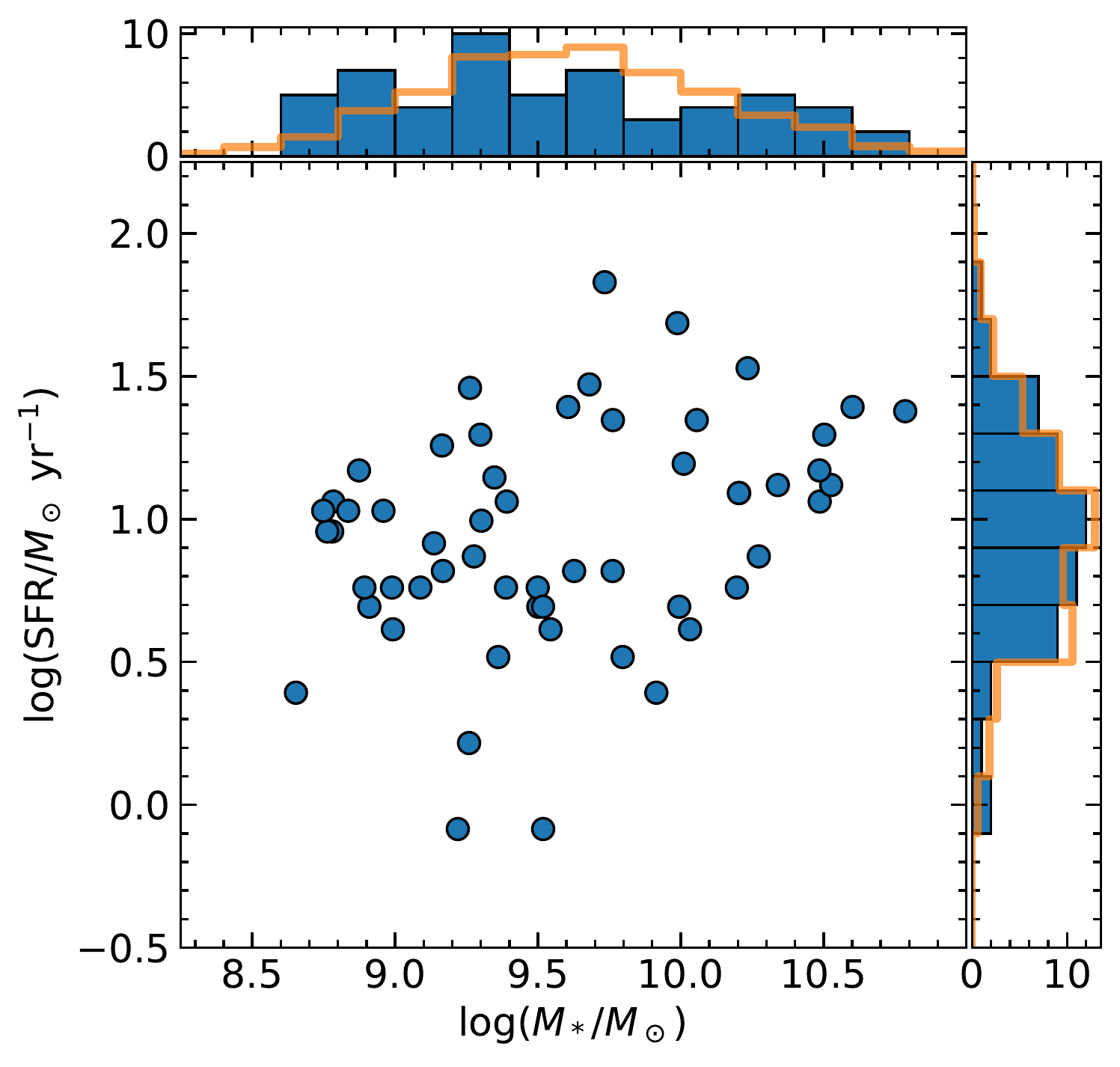}
\caption{ Distribution of SFR and $M_*$ of 56/59 galaxies in this sample; the remaining 3 galaxies have insufficient photometric measurements for reliable SED fitting. The normalised distributions of the parent KBSS sample are shown in the orange 1-D histograms. The SFR and $M_*$ of galaxies used in this work are similar to those of the parent KBSS sample; the values are all based on the BPASS-v2.2-binary spectral synthesis models \citep{stanway18}, assuming stellar metallicity $Z=0.002$, SMC extinction as described by \citet{theios19}, and  a \citet{chabrier03} stellar initial mass function.} 
\label{fig:mstar_sfr}
\end{figure}

{Reliable SED fits are available for 56 of the 59 galaxies using the BPASSv2.2 stellar population synthesis model \citep{stanway18} and SMC extinction curve. This choice of SED model has been shown to predict internally consistent stellar mass ($M_{\ast}$) and star-formation rate (SFR) for high-redshift galaxies having properties similar to those in our sample (see, e.g., \citealt{steidel16, strom17, theios19}), i.e., $8.5 \simlt  {\rm log}~(M_{\ast}/M_{\odot})\simlt 11$, and ${\rm 1 \simlt SFR/(M_{\odot}~yr^{-1}) \simlt 100}$. The distributions of $M_*$ and SFR (Figure \ref{fig:mstar_sfr}) are similar to those of the full KBSS galaxy sample, albeit with a slight over-representation of ${\rm log}~(M_{\ast}/M_{\odot}) \simlt 9$ galaxies
\footnote{For  direct comparison of our sample with the so-called ``star-formation main-sequence''  (SFMS), we note that SED fits that assume  solar metallicity SPS models from \citet{bruzual03} and attenuation curve from \citet{calzetti00} result in a distribution of SFR vs. $M_{\ast}$ entirely consistent  with the published SFMS at $z \sim 2$ (e.g., \citealt{whitaker14}).}}.

\begin{figure}
\includegraphics[width=8cm]{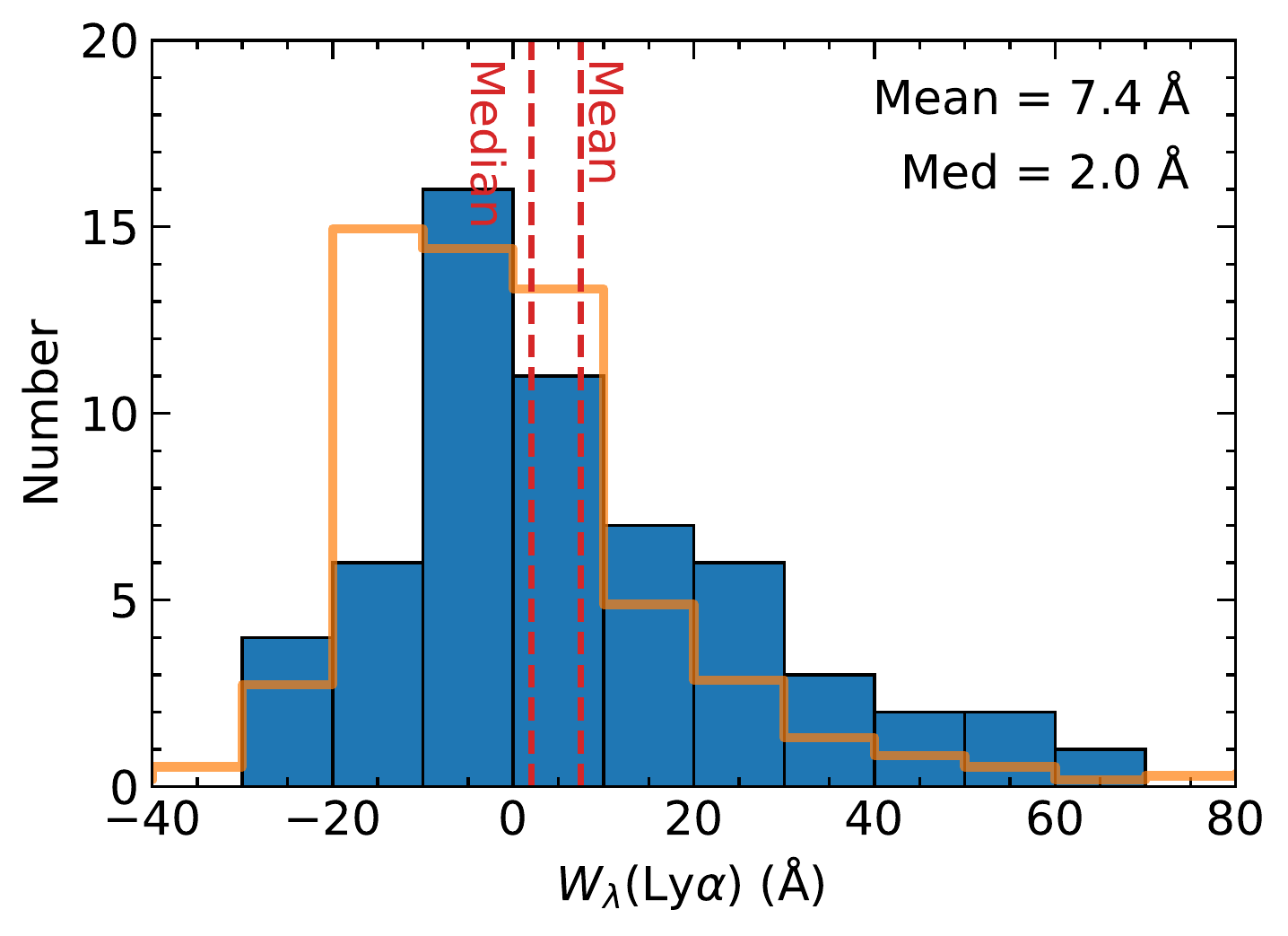}
\caption{ Distribution of \wlya\ for  the sample (blue histogram). The orange skeletal histogram shows the normalised \wlya\ distribution from \citet{reddy09}, which is a subset of the current KBSS sample large enough to be representative.  The sample discussed in this work is slightly biased toward  Ly$\alpha$-emitting galaxies compared to the parent sample of $z \sim 2-3$ KBSS galaxies.  } 
\label{fig:ew_distribution}
\end{figure} 
In order to facilitate comparison of the \lya\ emission line strength of sample galaxies with those in the literature,  Table~\ref{tab:sample} includes the rest-frame \lya\ equivalent width (\wlya) based on extraction of 1-D spectra from the KCWI data cubes over a spatial aperture defined by the extent of the UV continuum light of each galaxy\footnote{In general, the \lya\ emission evaluated within a central aperture represents only a fraction of the total that would be measured in a large aperture that accounts for the diffuse \lya\ halos with spatial extent well beyond that of the FUV continuum (see, e.g., \citealt{steidel11,wisotzki16}); however, the central \wlya\ is a closer approximation to values measured in most spectroscopic galaxy surveys.}.  The values of \wlya\ in Table~\ref{tab:sample} were measured using the method described in \citet{kornei10} (see also \citealt{reddy09}), where positive values indicate net emission and negative values net absorption. Aside from a slight over-representation of galaxies with the strongest \lya\ emission (\wlya$\simgt 40$ \AA), the sample in Table~\ref{tab:sample} is otherwise typical of UV-continuum-selected galaxies in KBSS,  by construction. The \wlya\ distribution for the sample in Table~\ref{tab:sample} is shown in Figure~\ref{fig:ew_distribution}. In addition to \wlya, we also measured the total \lya\ flux (\flya) and the ratio between the blue- and red-shifted components of  emission for the entire \lya\ halo ($\fblue / \fred$), which are discussed further in \S\ref{sec:az_halo}.

\begin{sidewaystable*}
\centering
\caption{Summary of the galaxy sample and the observations. \label{tab:sample}}
\begin{threeparttable}
\begin{tabular}{lccclrrrcrc}
\hline\hline
Object\tnote{a} & RA & DEC & $t_\mathrm{exp}$ (KCWI) & \multicolumn{1}{c}{Redshift\tnote{b}} & \multicolumn{1}{c}{\wlya\tnote{c}} & \multicolumn{1}{c}{$\flya(\mathrm{tot})$\tnote{d}} & \multirow{2}{*}{$\frac{\fblue}{\fred}$\tnote{e}} & Imaging\tnote{f} & \multicolumn{1}{c}{$\pa_0$\tnote{g}} & $\pa_0$\tnote{h} \\
Identifier & (J2000.0) & (J2000.0) & (hr) && \multicolumn{1}{c}{(\AA)} & \multicolumn{1}{c}{$10^{-17}~\mathrm{erg~s}^{-1}\mathrm{cm}^{-2}$} && Data & \multicolumn{1}{c}{(deg)} & Method \\
\hline
Q0100-BX210     & 01:03:12.02 & +13:16:18.5 & 5.4 & 2.2769      & $-3 \pm 2$ & $6.3 \pm 0.5$  & $0.08 \pm 0.05$  & F160W & 37 & (\rnum{1}), (\rnum{2}) \\
Q0100-BX212     & 01:03:12.51 & +13:16:23.2 & 5.2 & 2.1063      & $-4 \pm 3$ & $2.3 \pm 0.7$  & $0.5 \pm 0.29$   & F160W & -57 & (\rnum{1}), (\rnum{2}) \\
Q0100-C7        & 01:03:08.24 & +13:16:30.1 & 5.1 & 3.0408      & $12 \pm 1$ & $9.0 \pm 0.4$  & $0.15 \pm 0.03$  & F160W & 16 & (\rnum{1}), (\rnum{2}) \\
Q0100-D11       & 01:03:08.25 & +13:16:37.6 & 5.1 & 2.5865      & $6 \pm 1$  & $1.7 \pm 0.4$  & $0.11 \pm 0.14$  & F160W & -12 & (\rnum{1}), (\rnum{2}) \\
Q0142-BX165     & 01:45:16.87 & -09:46:03.5 & 5.0 & 2.3576      & $51 \pm 3$ & $37.5 \pm 0.5$ & $0.28 \pm 0.01$  & F160W & 14 & (\rnum{2}) \\
Q0142-BX188     & 01:45:17.79 & -09:45:05.6 & 5.3 & 2.0602      & $-6 \pm 1$ & $4.6 \pm 0.7$  & $0.47 \pm 0.15$  & F160W & 66 & (\rnum{1}), (\rnum{2}) \\
Q0142-BX195-CS10 & 01:45:17.11 & -09:45:06.0 & 4.5 & 2.7382 (UV) & $-1 \pm 3$ & $2.2 \pm 0.5$  & $0.47 \pm 0.24$  & F160W & 29 & (\rnum{2}) \\
Q0142-NB5859    & 01:45:17.54 & -09:45:01.2 & 5.8 & 2.7399 (UV) & $20 \pm 2$ & $3.7 \pm 0.4$  & $0.65 \pm 0.14$  & F160W & 42 & (\rnum{1}), (\rnum{2}) \\
Q0207-BX144     & 02:09:49.21 & -00:05:31.7 & 5.0 & 2.1682      & $25 \pm 3$ & $44.9 \pm 0.6$ & $0.29 \pm 0.01$  & F140W & -74 & (\rnum{1}), (\rnum{2}) \\
Q0207-MD60      & 02:09:53.69 & -00:04:39.8 & 4.3 & 2.5904      & $-27 \pm 2$ & $1.4 \pm 0.5$  & $-0.16 \pm 0.21$ & F140W & -64 & (\rnum{2}) \\
Q0449-BX88      & 04:52:14.94 & -16:40:49.3 & 6.3 & 2.0086      & $-7 \pm 3$ & $3.3 \pm 0.8$  & $0.16 \pm 0.16$  & F160W & 10 & (\rnum{2}) \\
Q0449-BX88-CS8  & 04:52:14.79 & -16:40:58.6 & 5.2 & 2.0957 (UV) & $4 \pm 1$ & $3.9 \pm 0.8$  & $0.34 \pm 0.17$  & F160W & -39 & (\rnum{1}), (\rnum{2}) \\
Q0449-BX89      & 04:52:14.80 & -16:40:51.1 & 6.3 & 2.2570 (UV) & $22 \pm 2$ & $7.6 \pm 0.4$  & $0.05 \pm 0.03$  & F160W & -32 & (\rnum{1}), (\rnum{2}) \\
Q0449-BX93      & 04:52:15.41 & -16:40:56.8 & 6.3 & 2.0070      & $-7 \pm 2$ & $7.4 \pm 0.8$  & $0.6 \pm 0.13$   & F160W, OSIRIS & -18 & (\rnum{1}), (\rnum{2}), (\rnum{3}) \\
Q0449-BX110     & 04:52:17.20 & -16:39:40.6 & 5.0 & 2.3355      & $35 \pm 2$ & $20.4 \pm 0.6$ & $0.47 \pm 0.03$  & F160W & -48 & (\rnum{1}), (\rnum{2}) \\
Q0821-MD36      & 08:21:11.41 & +31:08:29.4 & 2.7 & 2.583       & $75 \pm 10$ & $23.1 \pm 0.6$ & $0.16 \pm 0.02$  & F140W & 37 & (\rnum{1}), (\rnum{2}) \\
Q0821-MD40      & 08:21:06.96 & +31:07:22.8 & 4.3 & 3.3248      & $27 \pm 3$ & $11.0 \pm 0.4$ & $0.34 \pm 0.03$  & F140W & -7 & (\rnum{2}) \\
Q1009-BX215     & 10:11:58.71 & +29:41:55.9 & 4.0 & 2.5059      & $1 \pm 1$ &  $3.7 \pm 0.8$  & $0.34 \pm 0.15$  & F160W & -20 & (\rnum{1}), (\rnum{2}) \\
Q1009-BX218     & 10:11:58.96 & +29:42:07.5 & 5.3 & 2.1091      & $-6 \pm 3$ & $4.0 \pm 0.6$  & $0.18 \pm 0.11$  & F160W & -43 & (\rnum{1}), (\rnum{2}) \\
Q1009-BX222     & 10:11:59.09 & +29:42:00.5 & 5.3 & 2.2031      & $-4 \pm 1$ & $4.7 \pm 0.5$  & $0.28 \pm 0.08$  & F160W & -83 & (\rnum{1}), (\rnum{2}) \\
Q1009-BX222-CS9 & 10:11:58.92 & +29:42:02.6 & 5.3 & 2.6527 (UV) & $-1 \pm 4$ & $1.0 \pm 0.4$  & \multicolumn{1}{c}{---}  & F160W & 76 & (\rnum{1}), (\rnum{2}) \\
Q1009-D15       & 10:11:58.73 & +29:42:10.5 & 5.3 & 3.1028 (UV) & $-18 \pm 6$ & $4.2 \pm 0.4$  & $0.28 \pm 0.08$  & F160W & 28 & (\rnum{1}), (\rnum{2}) \\
Q1549-BX102     & 15:51:55.98 & +19:12:44.2 & 5.0 & 2.1934      & $50 \pm 3$ &  $19.5 \pm 0.5$ & $0.51 \pm 0.03$  & F606W & -87 & (\rnum{1}), (\rnum{2}) \\
Q1549-M17       & 15:51:56.06 & +19:12:52.7 & 3.3 & 3.2212 (UV) & $27 \pm 5$ & $4.3 \pm 0.5$  & $0.0 \pm 0.05$   & F606W & 72 & (\rnum{2}) \\
Q1623-BX432     & 16:25:48.74 & +26:46:47.1 & 3.6 & 2.1825      & $17 \pm 1$ & $10.8 \pm 0.7$ & $0.46 \pm 0.06$  & F160W & 16 & (\rnum{2}) \\
Q1623-BX436     & 16:25:49.10 & +26:46:53.4 & 3.6 & 2.0515 (UV) & $-12 \pm 2$ & $2.6 \pm 1.0$  & \multicolumn{1}{c}{---} & F160W & 12 & (\rnum{2}) \\
Q1623-BX453     & 16:25:50.85 & +26:49:31.2 & 4.8 & 2.1821      & $10 \pm 2$ & $2.8 \pm 0.5$  & $0.33 \pm 0.14$  & F160W, OSIRIS & 31 & (\rnum{3}) \\
Q1623-BX453-CS3 & 16:25:50.35 & +26:49:37.1 & 4.7 & 2.0244 (UV) & $17 \pm 2$ & $5.6 \pm 0.9$  & $0.3 \pm 0.11$   & F160W & -66 & (\rnum{1}), (\rnum{2}) \\
Q1623-C52       & 16:25:51.20 & +26:49:26.3 & 4.8 & 2.9700 (UV) & $4 \pm 1$ & $8.5 \pm 0.4$  & $0.22 \pm 0.03$  & F160W & -13 & (\rnum{2}) \\
Q1700-BX490     & 17:01:14.83 & +64:09:51.7 & 4.3 & 2.3958      & $-3 \pm 4$ & $12.8 \pm 0.5$ & $0.39 \pm 0.03$  & F814W, OSIRIS & 86 & (\rnum{1}), (\rnum{2}), (\rnum{3}) \\
Q1700-BX561     & 17:01:04.18 & +64:10:43.8 & 5.0 & 2.4328      & $-3 \pm 3$ & $8.5 \pm 0.6$  & $0.51 \pm 0.08$  & F814W & 9 & (\rnum{2}) \\
Q1700-BX575     & 17:01:03.34 & +64:10:50.9 & 5.0 & 2.4334      & $0 \pm 2$ & $5.2 \pm 0.6$  & $0.14 \pm 0.08$  & F814W & -34 & (\rnum{2}) \\
Q1700-BX581     & 17:01:02.73 & +64:10:51.3 & 4.7 & 2.4022      & $9 \pm 4$ & $10.6 \pm 0.7$ & $0.28 \pm 0.05$  & F814W & 27 & (\rnum{1}), (\rnum{2}) \\
Q1700-BX710     & 17:01:22.13 & +64:12:19.3 & 5.0 & 2.2946      & $-10 \pm 3$ & $8.1 \pm 0.6$  & $0.51 \pm 0.09$  & F814W & -19 & (\rnum{1}), (\rnum{2}) \\
Q1700-BX729     & 17:01:27.77 & +64:12:29.5 & 4.7 & 2.3993      & $14 \pm 2$ & $12.9 \pm 0.5$ & $0.17 \pm 0.03$  & F814W & -3 & (\rnum{1}), (\rnum{2}) \\
Q1700-BX729-CS4 & 17:01:28.95 & +64:12:32.4 & 3.0 & 2.2921 (UV) & $14 \pm 3$ & $5.8 \pm 1.1$  & $0.21 \pm 0.12$  & F814W & 40 & (\rnum{2}) \\
Q1700-BX729-CS9 & 17:01:27.49 & +64:12:25.1 & 4.7 & 2.4014 (UV) & $35 \pm 2$ & $1.1 \pm 0.5$  & $-0.19 \pm 0.21$ & F814W & -59 & (\rnum{1}), (\rnum{2}) \\
Q1700-MD103     & 17:01:00.21 & +64:11:55.6 & 5.0 & 2.3151      & $-24 \pm 1$ & $-0.5 \pm 0.5$ & \multicolumn{1}{c}{---} & F814W & 55 & (\rnum{2}) \\
Q1700-MD104     & 17:01:00.67 & +64:11:58.3 & 5.0 & 2.7465 (UV) & $6 \pm 2$ & $8.2 \pm 0.4$  & $0.11 \pm 0.03$  & F814W & 7 & (\rnum{1}), (\rnum{2}) \\
Q1700-MD115     & 17:01:26.68 & +64:12:31.7 & 4.7 & 2.9081 (UV) & $33 \pm 7$ & $7.8 \pm 0.5$  & $0.05 \pm 0.03$  & F814W & 3 & (\rnum{1}), (\rnum{2}) \\
Q2206-MD10      & 22:08:52.21 & -19:44:13.9 & 5.0 & 3.3269      & $5 \pm 2$ & $4.0 \pm 0.5$  & $0.07 \pm 0.08$  & F160W & 41 & (\rnum{1}), (\rnum{2}) \\
\hline
\end{tabular}
\end{threeparttable}
\end{sidewaystable*}

\setcounter{table}{0}
\begin{sidewaystable*}
\centering
\caption{----\textit{continued.}}
\begin{threeparttable}
\begin{tabular}{lccclrrrcrc}
\hline\hline
Object\tnote{a} & RA & DEC & $t_\mathrm{exp}$ (KCWI) & \multicolumn{1}{c}{Redshift\tnote{b}} & \multicolumn{1}{c}{\wlya\tnote{c}} & \multicolumn{1}{c}{$\flya(\mathrm{tot})$\tnote{d}} & \multirow{2}{*}{$\frac{\fblue}{\fred}$\tnote{e}} & Imaging\tnote{f} & \multicolumn{1}{c}{$\pa_0$\tnote{g}} & $\pa_0$\tnote{h} \\
Identifier & (J2000.0) & (J2000.0) & (hr) && \multicolumn{1}{c}{(\AA)} & \multicolumn{1}{c}{$10^{-17}~\mathrm{erg~s}^{-1}\mathrm{cm}^{-2}$} && Data & \multicolumn{1}{c}{(deg)} & Method \\
\hline
DSF2237b-MD38   & 22:39:35:64 & +11:50:27.5 & 3.7 & 3.3258      & $2 \pm 2$ & $3.0 \pm 0.5$  & $0.44 \pm 0.16$  & F606W & 52 & (\rnum{1}), (\rnum{2}) \\
Q2343-BX379     & 23:46:28.96 & +12:47:26.0 & 4.0 & 2.0427 (UV) & $-7 \pm 2$ & $6.4 \pm 1.3$  & $0.64 \pm 0.23$  & F140W & -60 & (\rnum{1}), (\rnum{2}) \\
Q2343-BX389     & 23:46:28.90 & +12:47:33.5 & 4.0 & 2.1712      & $-16 \pm 1$ & $1.6 \pm 0.6$  & $0.65 \pm 0.49$  & F140W & -50 & (\rnum{2}) \\
Q2343-BX391     & 23:46:28.07 & +12:47:31.8 & 4.0 & 2.1738      & $-16 \pm 1$ & $3.3 \pm 0.7$  & \multicolumn{1}{c}{---} & F140W & 21 & (\rnum{1}), (\rnum{2}) \\
Q2343-BX417     & 23:46:26.27 & +12:47:46.7 & 3.0 & 2.2231 (UV) & $-9 \pm 6$ & $4.8 \pm 0.7$  & $0.49 \pm 0.15$  & F160W & -38 & (\rnum{2}) \\
Q2343-BX418     & 23:46:18.57 & +12:47:47.4 & 4.9 & 2.3054      & $46 \pm 3$ & $40.5 \pm 0.6$ & $0.46 \pm 0.01$  & F140W, OSIRIS & 2 & (\rnum{1}), (\rnum{2}), (\rnum{3}) \\
Q2343-BX418-CS8 & 23:46:18.73 & +12:47:51.6 & 5.1 & 2.7234 (UV) & $66 \pm 15$ & $6.7 \pm 0.4$  & $0.15 \pm 0.04$  & F140W & -67 & (\rnum{1}), (\rnum{2}) \\
Q2343-BX429     & 23:46:25.26 & +12:47:51.2 & 5.2 & 2.1751      & $-8 \pm 5$ & $3.0 \pm 0.5$  & $0.31 \pm 0.15$  & F160W & 28 & (\rnum{1}), (\rnum{2}) \\
Q2343-BX442     & 23:46:19.36 & +12:47:59.7 & 4.6 & 2.1754      & $-18 \pm 4$ & $1.6 \pm 0.7$  & $0.38 \pm 0.52$  & OSIRIS & -12 & （\rnum{4}) \\
Q2343-BX513     & 23:46:11.13 & +12:48:32.1 & 4.8 & 2.1082      & $10 \pm 1$ & $18.6 \pm 0.7$ & $0.71 \pm 0.05$  & F140W, OSIRIS & -9 & (\rnum{1}), (\rnum{2}), (\rnum{3}) \\
Q2343-BX513-CS7 & 23:46:10.55 & +12:48:30.9 & 4.8 & 2.0144 (UV) & $-22 \pm 2$ & $4.2 \pm 1.0$  & $0.53 \pm 0.26$  & F140W & 71 & (\rnum{1}), (\rnum{2}) \\
Q2343-BX587     & 23:46:29.17 & +12:49:03.4 & 5.2 & 2.2427      & $-4 \pm 3$ & $7.2 \pm 0.5$  & $0.22 \pm 0.05$  & F160W & 44 & (\rnum{1}), (\rnum{2}) \\
Q2343-BX587-CS3 & 23:46:28.24 & +12:49:07.2 & 4.8 & 2.5727 (UV) & $2 \pm 2$ & $1.3 \pm 0.4$  & $0.17 \pm 0.26$  & F140W & -53 & (\rnum{1}), (\rnum{2}) \\
Q2343-BX587-CS4 & 23:46:28.62 & +12:49:04.8 & 5.2 & 2.8902 (UV) & $-13 \pm 1$ & $2.8 \pm 0.4$  & $0.08 \pm 0.08$  & F140W & -59 & (\rnum{1}), (\rnum{2}) \\
Q2343-BX610     & 23:46:09.43 & +12:49:19.2 & 5.3 & 2.2096      & $8 \pm 2$ & $13.5 \pm 0.7$ & $0.26 \pm 0.04$  & F140W & 22 & (\rnum{2}) \\
Q2343-BX660     & 23:46:29.43 & +12:49:45.6 & 5.0 & 2.1742      & $20 \pm 3$ & $25.7 \pm 0.6$ & $0.26 \pm 0.02$  & F140W, OSIRIS & 37 & (\rnum{1}), (\rnum{2}), (\rnum{3}) \\
Q2343-BX660-CS7 & 23:46:29.82 & +12:49:38.7 & 4.2 & 2.0788 (UV) & $41 \pm 1$ & $2.3 \pm 0.9$  & \multicolumn{1}{c}{---} & F140W & 89 & (\rnum{2}) \\
Q2343-MD80      & 23:46:10.80 & +12:48:33.2 & 4.8 & 2.0127      & $-26 \pm 3$ & $2.8 \pm 0.8$  & $0.22 \pm 0.24$  & F140W & 9 & (\rnum{1}), (\rnum{2}) \\
\hline\addlinespace[1ex]
\end{tabular}
\begin{tablenotes}\footnotesize
\item[a]{The ``CS'' objects are continuum serendipitous objects discussed in \S\ref{sec:sample}. Their naming follows nearby KBSS galaxies that are previously known, and may not be physically associated to the CS objects.}
\item[b]{If marked as ``UV'', the systemic redshift was estimated from features in the rest-UV spectra, calibrated as described by \citet{chen20}. The typical uncertainties on UV-estimated systemic redshifts are $\delta v \equiv c\delta z_{\textrm sys}/(1+z_{\rm sys})\simeq 100$ \kms.  Otherwise, $z_{\textrm sys}$ was measured from nebular emission lines in rest-optical (MOSFIRE) spectra, with $\delta v \simeq 20$ \kms. }
\item[c]{{Rest-frame \lya\ equivalent width. Details discussed in \S\ref{sec:sample}. }}
\item[d]{{Total \lya\ flux. See \S\ref{sec:az_halo} for more details. }}
\item[d]{{Flux ratio between the blueshifted and redshifted components of \lya\ emission. Details in \S\ref{sec:az_halo}}.}
\item[f]{F140W and F160W images obtained using HST-WFC3-IR, F606W and F814W images from HST-ACS. }
\item[g]{Typical uncertainty: $\pm 10^\circ$. }
\item[h]{Methods used to measure $\pa_0$: (\rnum{1}) using GALFIT on HST images; (\rnum{2}) using the pixel intensity second moment on HST images; (\rnum{3}) pixel intensity second moment on OSIRIS H$\alpha$ map; } (\rnum{4}) kinematics of H$\alpha$ emission. Details on the methods are shown in \S\ref{sec:pa_methods}. 
\end{tablenotes}
\end{threeparttable}
\end{sidewaystable*}

\section{Observations and Reductions} 
\label{sec:obs}
\subsection{KCWI}
\label{sec:kcwi}
The KCWI data discussed in the present work were obtained between 2017 September and 2020 November, in all cases using the medium-scale slicer made up of 24 slices of width 0\secpoint69 and length 20\secpoint3 on the sky. The instrumental setup uses the BL volume phase holographic (VPH) grating with an angle of incidence that optimises the diffraction efficiency near 4200 \AA, with the camera articulation angle set to record the spectra of each slice with a central wavelength of $\sim 4500$ \AA. A band-limiting filter was used to suppress wavelengths outside of the range 3500-5600 \AA\ to reduce scattered light; the useful common wavelength range recorded for all 24 slices in this mode is 3530-5530 \AA, with a spectral resolving power ranging from $R \simeq 1400$ at 3530 \AA\ to $R \simeq 2200$  at 5530 \AA. At the mean redshift of the sample ($\langle z \rangle = 2.42$), \lya\ falls at an observed wavelength of $\sim 4160$ \AA, where $R\sim 1650$.

The E2V 4k$\times$4k detector was binned 2$\times$2, which provides spatial sampling along slices of 0\secpoint29 pix$^{-1}$. Because each slice samples 0\secpoint68 in the dispersion direction, the effective spatial resolution element is rectangular on the sky, with an aspect ratio of $\sim 2.3:1$. We adopted the following approach to the observations, designed to ensure that the slicer geometry with respect to the sky is unique on each 1200s exposure so that the effective spatial resolution on the final stacked data cube is close to isotropic. Typically, a total integration of $\sim 5$ hours is obtained as a sequence of 15 exposures of 1200s, each obtained with the sky {position angle (PA)} of the instrument rotated by 10-90 degrees with respect to adjacent exposures. Each rotation of the instrument field of view is accompanied by a small offset of the telescope pointing before the guide star is reacquired. In this way, a given sky position is sampled in 15 different ways by the slicer.

\subsubsection{KCWI Data Reduction}
\label{sec:kcwi_reduction}
Each 1200s exposure with KCWI was initially reduced using the data reduction pipeline (DRP) maintained by the instrument team and available via the
Keck Observatory website\footnote{\href{https://github.com/Keck-DataReductionPipelines/KcwiDRP}{https://github.com/Keck-DataReductionPipelines/KcwiDRP}}. The DRP assembles the 2D spectra of all slices into a 3D data cube (with spaxels of 0\secpoint29$\times$0\secpoint69, the native scale) using a suite of procedures that can be customised to suit particular applications. The procedures include cosmic-ray removal, overscan subtraction and scattered light subtraction, wavelength calibration, flat-fielding, sky-subtraction, differential atmospheric refraction (DAR) correction, and flux-calibration. Wavelength calibration was achieved using ThAr arc spectra obtained using the internal calibration system during the afternoon prior to each observing night. 
Flat-fielding was accomplished using spectra of the twilight sky at the beginning or end of each night, after dividing by a b-spline model of the solar spectrum calculated using the information from all slices.  For each frame, the sky background was subtracted using the sky-modeling feature in the DRP, after which the sky-subtracted image (still in the 2-D format) is examined in order to mask pixels, in all 24 slices, that contain significant light from sources in the field. The frame was then used to make a new 2-D sky model using only unmasked pixels, and the sky-subtracted image is reassembled into a wavelength-calibrated (rebinned to 1 \AA\ per wavelength bin) data cube, at which time a variance cube, an exposure cube, and a mask cube are also produced.  

Next, we removed any remaining low frequency residuals from imperfect sky background subtraction by forming a median-filtered cube after masking obvious continuum and extended emission line sources using a running 3D boxcar filter. The typical dimensions of the filter are 100 \AA\ (100 pixels) in the wavelength direction, 16 pixels (4\secpoint6) along slices, and 1 pixel (0\secpoint69) perpendicular to the slices, with the last ensuring slice-to-slice independence. Minor adjustments to the filter dimensions were made as needed. If the running boxcar encounters a large region with too few unmasked pixels to allow a reliable median determination, then the pixel values in
the filtered cube were interpolated from the nearest adjacent regions for which the median was well-determined.  Finally, the median-filtered cube for each observed frame was subtracted from the data. We found that this method proved effective for removing scattered light along slices caused by bright objects within the KCWI field of view.

Because neither Keck \RNum{2} nor
KCWI has an atmospheric dispersion corrector, each cube is corrected for differential atmospheric refraction (DAR) (i.e., apparent position of an object as a function of wavelength) using the elevation and parallactic angle at the midpoint of the exposure and a model of the atmosphere above Maunakea. Finally, each cube was flux-calibrated using observations obtained with the same instrument configuration of one or more spectrophotometric standard stars selected from a list recommended by the KCWI documentation.

Prior to stacking reduced data cubes of individual exposures covering the same sky region, they must be aligned spatially and rotated to account for differences in the PA of the instrument with respect to the sky on different exposures. To accomplish this, we averaged the DAR-corrected cubes along the wavelength axis to create pseudo-white-light images, rotated each to the nominal sky orientation (N up and E left) based on the World Coordinate System (WCS) recorded in the header, and cross-correlated the results to determine the relative offsets in RA and Dec, which we found to have a precision of $\simeq 0.03$-arcsec ({root mean square,} RMS). The offsets were applied to the WCS in the header of each cube, and all cubes for a given pointing were then resampled to a common spatial grid with spatial sampling of $0\secpoint3 \times 0\secpoint3$ using a 2D \textit{drizzle} algorithm in the \textit{Montage} package\footnote{\href{http://montage.ipac.caltech.edu/}{http://montage.ipac.caltech.edu/}}, with drizzle factor of 0.7. If the wavelength grid is different among cubes, the spectrum in each spaxel was resampled to a common grid using cubic spline. Finally, we stacked the resampled cubes by averaging, weighted by exposure time. A white light image of the final stacked data cube was used to determine small corrections to the fiducial RA and Dec to align with other multiwavelength data of the same region.

\subsection{High Spatial Resolution Imaging}
\label{sec:image}

For galaxies in the mass range of our sample at $z\sim 2.5$, the typical half-light diameter is $\sim 4$ pkpc, which corresponds to $\simeq 0\secpoint5$ \citep{law12c}. To obtain reliable measurements of the orientation of the galaxy's projected major and minor axes (see \S\ref{sec:pa}), high resolution images are crucial. We gathered existing data from two sources: space-based optical or near-IR images from the HST/ACS or HST/WFC3; alternatively, H$\alpha$ maps obtained using the Keck/OSIRIS integral field spectrometer (\citealt{larkin06}) behind the WMKO laser guide star adaptive optics (LGSAO) system, which typically provides spatial resolution of $0\secpoint11-0\secpoint15$ (\citealt{law07,law09,law12b}).

\begin{table}
\caption{Summary of HST Observations  \label{tab:hst}}
\begin{threeparttable}	
\centering
\begin{tabular}{lccccc}
\hline\hline
Field & Inst. & Filter & Prog. ID & PI & $t_\mathrm{exp}$ (s)\tnote{a} \\\hline
DSF2237b & ACS  & F606W & 15287 & A. Shapley\tnote{b} & 6300 \\
Q0100    & WFC3 & F160W & 11694 & D. Law\tnote{c}    & $~8100$ \\
Q0142    & WFC3 & F160W & 11694 & D. Law\tnote{c}     & $~8100$ \\
Q0207    & WFC3 & F140W & 12471 & D. Erb     & $~~800$ \\
Q0449    & WFC3 & F160W & 11694 & D. Law\tnote{c}     & $~8100$ \\
Q0821    & WFC3 & F140W & 12471 & D. Erb     & $~~800$ \\
Q1009    & WFC3 & F160W & 11694 & D. Law\tnote{c}     & $~8100$ \\
Q1549    & ACS  & F606W & 12959 & A. Shapley\tnote{d} & $12000$ \\
Q1623    & WFC3 & F160W & 11694 & D. Law\tnote{c}     & $~8100$ \\
Q1700    & ACS  & F814W & 10581 & A. Shapley\tnote{e} & $12500$ \\
Q2206    & WFC3 & F160W & 11694 & D. Law\tnote{c}     & $~8100$ \\
Q2343    & WFC3 & F140W & 14620 & R. Trainor & $~5200$ \\
Q2343    & WFC3 & F160W & 11694 & D. Law\tnote{c}    & $~8100$ \\\hline
\end{tabular}
\begin{tablenotes}\footnotesize
\item[a]{Defined as the median exposure time across the whole FoV.}
\item[b]{\citet{pahl21}.}
\item[c]{\citet{law12a,law12c}.}
\item[d]{\citet{mostardi15}.}
\item[e]{\citet{peter07}.}
\end{tablenotes}
\end{threeparttable}
\end{table}

The {\it HST/WFC3-IR} images in this work were obtained using either the F140W or F160W filters, from programs listed in Table~\ref{tab:hst}, with spatial resolution of $\sim 0\secpoint16$ and$\sim 0\secpoint 18$, respectively. The HST/ACS images were taken in either the F606W or F814W filters, with spatial resolution of $\sim 0\secpoint 09$ {(full width at half maximum; FWHM)}. In all cases, overlapping exposures were aligned and combined using \textit{DrizzlePac}\footnote{\href{https://www.stsci.edu/scientific-community/software/drizzlepac}{https://www.stsci.edu/scientific-community/software/drizzlepac}}. For the Q2343 field,  where galaxies have been observed with comparably deep observations in two filters, we selected the image with the smaller estimated uncertainty in the measured position angle (see \S\ref{sec:pa_methods}). The Keck/OSIRIS \Ha\ maps of six galaxies included in the sample are presented by \citet{law07, law09, law12b,law18};  details of the observations and data reduction can be found in those references.

\section{Galaxy Azimuthal Angle}
\label{sec:pa}

\subsection{Motivation}

\begin{figure}
\centering
\includegraphics[width=6cm]{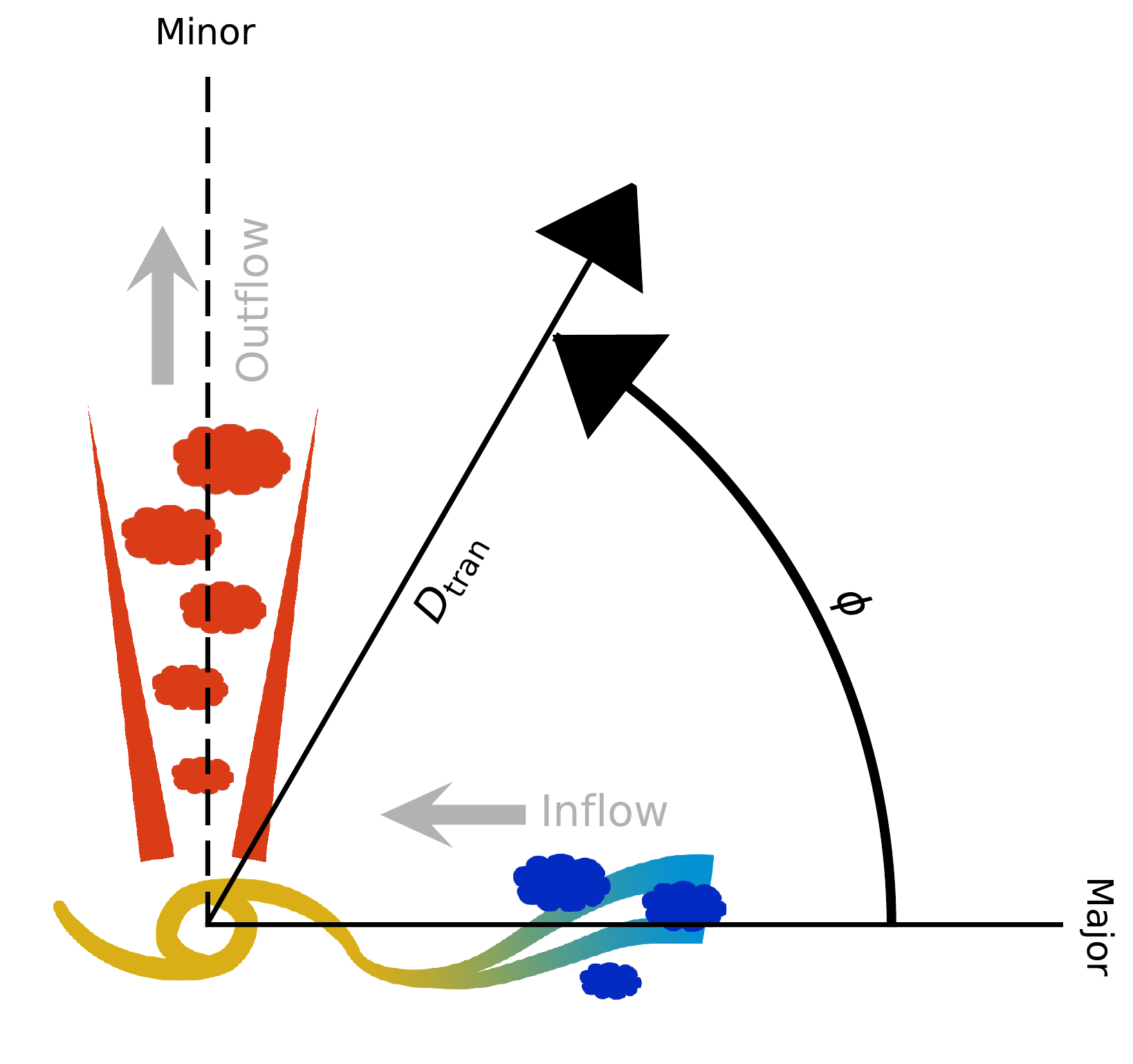}
\caption{ A schematic diagram of how the galaxy azimuthal angle ($\phi$) is defined in this work and how it might be related to the origin and kinematics of gas in the CGM under common assumptions of a bi-conical outflow with accretion along the disk plane. Suppose that we are viewing a galaxy projected on the sky in this diagram, naively, one would expect to see inflow aligning with the projected galaxy major axis, and outflow aligning with the minor axis. Impact parameter, $D_\mathrm{tran}$ is defined as the projected distance from the center of the galaxy. The galaxy azimuthal angle, $\phi$ is defined as the projected angle on the sky with respect to the center of the galaxy, and starts from the projected galaxy major axis. }
\label{fig:def_az}
\end{figure} 

The physical state of the CGM is controlled by the competition between accretion of new material onto the galaxy, and outflows driven by processes originating at small galactocentric radii. In what one might call the ``classic'' picture {(see \citealt{veilleux05, tumlinson17} for review articles)}, based on observations of nearby starburst galaxies, inflowing and outflowing gas occurs preferentially along the major and minor axes, respectively, as in the schematic diagram in Figure~\ref{fig:def_az}. 
It is well-established around star-forming galaxies at $z \lesssim 1$ that outflows driven by energy or momentum from stellar feedback (radiation pressure from massive stars, supernovae) originating near the galaxy center tend to escape along the direction that minimizes the thickness of the ambient {interstellar medium (ISM)} that would tend to slow or prevent outflows from escaping the galactic disk. Meanwhile, accretion of cool gas from the {intergalactic medium (IGM)} is believed to occur in quasi-collimated filamentary flows that carry significant angular momentum and thus would tend to approach the inner galaxy in a direction parallel to the disk plane {\citep{nelson19, peroux20}}.  If feedback processes are sufficiently vigorous, the accretion is also most likely along directions that do not encounter strong outflows.  When projected onto the plane of the sky, these considerations would tend to intersect outflowing material along the polar direction (the projected minor axis) and accreting material when the azimuthal angle (see Figure~\ref{fig:def_az}) is closer to the PA of the projected major axis. 
 
At redshifts $z \simlt 1$, there is strong support for this general geometric picture from the statistics of the incidence and strength of rest-UV absorption lines observed in the spectra of background sources, as a function of the azimuthal angle of the vector connecting the sightline and the center of the galaxy. As might be expected from a picture similar to Fig.~\ref{fig:def_az}, sightlines with $\phi$ close to the minor axis intersect gas with a large range of velocities, which kinematically broadens the observed absorption complexes comprised of many saturated components (e.g., \citealt{bordoloi11, bouche12, kacprzak12, schroetter19}). For azimuthal angles $\phi$ close to the major axis, the absorption features are strong due to the high covering fraction and column density of low-ionization gas near the disk plane, and broadened by the kinematics of differential rotation. The same picture has been supported by cosmological simulations where the gas metallicity, radial velocity, and the amount of outflowing mass all show significant differences along the galaxy major and minor axes \citep{nelson19, peroux20}. 
 
It remains unclear, however, whether this geometric picture should be applied to galaxies at $z \sim 2$, where a large fraction of galaxies, particularly those with ${\rm log}(M_{\ast}/M_{\odot}) \simlt 10$, appear to be ``dispersion dominated'', where the rotational component of dynamical support ($V_{\rm rot}$) is significantly smaller than that of apparently random motion ($\sigma$) (\citealt{law09, forster09}), and where the central dynamical mass of the galaxy may be dominated by cold gas rather than stars -- and therefore any disk would be highly unstable.  For such galaxies, there is not always a clear connection between the morphology of starlight and the principal kinematic axes 
(e.g., \citealt{erb04, law12c}). Meanwhile, a WFC3 survey conducted by \citet{law12c} for similar galaxies strongly supports 3D triaxial morphology, instead of inclined disk, to be the most suitable model to describe these galaxies. The fact that the vast majority of galaxies with ``down the barrel'' spectra at $z > 2$ have systematically blueshifted interstellar absorption lines and systematically redshifted \lya\ emission suggests that outflows cannot be confined to a small range of azimuth (\citealt{shapley03, steidel10, jones12} etc.).

The spatial and spectral distribution of the \lya\ emission surrounding galaxies may provide crucial insight into the degree of axisymmetry and the dominant magnitude and direction of gas flows in the CGM.  In any case, IFU observations of \lya,  where both the geometry and kinematics of the extended emission can be mapped, complement information available from absorption line studies.

In the analysis below, we define the galaxy azimuthal angle ($\phi$) as the absolute angular difference between the vector connecting the galaxy centroid and a sky position at projected angular distance of $\theta_{\rm tran}$ (or projected physical distance $D_{\rm tran}$) and $\pa_0$, the PA of the galaxy major axis measured from the galaxy stellar continuum light, as illustrated schematically in Figure \ref{fig:def_az}, 
\begin{eqnarray}
\label{eq:azimuthal_angle}
	\phi = |\mathrm{PA} - \pa_0|. 
\end{eqnarray}
The zero point for measurements of position angle is arbitrary, but for definiteness we measure PA as degrees {east (E) of north (N)}, so that angles increase in the counter-clockwise direction when N is up and E to the left. The use of equation~\ref{eq:azimuthal_angle} to define $\phi$ implies that $0 \le \phi/{\rm deg} \le 90$, and that $\phi = 0$ (90) deg corresponds to the projected major (minor) axis.

\begin{figure*}
\includegraphics[width=18cm]{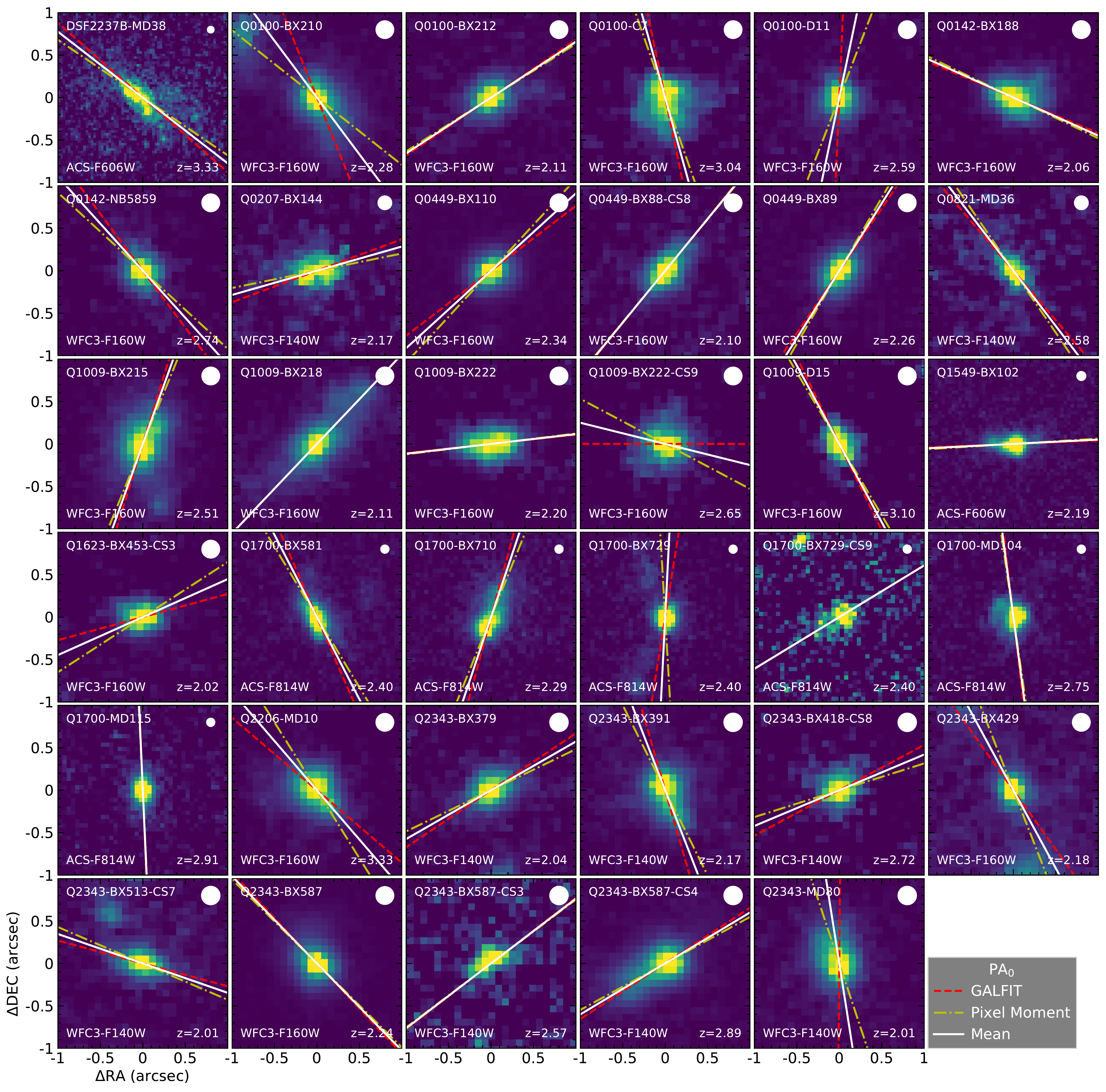}
\caption{  HST images of the 35 galaxies whose $\pa_0$ were determined from Methods \rnum{1} and \rnum{2}. For each image, shown in the four corners, from top-left in clockwise direction, are the KBSS identifier, the FWHM of PSF, redshift, and the instrument and filter that the image was taken with. The dashed red line, dash-dotted yellow line, and the solid white line indicate the direction of $\pa_0$ measured from GALFIT, pixel moment, and the average between the two.} 
\label{fig:hst_gal_mmt}
\end{figure*} 

\begin{figure*}
\includegraphics[width=18cm]{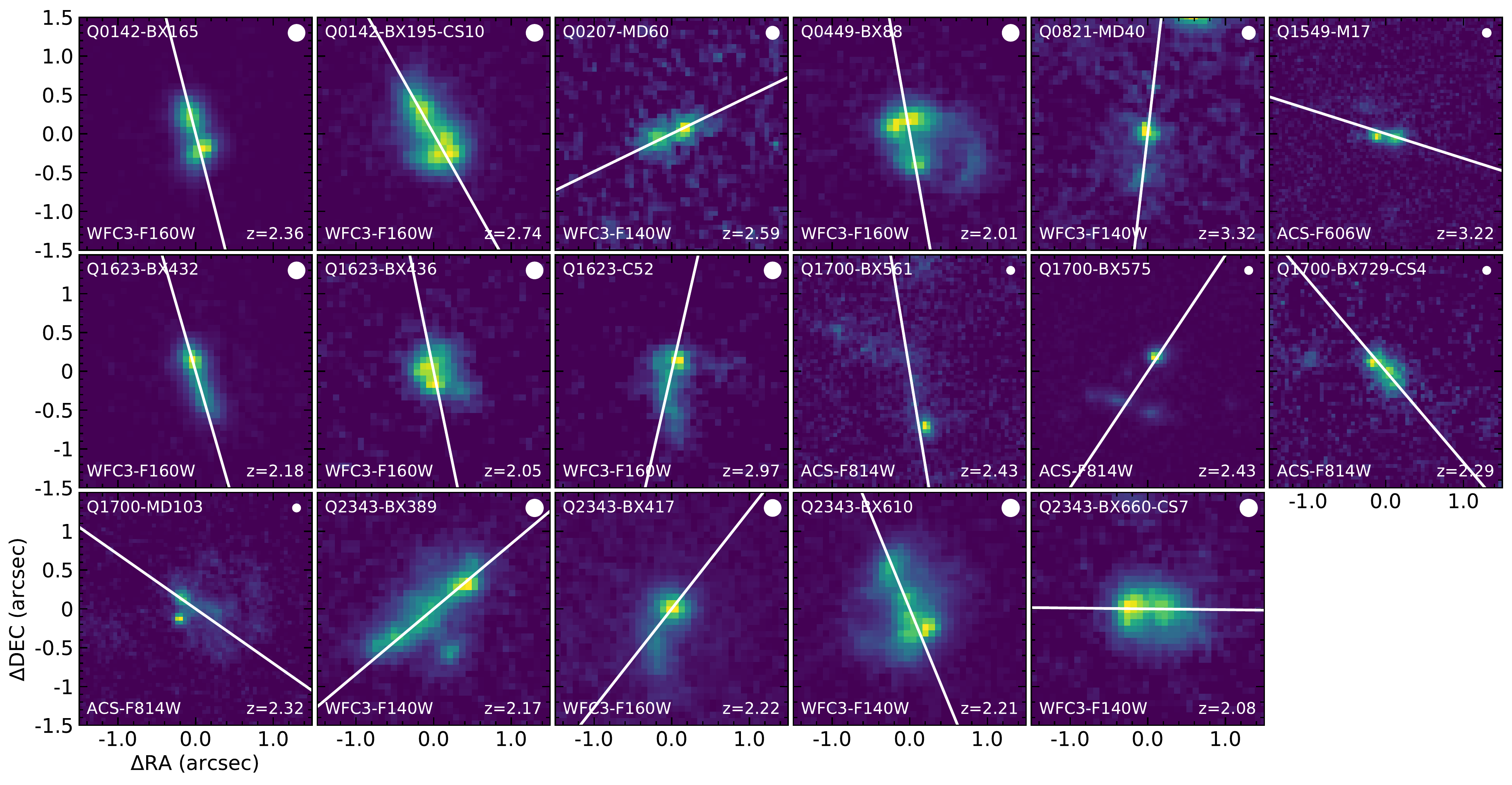}
\caption{   Same as Figure \ref{fig:hst_gal_mmt}, except that the galaxies do not have a clear central SB peak. Therefore, their $\pa_0$ were determined only from the pixel moment (white solid line). } 
\label{fig:hst_mmt_only}
\end{figure*} 
\begin{figure*}
\includegraphics[width=14cm]{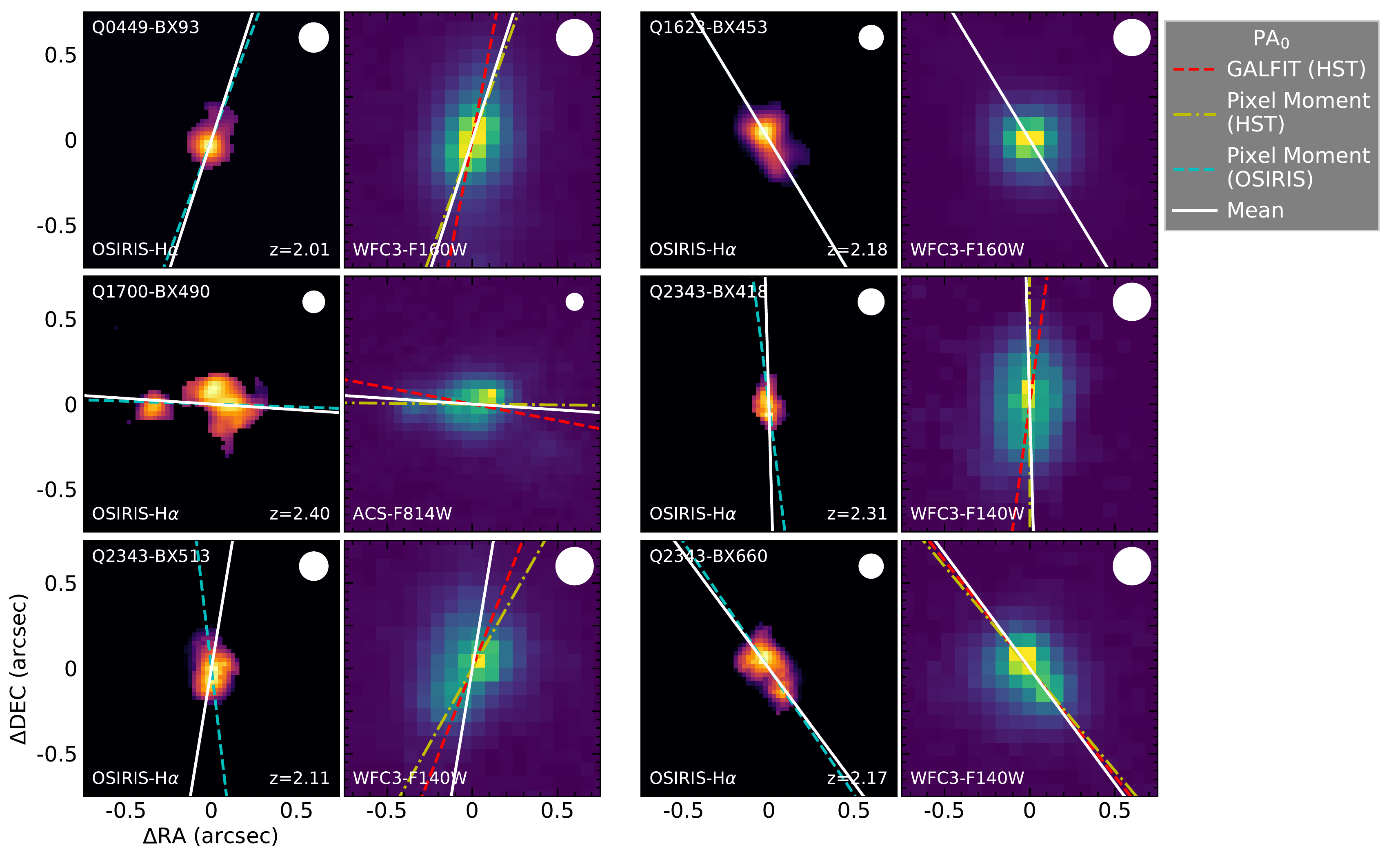}
\caption{ Similar to Figure \ref{fig:hst_gal_mmt}, showing the $\pa_0$ of galaxies with both Keck/OSIRIS H$\alpha$ maps  from \citet{law09} and HST continuum images. For each panel, the left image shows the OSIRIS H$\alpha$ map, in which the cyan dashed line is $\pa_0$ measured from this map using second pixel moment. The right image shows the HST image, where the red dashed line and the yellow dash-dotted line show the $\pa_0$ measured from GALFIT fitting and second moment from this image when present.  The white solid lines are the final $\pa_0$ determined for each galaxy by averaging the OSIRIS and HST measurements\protect\footnotemark.  } 
\label{fig:hst_osiris}
\end{figure*} 
\footnotetext{Since the HST WFC3-IR/F160W image of Q1623-BX453 is unresolved, its $\pa_0$ is only derived from the OSIRIS H$\alpha$ map.}

\begin{figure}
\centering
\includegraphics[width=8cm]{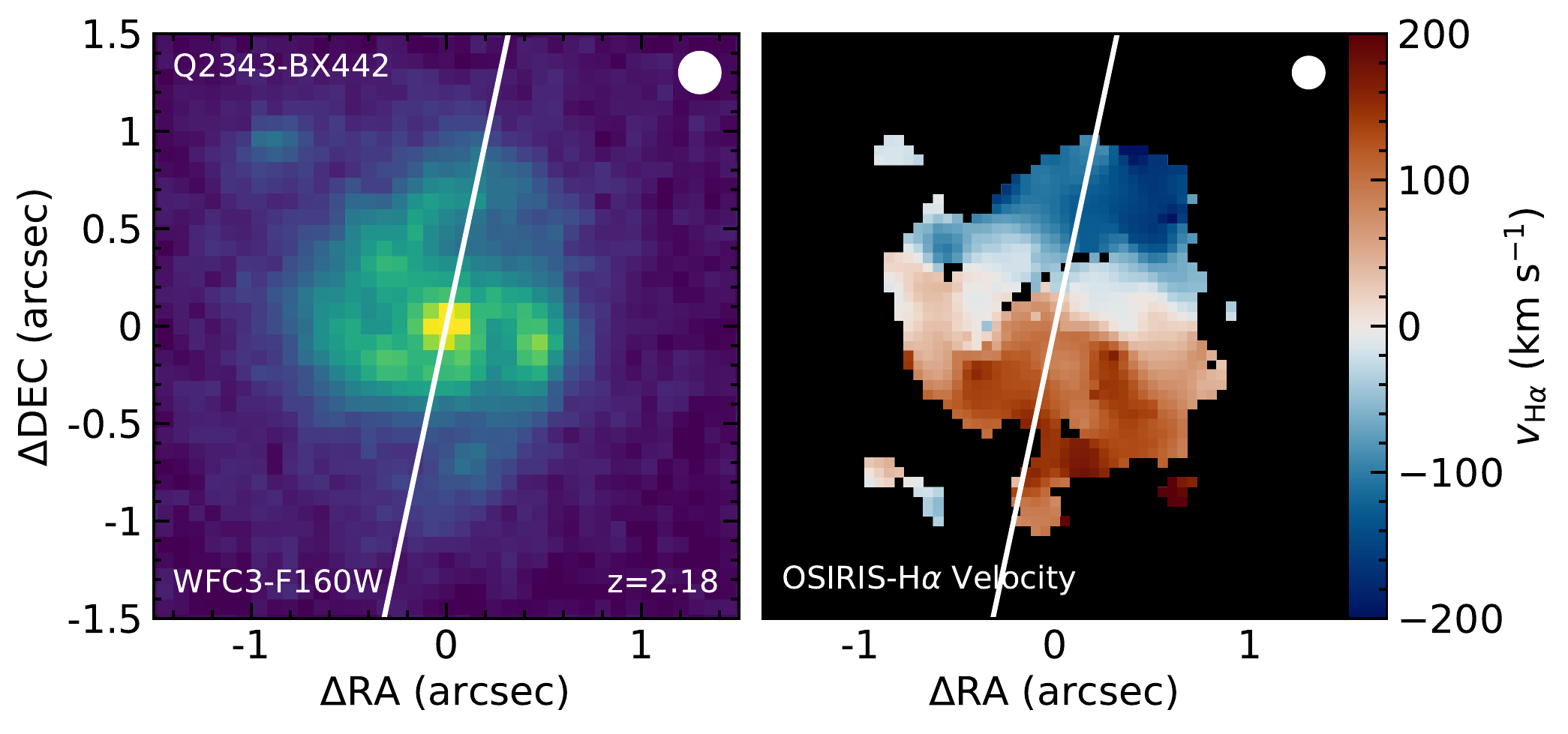}
\caption{  Left: HST WFC3-F160W image of Q2343-BX442. Right: H$\alpha$ velocity map of Q2343-BX442 by \citet{law12b}. The $\pa_0$ (white solid line) is defined to be perpendicular to its rotational axis. } 
\label{fig:q2343_bx442}
\end{figure} 

\subsection{Methods}
\label{sec:pa_methods}
In order to obtain reliable measurements of $\phi$, it is important to measure $\pa_0$ accurately and consistently. We used up to four different methods to determine $\pa_0$ for each galaxy. The choice of method depends on the information available; these are briefly summarised as follows:

\begin{itemize}

\item { (\rnum{1}) \it S\'ersic profile fitting of HST images:} 
In this method, we fit a 2D S\'ersic profile \citep{sersic63} to the host galaxy by using \textit{GALFIT} \citep{peng02, peng10}, and determined $\pa_0$ from the best-fit model parameters. The point-spread function (PSF) was measured by selecting stellar sources over the full HST pointing using the star classifier in \textit{SExtractor}, which calculates a ``stellarity index''for each object based on a neural network. We then examined sources with the highest 3\% stellarity indices by eye, and normalised and stacked them to form an empirical PSF. {Our fiducial model consists of a 2D elliptical S\'ersic profile convolved with the PSF. We also included the first-order Fourier mode to handle the asymmetric morphology in most cases. However, over- or under-fitting can cause failure of convergence or unreasonably large fitting errors and residuals. In most cases, the cause of the failure and the required adjustment are obvious in the original galaxy image and the model residual. For example, if the residual reveals an additional source, we would add an additional S\'ersic component or a simple scaled PSF to the model, depending on the size of the additional source. Meanwhile, if the primary source shows a triangular morphology, we would add the third-order Fourier mode\footnote{The second-order Fourier mode is degenerate with the ellipticity.} associated with the S\'ersic profile. However, in certain cases, obtaining a successful fit requires experimenting with the model by adding or removing certain degrees of freedom. The rule of thumb is that we add or remove degrees of freedom one at a time, and adopt the adjustment if it makes the fit converge, or significantly diminishes the reduced $\chi^2$ and the fitted error of $\pa_0$. In the end, 23 galaxies were fit with the fiducial model. Nine galaxies (Q0100-C7, Q0821-MD36, Q1009-BX222, Q1009-D15, Q1623-BX453-CS3, Q1700-BX729-CS9, Q2343-BX418, Q2343-BX418-CS8, Q2343-BX660) were fit without the Fourier modes. Five galaxies (Q1009-BX218, Q1700-BX490, Q1700-BX710, Q1700-BX729, Q2343-BX513-CS7) were fit with additional sources. Three galaxies (Q2343-BX391, Q2343-BX587-CS3, Q2343-BX587-CS4) were fit with third-order Fourier modes.} Galaxies with unsuccessful fits require using alternative methods, detailed below. Successful S\'ersic fits were obtained for 40 of 59 galaxies, shown in Figures~\ref{fig:hst_gal_mmt} and \ref{fig:hst_osiris};  galaxies with successful applications of this method tend to be isolated and to have a dominant central high-surface-brightness component.\\

\item {  (\rnum{2}) \it Second moment of pixel intensity on HST images:} 
This method calculates the flux-weighted pixel PA arithmetically derived from the second moment of the pixel intensity, 
\begin{eqnarray}
	\tan (2 {\pa_0}) = \frac{2 \langle xy \rangle}{\langle x^2 \rangle - \langle y^2 \rangle}, 
\end{eqnarray}
where $x$ and $y$ are the pixel positions relative to the center of the galaxy in the x- and y-directions, and ``$\langle\textrm{...}\rangle$'' indicates the arithmetic mean of the pixel coordinates weighted by pixel flux. Galaxy centers were defined as the point where $\langle x \rangle = 0$ and $\langle y \rangle = 0$. Especially for the galaxies that are morphologically complex, we found that the $\pa_0$ measurements using this method are sensitive to the surface brightness threshold used to define the outer isophotes, and to the spatial resolution of the HST image. Therefore, all ACS images (which have higher spatial resolution than those taken with WFC3-IR) were convolved with a 2D Gaussian kernel to match the PSF to those of the WFC3-IR/F160W images.  We tested various surface brightness thresholds, and found that a threshold of 50\% of the peak SB after convolution provides the most consistent measurements between the $\pa_0$ measured using (\rnum{1}) and (\rnum{2}) has an RMS of 10.8 degrees.  \\

\item { (\rnum{3}) \it Second moment of pixel intensity of OSIRIS H$\alpha$ maps:} 
This method uses the same algorithm as in (\rnum{2}), applied to Keck/OSIRIS H$\alpha$ maps rather than HST continuum images. The SB threshold follows \citet{law09}. There are 6 galaxies whose major axes were determined using this method, including one (Q1623-BX453) whose F160W image is unresolved. For the remaining 5 galaxies, the RMS between $\pa_0$ measured from the HST images and this method is $\sim 15$ degrees. All 6 galaxies are shown in Figure \ref{fig:hst_osiris}.  \\

\item {(\rnum{4}) \it  H$\alpha$ Kinematics:} 
As the only galaxy in this sample that demonstrates not only rotational kinematics from H$\alpha$ emission, but also clear disk morphology \citep{law12b}, the galaxy major axis of Q2343-BX442 is defined to be perpendicular to its rotational axis (Figure \ref{fig:q2343_bx442}). Because of its complex morphology, we did not apply method (\rnum{1}) to this galaxy. Method (\rnum{2}) is likely dominated by the inner spiral structure and leads to a $\pa_0$ that is $\sim 90$ deg apart from that determined from kinematics, while method (\rnum{3}) is consistent with this method within 15 deg. 
\end{itemize}

{Table~\ref{tab:sample} lists the adopted measurement of $\pa_0$ for each galaxy in the sample, as well as the methods used to determine it.  In cases where multiple methods were applied, we used the average $\pa_0$ of the OSIRIS and HST measurements, where  the latter value is an average of the results obtained using methods (\rnum{1}) and (\rnum{2}). This way, more weight is given  to method (\rnum{3}) results when available, since it utilises entirely independent data from a different instrument.  The robustness of $\pa_0$ measurements is discussed in the following two subsections.} 

\subsection{Relationship between the Kinematic and Morphological Major Axes}

An important issue for the interpretation of morphological measurement of $\pa_0$ is the extent to which it is likely to be a proxy for the {\it kinematic} major axis.  Two of the 17 galaxies in Figure~\ref{fig:hst_mmt_only} (Q2343-BX389 and Q2343-BX610) were studied as part of the SINS IFU survey by \citet[hereafter FS18]{fs18}, and were subsequently re-observed with the benefit of adaptive optics by FS18; the latter found that both have clear rotational signatures and that the inferred difference between their morphological major axis $\pa_0$ and the kinematic major axis $\pa_{\rm kin}$ are $\Delta\pa \equiv |\pa_0 - \pa_{\rm kin}| = 2\pm5$ deg and $27\pm10$ deg for Q2343-BX389 and Q2343-BX610, respectively. From their full sample of 38 galaxies observed at AO resolution, FS18 found $\langle \Delta {\rm PA} \rangle = 23$ deg (mean), with median $\Delta {\rm PA}_{\rm med} =13$ deg. 

Among the 6 galaxies which also have H$\alpha$ velocity maps from \citet{law09} (used for method (\rnum{3}) above), two (Q0449-BX93 and Q1623-BX453) show no clear rotational signature, so that $\pa_{{\rm kin}}$ is indeterminate. Small amounts of velocity shear were observed for Q2343-BX513 and Q2343-BX660, both having $\pa_{\rm kin}$ consistent with $\pa_0$\footnote{Q2343-BX513 was also observed by FS18, who found $\pa_{\rm kin} = -35$ deg and $\Delta {\rm PA}=40\pm12$ deg; our adopted $\pa_0=-9$ deg differs by 26 deg from their $\pa_{\rm kin}$ value}. However, for Q2343-BX418, the implied $\pa_{\rm kin}$ is nearly perpendicular to its morphology-based $\pa_0$. For Q1700-BX490, the kinematic structure is complicated by the presence of two distinct components: the brighter, western component appears to have $\pa_{\rm kin} \simeq 20$ deg, which would be consistent with $\pa_0$ measured from its  \Ha\ intensity map; however, if the eastern component, which has a slightly blue-shifted velocity of $\sim 100 \kms$, is included as part of the same galaxy, we found $\pa_0 = 86$ deg.

As summarised in Table~\ref{tab:sample} and Figures~\ref{fig:hst_gal_mmt}-\ref{fig:hst_osiris}, most of the galaxy sample has $\pa_0$ measured using 2 or more of the methods described above, and generally the morphologically-determined major axis $\pa_0$ agree with one another to within $\sim 10$ deg. We caution that no high-spatial-resolution {\it kinematic} information is available for most of the sample; based on the subset of 9 that do have such measurements, approximately two-thirds show reasonable agreement between $\pa_0$ and $\pa_{\rm kin}$.

\subsection{Distribution and robustness of $\pa_0$ measurements}
\label{sec:pa_discussion}

\begin{figure}
\centering
\includegraphics[width=8cm]{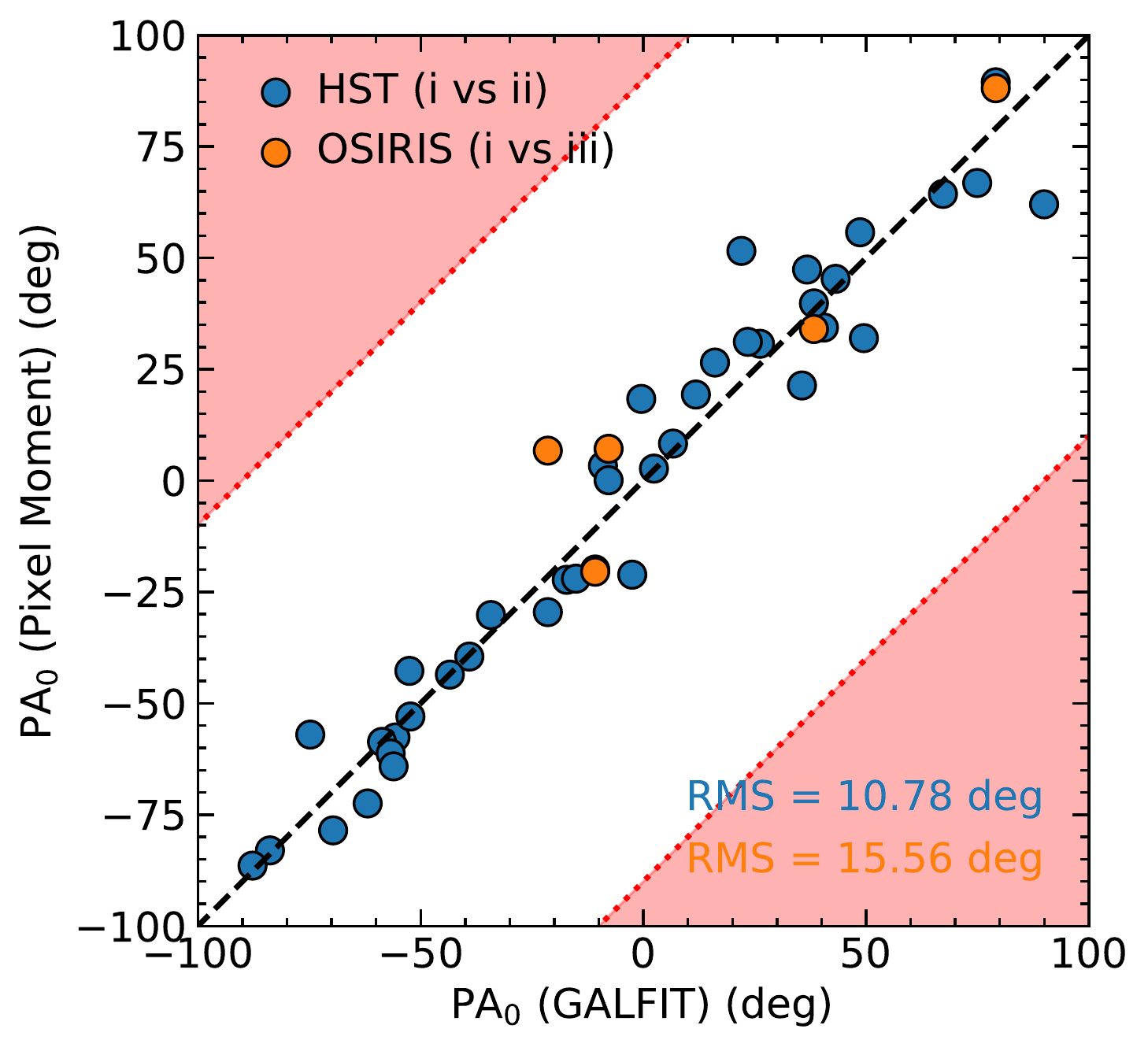}
\caption{   {Comparison of $\pa_0$ values measured using different methods. Blue points compare methods (\rnum{1}) and (\rnum{2}), while orange points compare methods (\rnum{1}) and (\rnum{3}). The red shaded regions indicates where the absolute differences between the two measurements are greater than 90 degrees, arithmetically forbidden because of the rotational symmetry. The overall RMS $= 11.4$ deg.  }} 
\label{fig:pa_correlation}
\end{figure} 

{To estimate the systematic uncertainty of the $\pa_0$ measurements, we compare values measured using different methods for the same objects in Figure \ref{fig:pa_correlation}. Between methods (\rnum{1}) and (\rnum{2}), which are both based on HST images, the measured $\pa_0$ for 40 galaxies are well-centred along the 1-to-1 ratio, with RMS $\simeq 10.8$ deg. Values of $\pa0$ measured using method (\rnum{3}) are based on spectral line maps from an IFU rather than continuum light in a direct image, so that the 5 $\pa_0$ values measured using methods (\rnum{1}) and (\rnum{3}) exhibit larger scatter, with RMS$\simeq 15.6$ deg relative to a 1:1 ratio. Therefore, we conclude that the uncertainty in final $\pa_0$ measurements is $\lesssim 15$ deg. }

\begin{figure}
\centering
\includegraphics[width=8cm]{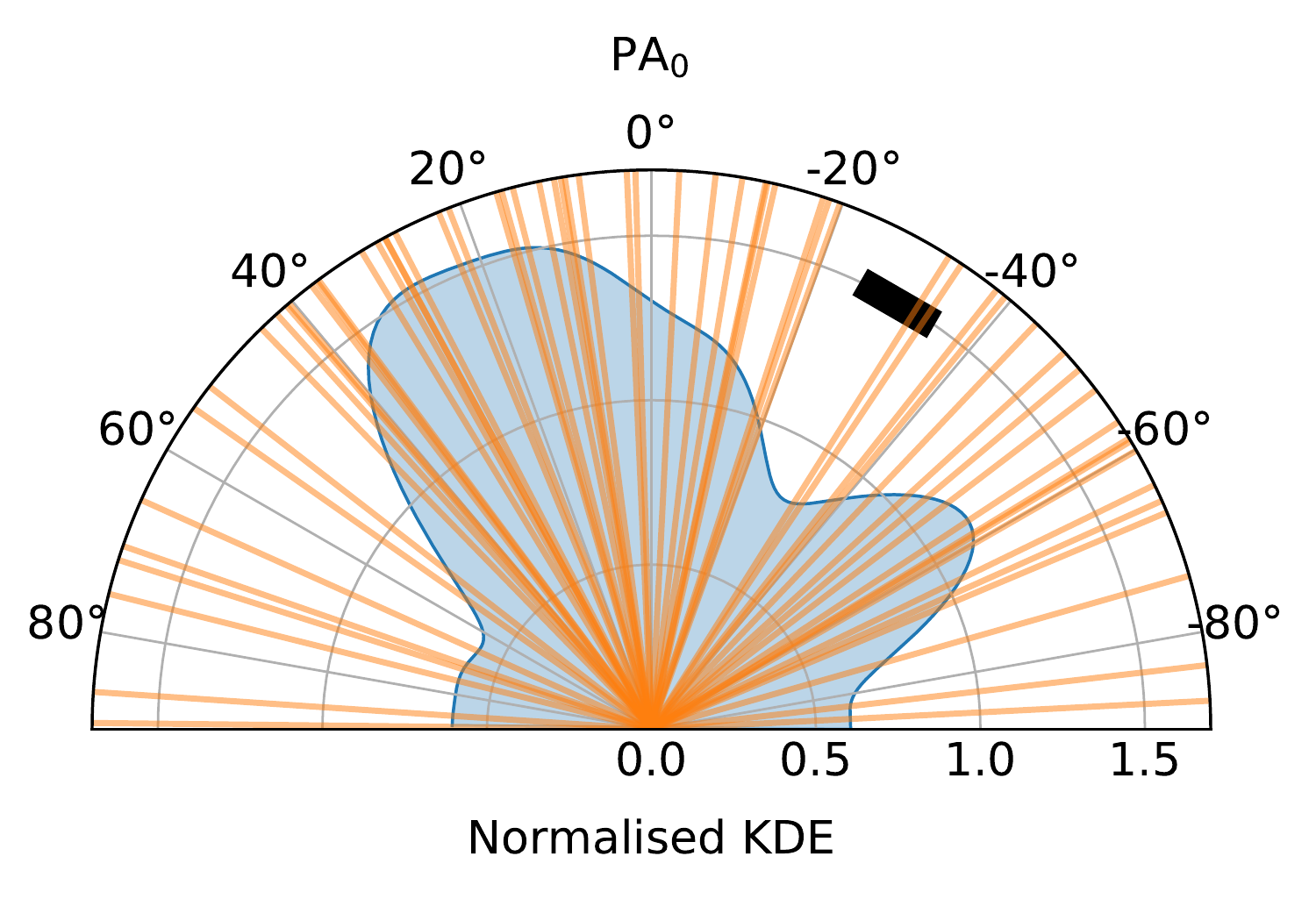}
\caption{   The kernel density estimate (KDE; blue shaded region) of $\pa_0$ for the galaxy sample, normalised so that a uniform distribution would have a constant $\mathrm{KDE} = 1$. 
The KDE was constructed using Gaussian kernels of fixed $\sigma = 10^\circ$, corresponding to an opening angle represented by the black block at the top-right. The orange solid lines indicate the values of $\pa_0$ for the individual galaxies. There is an apparent excess in the KDE  of galaxies with $\pa_0 \simeq 10-40^\circ$, which we attribute to sample variance.} 
\label{fig:pa_kde}
\end{figure}

Figure \ref{fig:pa_kde} shows the normalised Kernel Density Estimator (KDE) and the individual measurements of $\pa_0$. There is an apparent excess in the occurrence rate of values between  $\pa_0 \sim 10-40^\circ$. To evaluate its possible significance, we conducted 1000 Monte-Carlo realisations of a sample of 58 galaxies with randomly assigned $\pa_0$;  we find that there is a $\simeq 5$\% probability that a similar excess is caused by chance, thus is not statistically significant, and is consistent with that expected from sample variance.  

Furthermore, because the HST and OSIRIS images used to measure $\pa_0$ were rotated to the nominal North up and East left orientation prior to measurement, we tested whether significant bias might result from the choice of pixel grid by re-sampling all images with pixel grids oriented in five different directions. We found that the values measured for $\pa_0$ were consistent to within 10 deg (RMS).

\begin{figure*}
\centering
\subfloat[\label{fig:pa_slicer}]{\includegraphics[width=4.5cm]{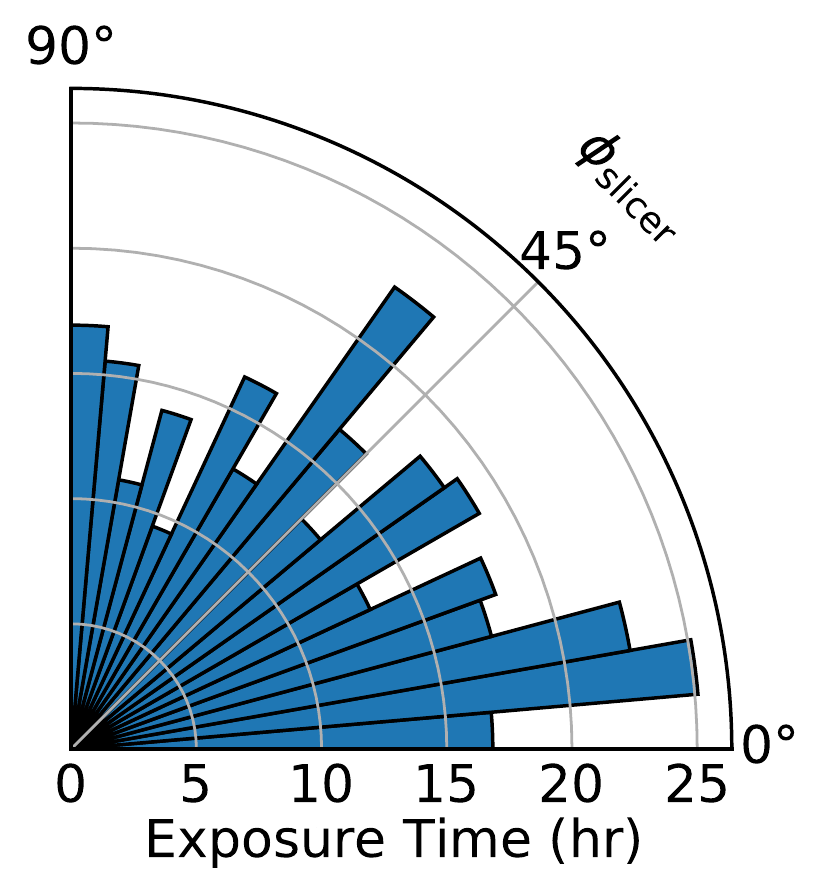}}
\subfloat[\label{fig:kcwi_hst_profile}]{\includegraphics[width=11.5cm]{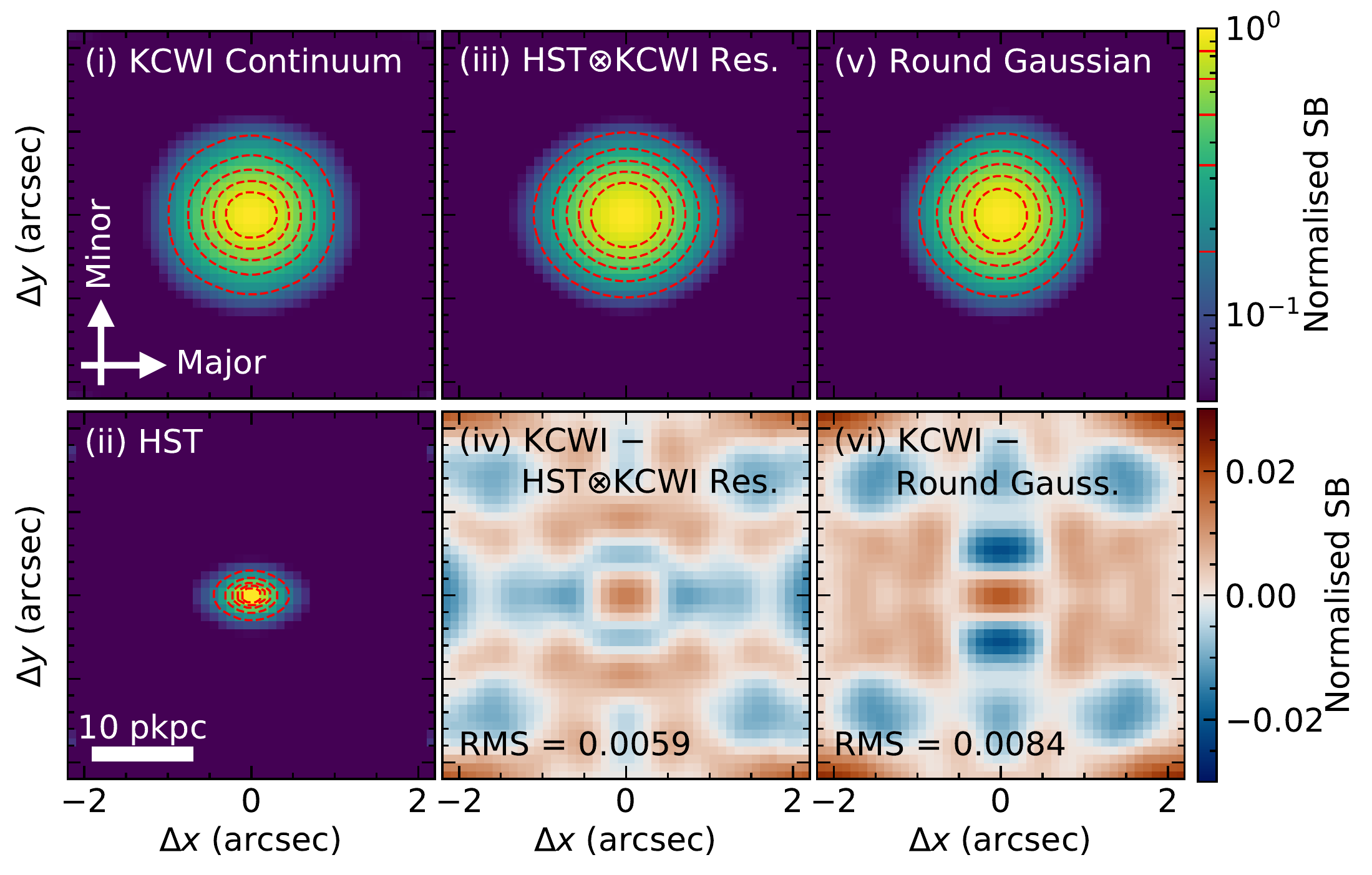}}
\caption{ (a) Histogram of the relative contribution of measurements made at different slicer azimuthal angles ($\phi_{\rm slicer}$) in units of total exposure time. The distribution of $\phi_\mathrm{Slicer}$ is relatively uniform, with a small excess near $\phi_\mathrm{Slicer} \sim 10^\circ$. (b) Stacks of the galaxy continuum images for which the major and minor axes of each galaxy were aligned with the X and Y axes prior to averaging. Each panel shows (\rnum{1}) the pseudo-narrow-band image (rest frame $1230\pm6$ \AA) of the KCWI galaxy continuum, (\rnum{2}) the stacked HST continuum image, after aligning the principal axes in the same way, (\rnum{3}) the HST image convolved with a Gaussian kernel of $\mathrm{FWHM} = 1\secpoint02$ to match the KCWI continuum, (\rnum{4}) the residual between the KCWI continuum and the HST image convolved with the KCWI PSF, (\rnum{5}) a 2D circular Gaussian profile with $\mathrm{FWHM} = 1.21~\mathrm{arcsec}$ as the best symmetric Gaussian profile from a direct fit of the KCWI continuum image, and (\rnum{6}) the residual between the KCWI continuum and the model in (\rnum{5}) isotropic Gaussian profile. {The colour map of (\rnum{1}), (\rnum{2}), (\rnum{3}), and (\rnum{5}) is in log scale, with linear red contours in the dcrement of 0.17. The colour map of (\rnum{4}) and (\rnum{6}) is in linear scale.} The residual map in panel (\rnum{6}) shows a clear dipole residual in the Y (minor axis) direction that is not present in (\rnum{4}). The RMS values in panels (\rnum{4}) and (\rnum{6}) were calculated within $|\Delta x| < 1~\mathrm{arcsec}$ and $|\Delta y| < 1~\mathrm{arcsec}$ to reflect the dipole residual. The ``boxiness'' of the KCWI stack is likely due to the undersampling of KCWI in the spatial direction. Taken together, (b) demonstrates that the KCWI PSF is axisymmetric (with ${\rm FWHM} = 1\secpoint02$) , and that the KCWI continuum image is capable of distinguishing the galaxy major and minor axes. } 
\label{fig:symmetry}
\end{figure*} 

The possible systematic bias introduced by an uneven $\pa_0$ distribution is mitigated by our observational strategy of rotating the KCWI instrument PA between individual exposures. We define $\pa_\mathrm{Slicer}$ as the position angle along the slices for each 1200 s exposure, and
\begin{eqnarray}
	\phi_\mathrm{slicer} = |\pa_\mathrm{Slicer} - \pa_0|.
\end{eqnarray}
Figure \ref{fig:pa_slicer} shows the distribution of $\phi_\mathrm{slicer}$ in units of total exposure time. There is a slight tendency for the observations to align with the galaxy major axis (i.e., $\phi_{\rm slicer} \sim 0$) that results from our usual practice of beginning a sequence of exposures of a given pointing with one at $\pa_{\rm slicer} = 0$ deg and the aforementioned excess of galaxies with $\pa_0 \sim 20^\circ$.

In order to characterise the PSF of our final KCWI data cubes, and to show that it is effectively axisymmetric, Figure~\ref{fig:symmetry}b shows various composite images of the galaxy sample.  Prior to forming the stack, each individual data cube was rotated to align the galaxy major (minor) axis with the X (Y) coordinate of the stacked image. The KCWI stacked galaxy continuum image [panel (\rnum{1})] was made by integrating each aligned data cube along the wavelength axis over the range $1224 \le \lambda_0/{\rm \AA} \le 1236$ in the galaxy rest frame (i.e., $2000 < \Delta v/\kms \le 5000$).  This integration window was chosen to be representative of the UV continuum near \lya\ without including the \lya\ line itself, and to be unaffected by \lya\ absorption from the IGM. {The center of the galaxy is determined with high confidence (see \S\ref{sec:spatial_profile}) by fitting a 2D Gaussian function to the individual galaxy continuum image.} A similar approach -- aligning the principal axes in the high-resolution HST images prior to stacking -- was used to produce the HST stacked continuum image shown in panel (\rnum{2}){, except that the centers in the HST images were measured using the first moment of pixel intensity.} Both stacks were conducted in units of observed surface brightness. The FWHM of the stacked HST image along the major and minor axes are $0\secpoint55$ and $0\secpoint35$ respectively. The stacked KCWI continuum image is well reproduced by the convolution of the $\pa_0$-aligned stacked HST image [panel (\rnum{2}) of Fig.~\ref{fig:symmetry}b] with an axisymmetric 2-D Gaussian profile with $\mathrm{FWHM} = 1\secpoint02$ [see panels (\rnum{3}) and (\rnum{4}) of Fig.~\ref{fig:symmetry}b]. 

Comparing the convolution of HST and KCWI [panel (\rnum{3})]  with the best direct fit of a symmetric Gaussian profile to the KCWI continuum image ($\rm FWHM = 1\secpoint21$) [panel (\rnum{5})] show that even at the $\simeq 1\secpoint02$ resolution of KCWI one can clearly distinguish the major axis elongation. {The residual map assuming a symmetric Gaussian profile [panel (\rnum{6})] shows a clear dipole residual compared to panel (\rnum{4}). }  Thus, Figure~\ref{fig:symmetry} shows that (1) the PSF of the KCWI cubes is axisymmetric and thus has not introduced a bias to the azimuthal light distribution measurements and (2) the spatial resolution is sufficient to recognize non-axisymmetry even on sub-arcsec angular scales of the continuum light.

\section{Analyses}
\label{sec:analyses}

\subsection{\lya\ Spatial Profile}
\label{sec:spatial_profile}

\begin{figure}
\centering
\includegraphics[width=8cm]{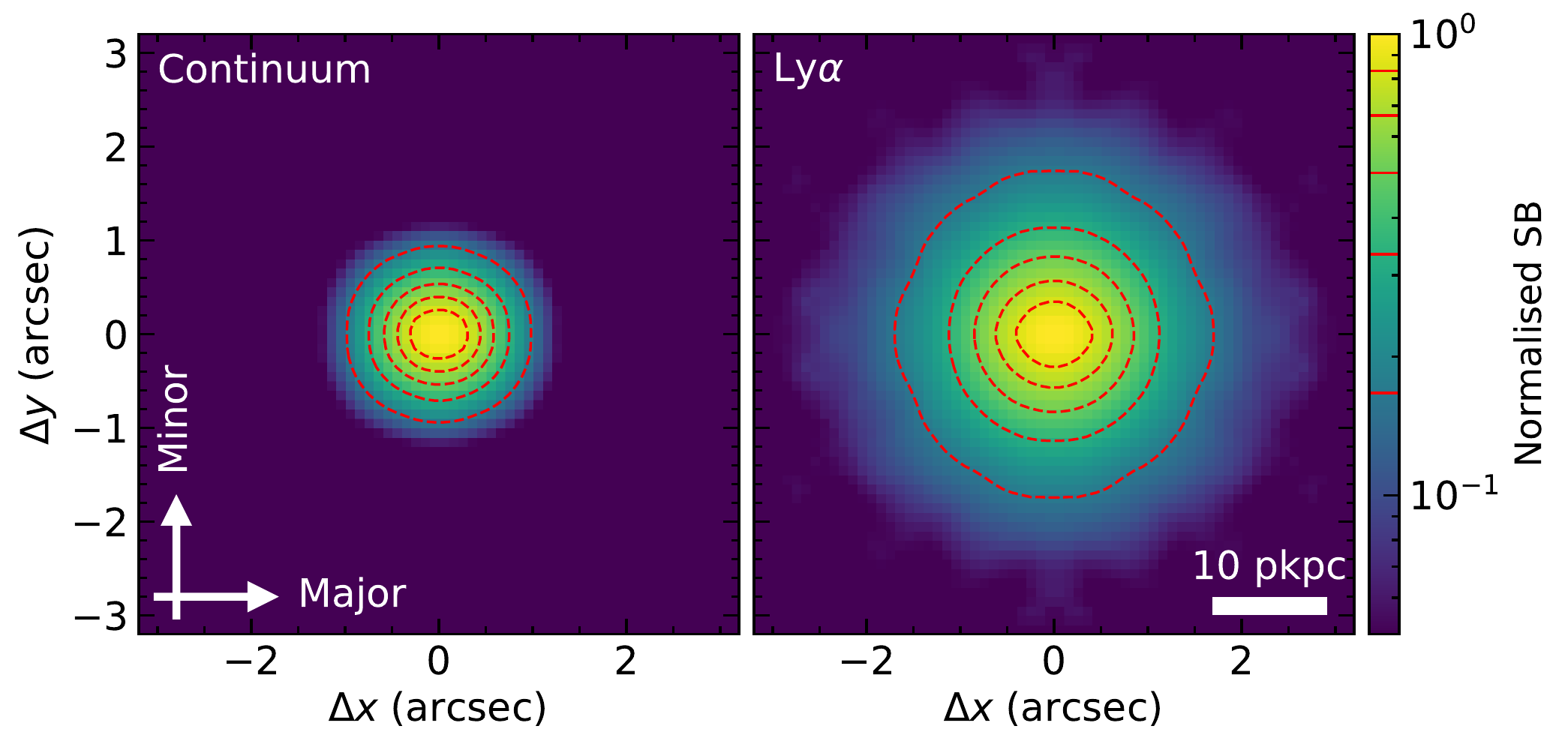}
\caption{   Stacked images of the galaxy continuum (Left) and the continuum-subtracted \lya\ emission (Right)  with the X- and Y-axes aligned with the galaxy major and minor axes, respectively. The color coding is on a log scale, while the contours are linear. The intensity scales have been normalized to have the same peak surface brightness intensity at the center. The \lya\ emission is more extended than the continuum emission. }
\label{fig:img_cont_lya}
\end{figure} 

\begin{figure}
\centering
\includegraphics[width=8cm]{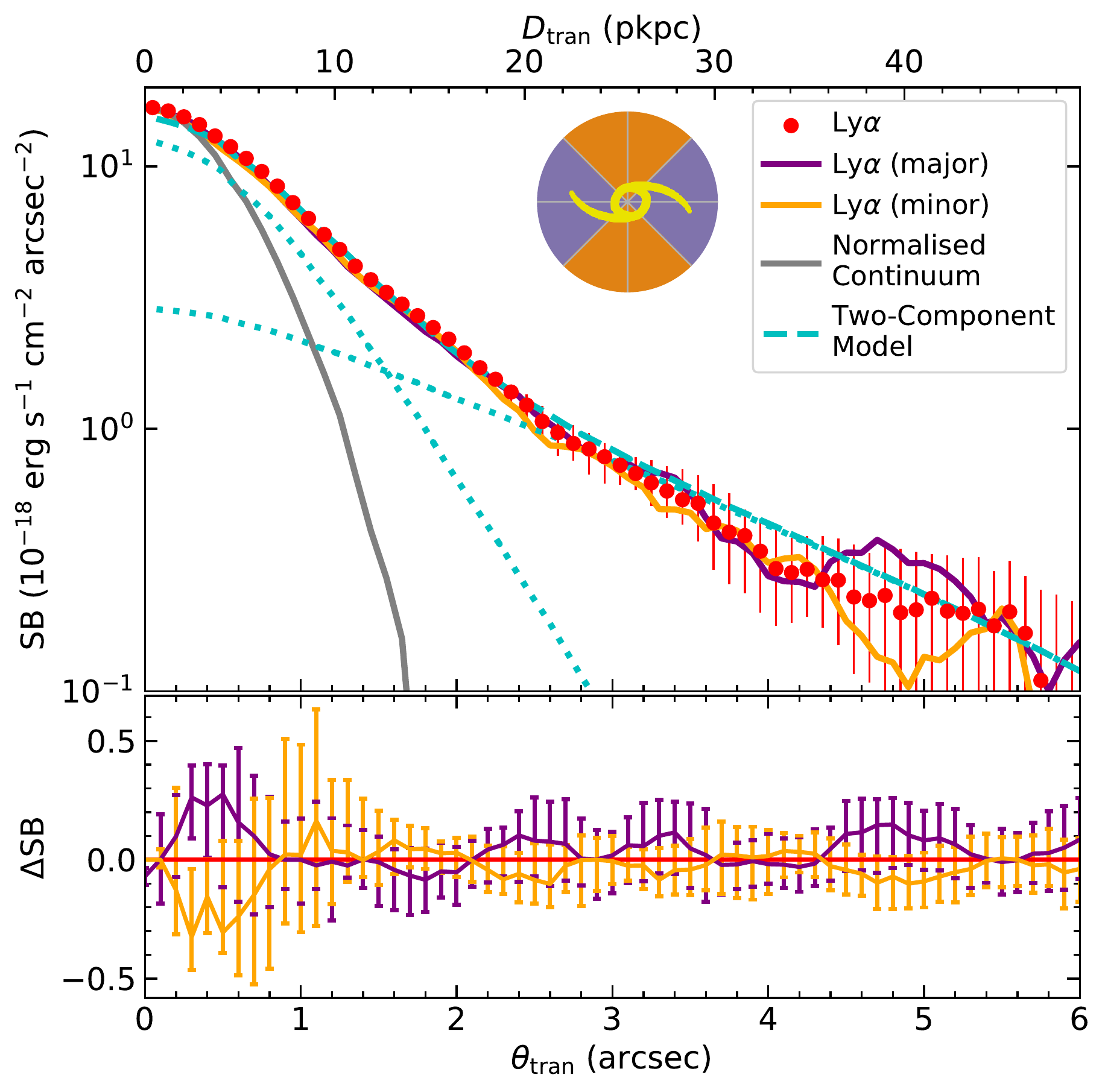}
\caption{  Top panel: The average \lya\ surface brightness profile of the continuum-subtracted composite \lya\ image shown the righthand panel of Figure \ref{fig:img_cont_lya}. Red points represent the median surface brightness evaluated over all azimuthal angles ($0^\circ < \phi \le 90^\circ$) as a function of projected distance from the galaxy center. Orange and purple curves show the profiles evaluated over $0^\circ < \phi \le 45^\circ$ ( major axis) and $45^\circ < \phi < 90^\circ$ (minor axis) azimuthal angles. {Dashed cyan curve shows the best-fit profile of the two-component exponential model. Dotted cyan curves show the two component separately.} The grey profile shows the normalised continuum for comparison. 
Bottom: The residual surface brightness profile formed by subtracting the all-azimuth average from the major and minor axis profiles. The residuals are consistent with zero aside from a marginally-significant difference at $\theta_{\rm tran} < 1\secpoint0$, where the the \lya\ emission is slightly stronger along the major axis. Unless otherwise noted, the conversion between $\theta_\mathrm{tran}$ and $D_\mathrm{tran}$ for this and later figures assumes a redshift of 2.3, the median redshift of the sample.}
\label{fig:lya_profile}
\end{figure} 

To study the dependence of the  \lya\ emission profile on galaxy azimuthal angle, we first analyse the \lya\ surface brightness (SB) profile as a function of the impact parameter (or transverse distance, $D_\mathrm{tran}$). Figure \ref{fig:img_cont_lya} compares the stacked continuum and continuum-subtracted narrow-band \lya\ emission, composed in the same way as in Figure \ref{fig:symmetry}. The integration window of the \lya\ image is $-700 < \Delta v / (\mathrm{km s}^{-1}) \le 1000$ (1213 \AA\ -- 1220 \AA) to include most of the \lya\ emission (as shown later in Figure \ref{fig:intro_cylindrical}). Continuum subtraction throughout this work was done spaxel-by-spaxel in the data cube of each target galaxy by subtracting the average flux density in two windows flanking the position of \lya, with $2000 < |c\Delta v_{\rm sys} / \mathrm{km~s}^{-1}| < 5000$, where $\Delta v_{\rm sys}$ is the velocity separation relative to rest-frame \lya\ (i.e., two windows each of width $\simeq 12$ \AA\ in the rest frame, [1195-1207] and [1224-1236] \AA. {Similar to \S\ref{sec:pa_discussion}, the center of the galaxy in the KCWI data was determined by fitting a 2D Gaussian profile to the KCWI continuum image. The fitting box was chosen by eye to include all of the signal from the galaxy, but excluding nearby contamination. Despite the arbitrary box size, we found that the derived centroid is very robust:  varying the box size by 4 pixels ($1\secpoint2$), the fit result does not change by more than $0\secpoint01$. The typical fitting error propagated from the reduced $\chi^2$ is also $\sim 0\secpoint01$ (median), i.e., much smaller than the seeing disk. }

A 2D Gaussian fit to the profiles finds that the FWHM values are {$1\secpoint279 \pm 0\secpoint003 (\textrm{major axis}) \times 1\secpoint182 \pm 0\secpoint003 (\textrm{minor axis})$ (or a $\sim 8\%$ difference) for the continuum emission and $1\secpoint689 \pm 0\secpoint005 \times 1\secpoint705 \pm 0\secpoint005$ ($< 1\%$ difference between major and minor axes)} for the \lya\ emission. Therefore, the \lya\ emission in the stacked image is both more symmetric and more spatially extended than the continuum emission.

Figure \ref{fig:lya_profile} shows the median \lya\ SB as a function of $D_\mathrm{tran}$ (red). Each point represents the median SB of a bin of pixels with $\Delta D_\mathrm{tran} = 0.1~\mathrm{pkpc}$. The \lya\ surface brightness profile falls off much more slowly than the continuum (grey). {Following \citet{wisotzki16}, we fit the \lya\ SB profile with a two-component model -- a compact ``core'' component and an extended ``halo'' component.  Both components are exponential profiles convolved with the KCWI PSF with the amplitudes, exponential radii, and a uniform background term as free parameters. Further details of the model fitting will be described in a future work (R. Trainor \& N. Lamb, in prep.). The best-fit exponential radii $r_\mathrm{exp} = 3.71^{+0.06}_{-0.04}$ pkpc and $15.6^{+0.5}_{-0.4}$ pkpc. The $r_\mathrm{exp}$ of the halo component is close to \citet{steidel11} (for KBSS galaxies observed with narrow-band \lya\ imaging), which found the median-stacked \lya\ profile has $r_\mathrm{exp} = 17.5$ pkpc, but slightly more extended than \citet{wisotzki18} for SF galaxies at $z>3$ (see also \citealt{matsuda12, momose14, leclercq17}).} Dividing the SB profiles into two subsamples with $0^\circ \le \phi < 45^\circ$ (purple) and  $45^\circ \le \phi \le 90^\circ$ (orange) that represent the galaxy major and major axes respectively, one can see that the resulting profiles are consistent with one another to within 1$\sigma$,  or within $\lesssim 2\times 10^{-19} \mathrm{~erg~s}^{-1}\mathrm{cm}^{-2}\mathrm{arcsec}^{-2}$. The possible exception is at the smallest projected distances ($D_\mathrm{tran} < 1~\mathrm{arcsec}$, or $\lesssim 8$ pkpc), where the \lya\ emission is marginally enhanced; if real, the difference in profiles (the asymmetry) represents $< 2\%$ of the total \lya\ flux. Thus, the composite \lya\ intensity is remarkably symmetric, suggesting an overall lack of a strong statistical connection between the morphology of the starlight and that of the extended \lya\ emission surrounding individual star-forming galaxies at $z \sim 2-3$.

\subsection{Cylindrical Projection of 2D Spectra}
\label{sec:2dspec}

\begin{figure*}
\centering
\includegraphics[width=6.592cm]{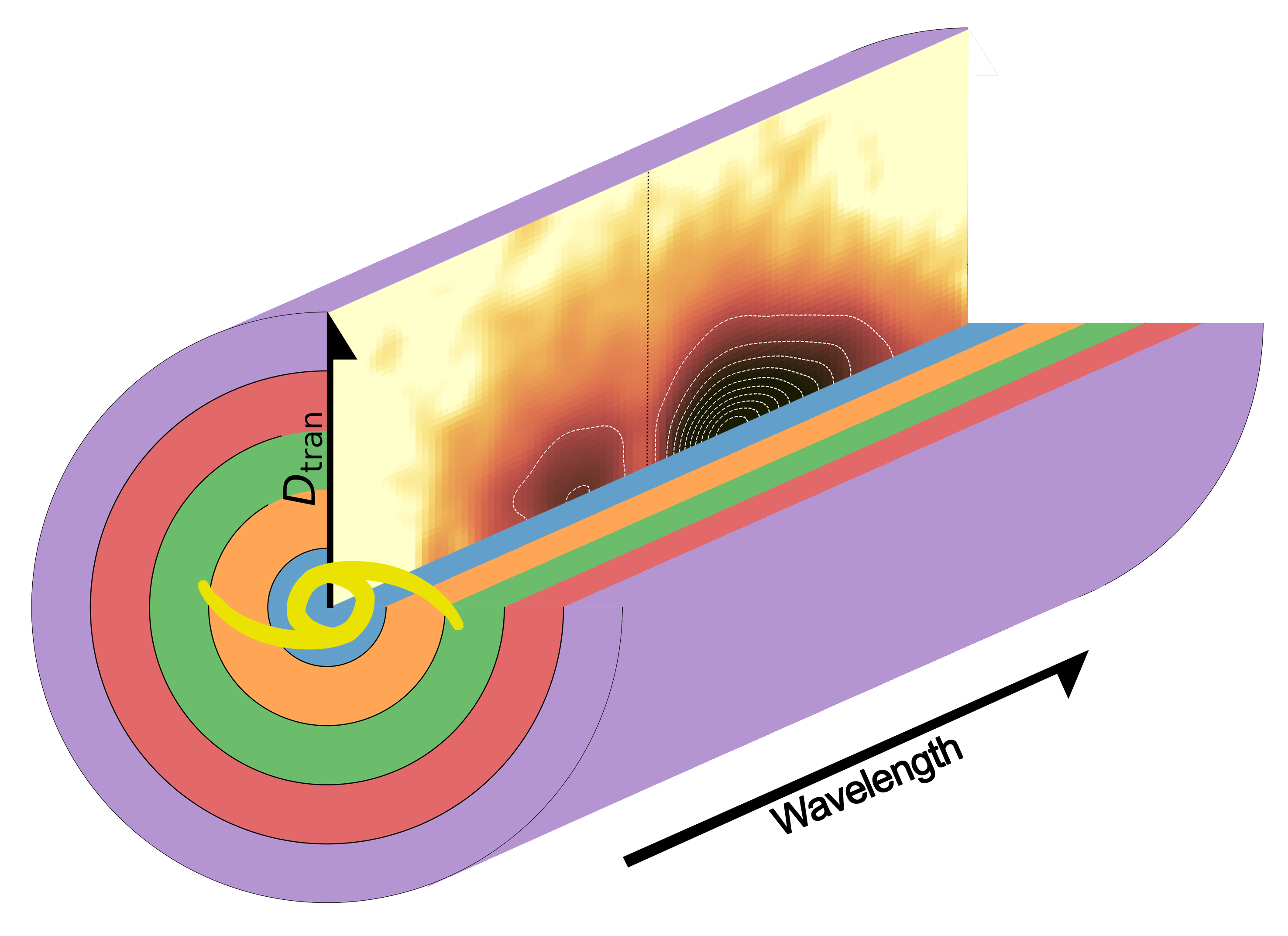}
\includegraphics[width=9.408cm]{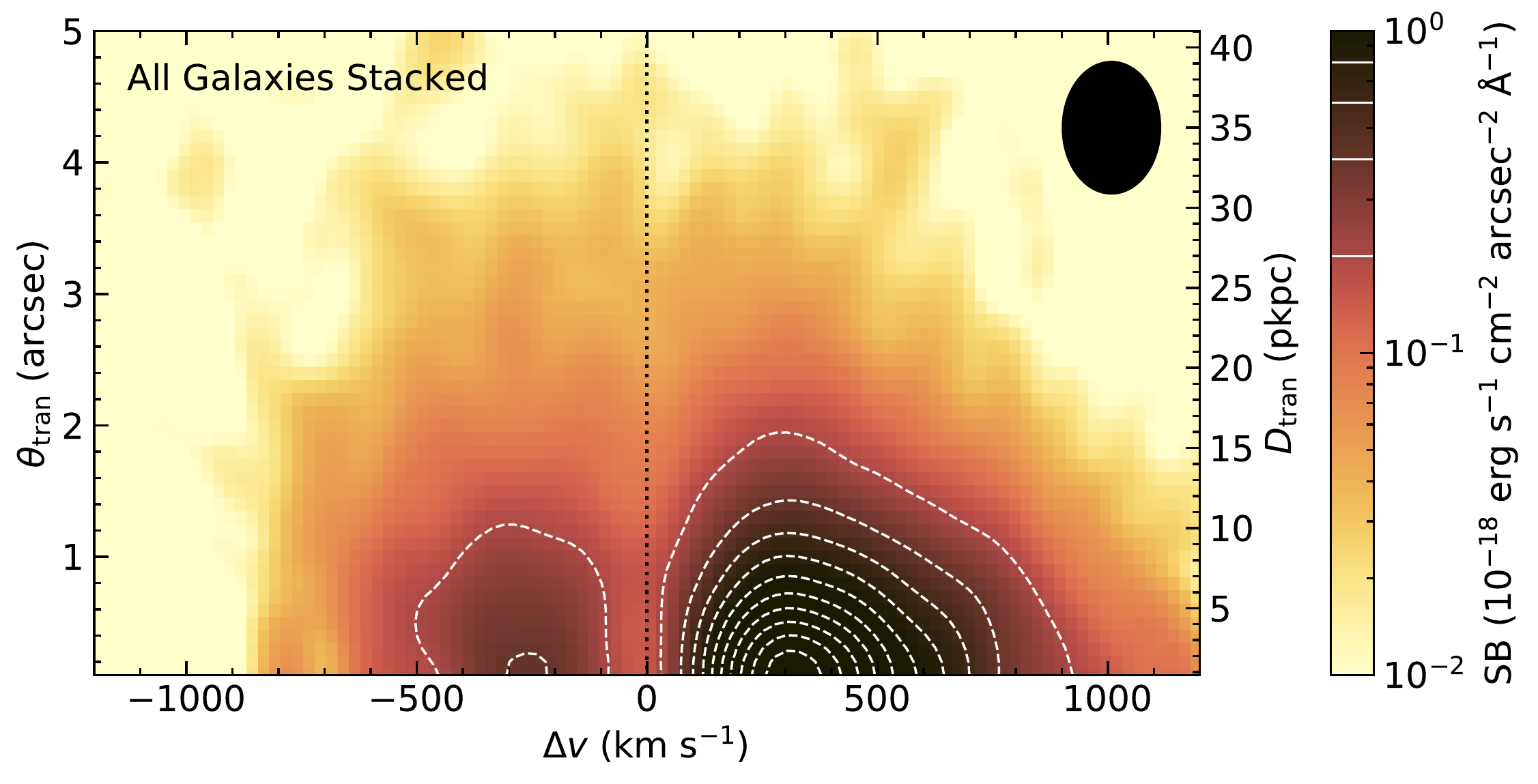}
\caption{  {\it Left}: A schematic diagram explaining cylindrically projected 2D (CP2D) spectra. Spaxels with similar $D_\mathrm{tran}$ are averaged to create the emission map in $D_\mathrm{tran}$-$\Delta v$ space. {\it Right}: The composite CP2D spectra of the continuum-subtracted \lya\ emission line map averaged over all 59 galaxies, at all azimuthal angles ($\phi$). The colour-coding of the \lya\ surface intensity is on a log scale to show the full extent of the emission, whereas the contours are spaced linearly and marked as white lines in the colourbar. The stack was formed by shifting the wavelengths of each galaxy data cube to the rest frame, leaving the surface brightness in observed units. 
The black ellipse at the top right shows the effective resolution of the stacked maps, with principal axes corresponding to the spectral resolution FWHM and the spatial resolution FWHM (see \S\ref{sec:pa_discussion}). Pixels with $\theta_\mathrm{tran} < 0.1~\mathrm{arcsec}$ have been omitted to suppress artifacts owing to the singularity in the cylindrical projection.}   
\label{fig:intro_cylindrical}
\end{figure*}

The similarity of the \lya\ surface brightness profile along galaxy major and minor axes suggests that extended \lya\ emission depends little on the galaxy orientation. However, the KCWI data cubes allow for potentially finer discrimination through examination of both the surface brightness and kinematics of \lya\ emission as a function of projected galactocentric distance. To facilitate such comparison, we introduce ``cylindrical projections'' of 2D \lya\ emission. The basic idea behind cylindrical projection, illustrated in Figure~\ref{fig:intro_cylindrical}, is to provide an intuitive visualisation of spatial and spectral information simultaneously. 

Compared to the standard 2D spectrum one obtains from slit spectroscopy, the cylindrical 2D spectrum replaces the 1D spatial axis  (i.e., distance along a slit) with projected distance,  by averaging spaxels in bins of $D_\mathrm{tran}$ or, equivalently, $\theta_{\rm tran}$. When projected as in the righthand panel of Figure~\ref{fig:intro_cylindrical}, it can also be viewed as the \lya\ spectrum at every projected radial distance (averaged, in this case, over all azimuthal angles) or as the average radial profile at each slice of wavelength or velocity. 

Figure \ref{fig:intro_cylindrical} shows the stacked cylindrical 2D spectrum formed by averaging the continuum subtracted data cubes at wavelengths near rest-frame \lya\ for all 59 galaxies in Table~\ref{tab:sample}. This composite cylindrical 2D spectrum, analogous to a ``down-the-barrel'' \lya\ spectrum in 1D, but evaluated as a function of galactocentric distance, shows that the \lya\ emission line is comprised of distinct redshifted and blueshifted components extending to $\pm 1000$ \kms\ with respect to $v_{\rm sys} = 0$, with a minimum close to $v_{\rm sys} = 0$. The vast majority of individual galaxies, and therefore also the average in the stacked profile, has $F_{\lya}(\mathrm{blue}) / F_{\lya}(\mathrm{red}) \simeq 0.3$, and is thus ``red peak dominated''.  The two-component spectral morphology extends to at least $\theta_{\rm tran}\simeq 3~\mathrm{arcsec}$ or $D_{\rm tran} \simeq 25~\mathrm{pkpc}$. This overall spectral morphology is most readily explained as \lya\ photons being resonantly scattered by outflowing material, whereby redshifted photons are scattered from the receding (opposite) side are more likely to escape in the observer's direction than blue-shifted photons \cite[e.g.][]{pettini01,steidel10,dijkstra14}. 

As $D_\mathrm{tran}$ increases and the \lya\ SB decreases exponentially, the two \lya\ peaks become less distinct and merge into a symmetric ``halo'' centered on $\Delta v = 0$. The vast majority of the \lya\ emission is within $-700 < \Delta v / \mathrm{km~s}^{-1} < 1000$. However, we caution that the apparent blue edge at $\Delta v \sim -700~\mathrm{km~s}^{-1}$ of \lya\ emission in this composite 2D spectrum is likely caused by continuum over-subtraction resulting from the relatively simple technique that we used in this work. The continuum subtraction assumed a linear interpolation of the continuum spectrum underneath the \lya\ emission (see \S\ref{sec:spatial_profile}), which tends to over-estimate the continuum flux blueward of the systemic redshift of the galaxy due to intrinsic \lya\ absorption in the stellar spectrum and residual effects of the often-strong \lya\ absorption damping wings on which \lya\ emission is superposed.
To improve on this would require a more sophisticated continuum-subtraction method in the inner $\simeq 1\secpoint0$ of the galaxy profile; however, since most of the remainder of this work will involve comparison of 2D cylindrical projections with one another, the imperfections in continuum subtraction at small $\theta_{\rm tran}$ are unlikely to affect the results.

\subsection{Dependence on azimuthal angle of cylindrically projected 2D (CP2D) spectra}
\label{sec:2dspec_azimuthal}

\begin{figure*}
\centering
\includegraphics[width=16cm]{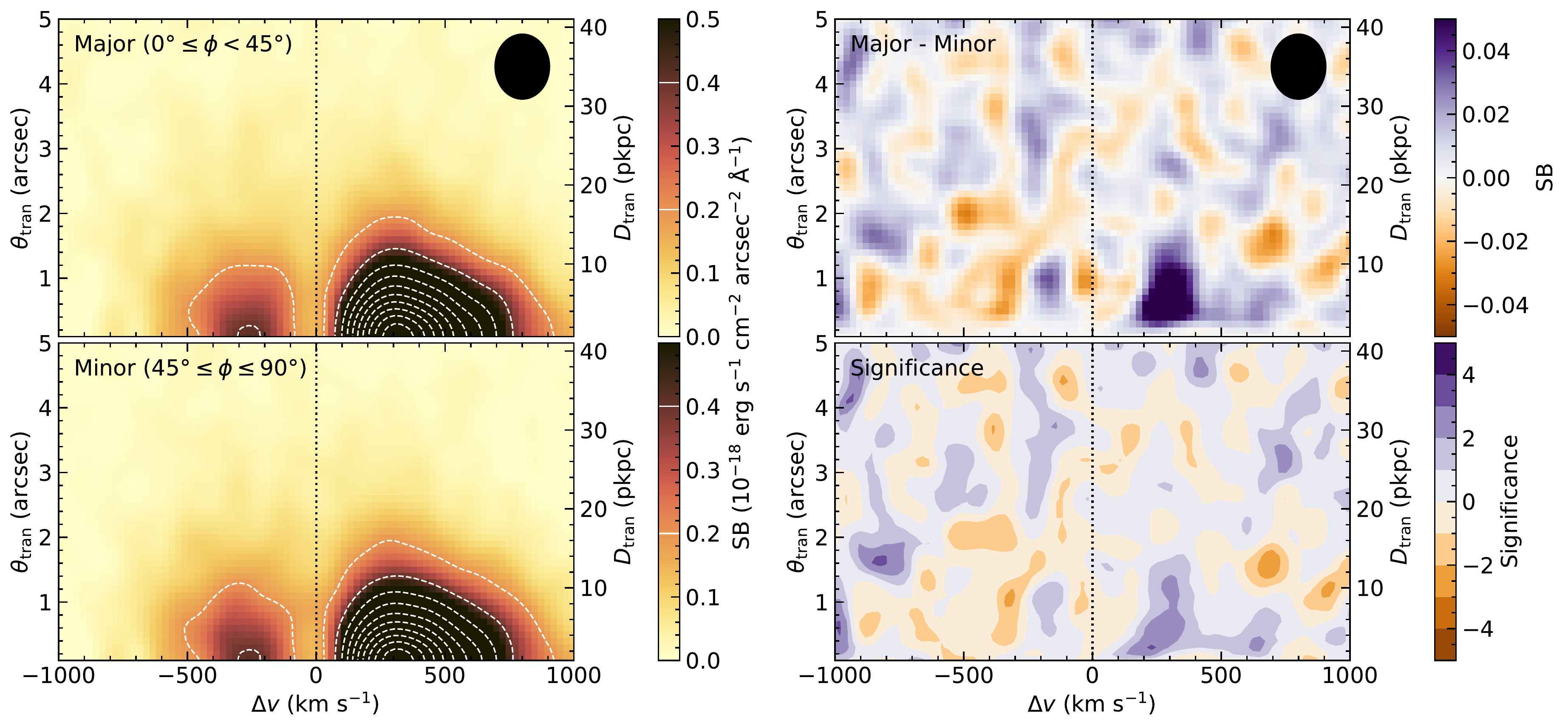}
\caption{ {\it Left}: The stacked CP2D spectra along the galaxy major ($0^\circ \le \phi < 45^\circ$; top) and minor ($45^\circ \le \phi \le 90^\circ$; bottom) axes. Both the colour-coding and the contours are on linear scales.  {\it Right}: The residual CP2D maps: the top panel shows the difference between the Major axis and Minor axis maps, in the same units of surface intensity as in the lefthand panels, where blue colours indicate regions with excess \lya\ surface intensity along the Major axis; orange colours indicate regions where \lya\ is brighter in the Minor axis map. The bottom panel shows the same residual map in units of the local noise level.  The most prominent feature is excess \lya\ emission along the galaxy major axis relative to that along the minor axis),  at $\Delta v \sim +300~\mathrm{km s}^{-1}$, extending to $\theta_{\rm tran} \sim 2\secpoint0$ or $D_{\rm tran} \sim 15$ pkpc.    } 
\label{fig:2dspec}
\end{figure*}

To investigate how \lya\ emission depends on the galaxy azimuthal angle, we split each CP2D spectrum of individual galaxies averaged over two independent bins of azimuthal angle ($\phi$) with respect to the galaxy's major axis:  $0^\circ \le \phi < 45^\circ$ (``major axis'') and  $45^\circ \le \phi \le 90^\circ$ (``minor axis''), as in \S\ref{sec:spatial_profile} and Figure \ref{fig:lya_profile}. The CP2D stacks covering these azimuth ranges were combined separately to form CP2D composites that we refer to as ``Major Axis'' and ``Minor Axis''. To reveal subtle differences in surface brightness and/or velocity along these two directions, we subtracted one from the other --  Figure \ref{fig:2dspec} shows the result. 

The difference between the CP2D spectra along the major and minor axes (top right of Figure \ref{fig:2dspec}) shows excess emission along the galaxy major axis at $\Delta v \simeq +300~\mathrm{km s}^{-1}$ -- consistent with the velocity of the peak of the redshifted component in the full composite CP2D spectrum --  within $\theta_{\rm tran} \simlt 2$\arcs\ ($D_\mathrm{tran} \lesssim 15~\mathrm{pkpc}$). The  \lya\ flux of this asymmetric component of \lya\ amounts to $\simlt 2$\% of the total, and has a peak intensity $\simeq 5$\% that of the peak of the redshifted \lya\ component shown in Fig.~\ref{fig:lya_profile}.

The significance map in the bottom-right panel of Figure~\ref{fig:2dspec} is based on the standard deviation of 100 independent mock CP2D stacks, each made by assigning random $\pa_0$ to the galaxies in our sample before combining. 
The mock stacks were then used to produce a 2D map of the RMS residuals evaluated in the same way as the observed data. Considering the effective resolution, the overall significance (compared to the standard deviation) of the most prominent feature in the top-right panel of Figure~\ref{fig:2dspec} is $\simeq 2-2.5 \sigma$ per resolution element. Thus, while the residual (excess) feature may be marginally significant statistically, the level of asymmetry relative to the total \lya\ flux is in fact very small.

\subsection{The robustness of residual \lya\ aysmmetry}
\label{sec:lya_robustness}

Despite the marginally significant detection of the excess \lya\ emission along galaxy major axes, its robustness is subject to scrutiny. In particular, we would like to determine whether the apparent detection is typical of the population or is caused by a few outlier objects having very asymmetric \lya\ as a function of azimuthal angle. 

\begin{figure}
\centering
\includegraphics[width=8cm]{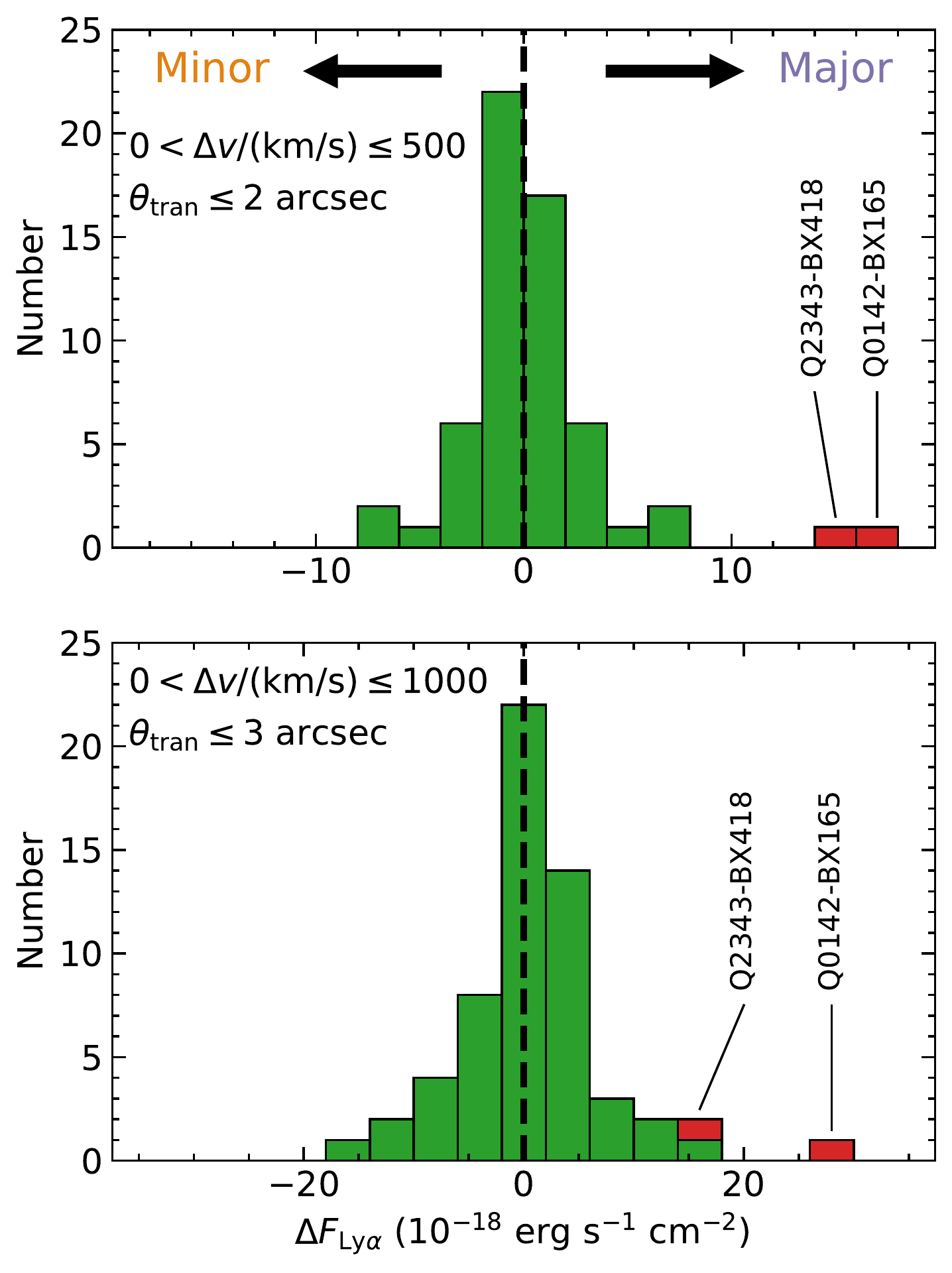}
\caption{ Distribution of the difference in \lya\ flux integrated over velocity and angular distance in the bins of azimuthal angle corresponding to ``major'' and ``minor'' axes. Positive (negative) values indicate that \lya\ emission is stronger along the major (minor) axis. The integration is conducted within $0 < \Delta v / (\mathrm{km~s}^{-1}) < 500$ and $\theta_\mathrm{tran} \le 2~\mathrm{arcsec}$ (top) and $0 < \Delta v / (\mathrm{km~s}^{-1}) < 1000$ and $\theta_\mathrm{tran} \le 3~\mathrm{arcsec}$ (bottom). There are two outliers in the first
integration (top panel), while one remains in the second (bottom panel).  }
\label{fig:hist_excess_lya}
\end{figure}

Figure \ref{fig:hist_excess_lya} shows histograms of the difference in integrated \lya\ flux between the major and minor axis bins of azimuthal angle. We calculated the differences integrated over two different ranges of $\Delta v$ and $\theta_{\rm tran}$; (1) the range where the excess shown in Figure~\ref{fig:2dspec} is most prominent, $\theta_{\rm tran} \le 2$\arcs\ and $0 \le (\Delta v)/\kms \le 500$, shown in the top panel, and (2) the range which encapsulates most of the redshifted component of \lya\, $\theta_{\rm tran} \le 3$\arcs\ and $0 \le (\Delta v)/\kms \le 1000$, shown in the bottom panel.
Two galaxies -- Q0142-BX165 and Q2343-BX418 -- are clearly outliers in (1), while only Q0142-BX165 stands out in (2). The distribution of $\Delta F_{\lya}$ for the other 56 galaxies in the sample is relatively symmetric around $\Delta F_{\mathrm{Ly}\alpha} = 0$. 

\begin{figure*}
\centering
\includegraphics[width=16cm]{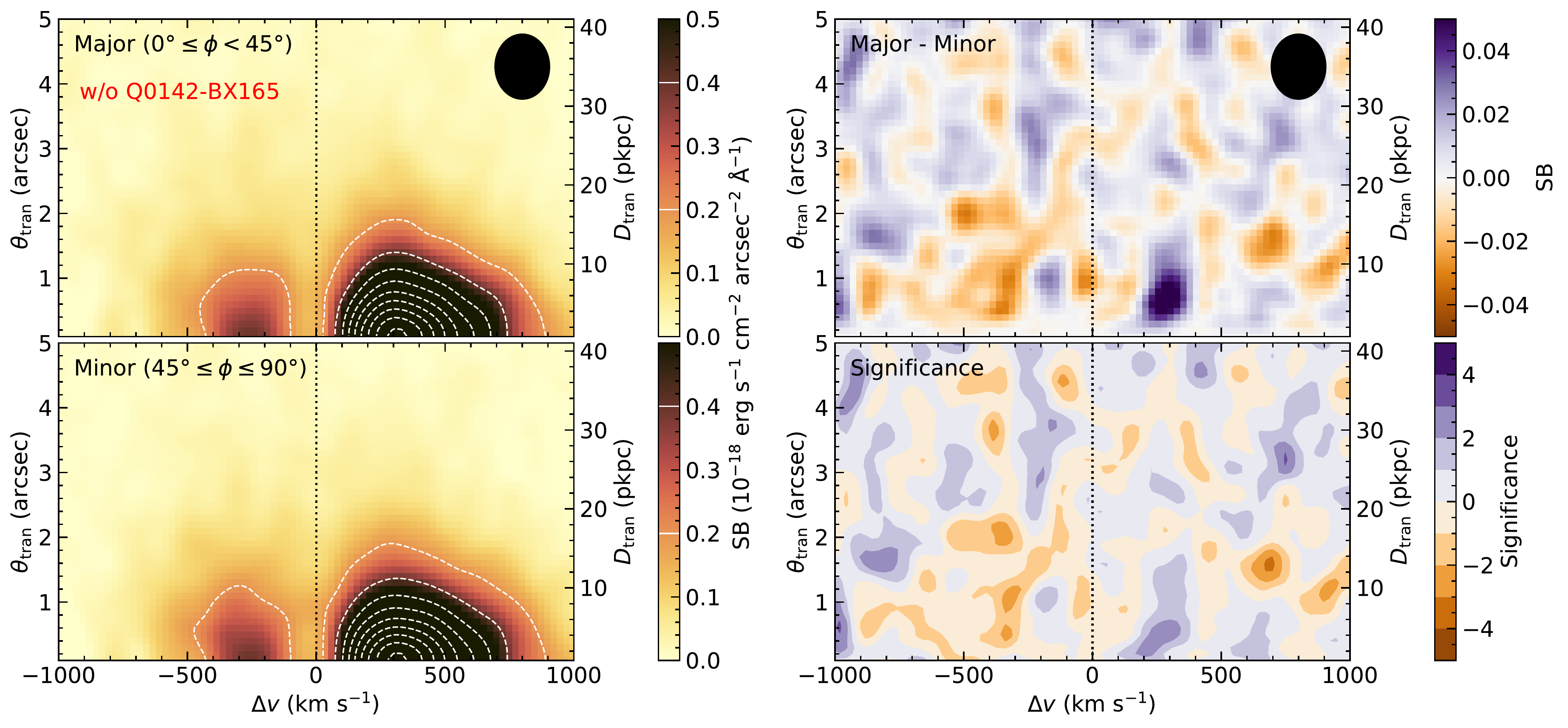}
\caption{  Same as Figure \ref{fig:2dspec}, but without Q0142-BX165, which is the strongest outlier in terms of  excess \lya\ emission along the galaxy major axis.  }
\label{fig:2dspec_nooutlier1}
\end{figure*} 
\begin{figure*}
\centering
\includegraphics[width=16cm]{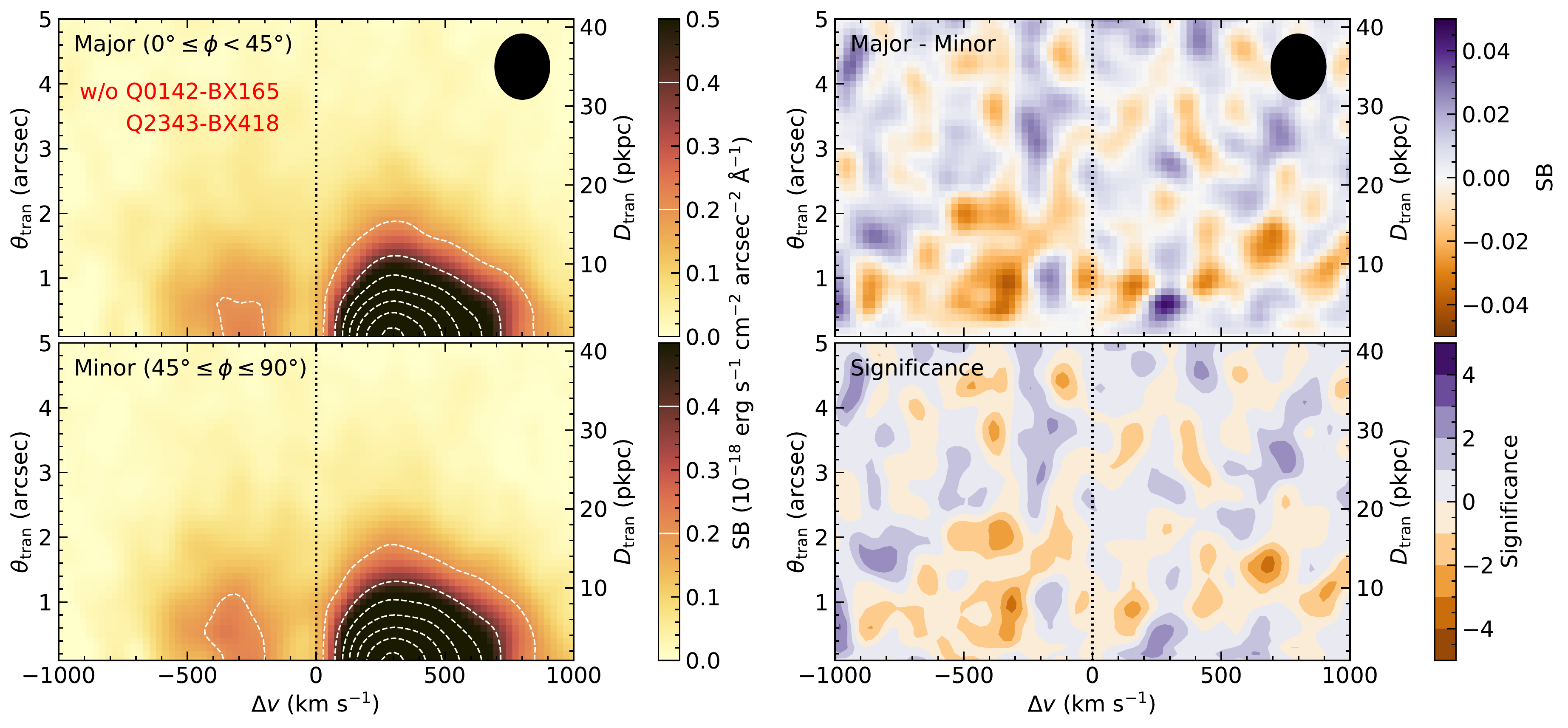}
\caption{  Same as Figure \ref{fig:2dspec}, but without Q0142-BX165 and Q2343-BX418, the two most significant outliers in the top panel of Figure~\ref{fig:hist_excess_lya}. Significant excess emission that is larger than a resolution element for the redshifted peak no longer exists.} 
\label{fig:2dspec_nooutlier2}
\end{figure*} 

Figures~\ref{fig:2dspec_nooutlier1}~and~\ref{fig:2dspec_nooutlier2} show the stacked \lya\ profiles
as in Figure~\ref{fig:2dspec}, but with the strongest outliers removed from the stack. 
After removing both outliers (Figure~\ref{fig:2dspec_nooutlier2}), the excess \lya\ emission along the galaxy major axis at $\Delta v \simeq 300~\mathrm{km~s}^{-1}$ becomes consistent with noise.
Although Q2343-BX418 is not an extreme outlier in terms of the overall \lya\ asymmetry of the integrated redshifted component of \lya\ emission (bottom panel of Fig.~\ref{fig:hist_excess_lya}), when only Q0142-BX165 is removed from the stack (Fig.~\ref{fig:2dspec_nooutlier1}) the composite 2D spectra still show obvious excess \lya\ emission along the galaxy major axis, albeit with slightly reduced significance. Meanwhile, when both outliers are removed from the stack (Figure~\ref{fig:2dspec_nooutlier2}), one can see an emerging excess \lya\ emission along the {\it minor} axis,  for the {\it blue} peak, with $-700 \lesssim \Delta v / (\mathrm{km~s}^{-1}) \lesssim -200$ and $\theta_{\rm tran} \simlt 2\secpoint5$ with integrated significance $\sim 2 \sigma$. The flux of the excess blueshifted emission comprises $\sim 10$\% of the total blueshifted \lya\ flux, with a peak amplitude $\sim 1$\% of the peak \lya\ intensity (i.e., the redshifted peak).

\begin{figure}
\centering
\includegraphics[width=8cm]{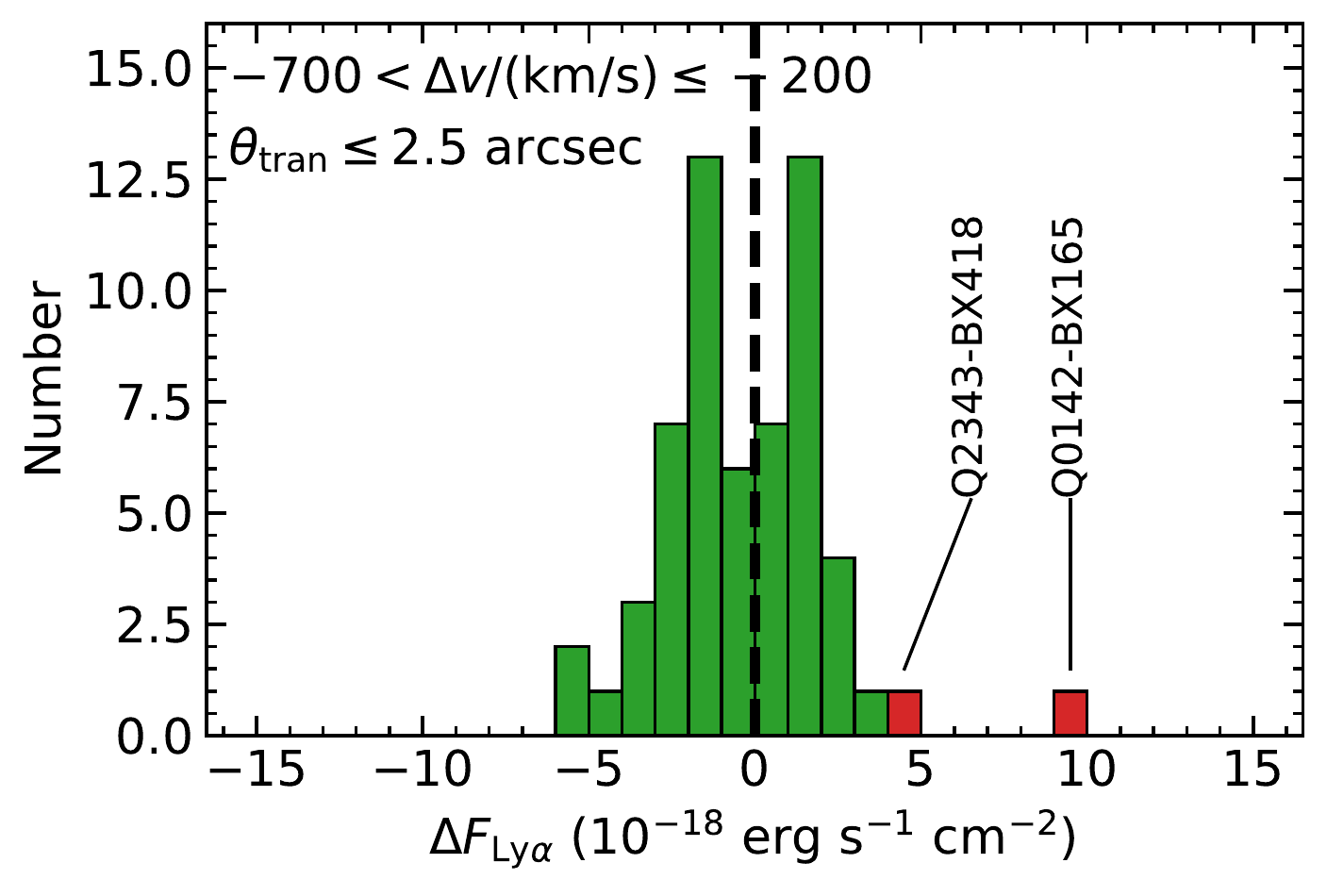}
\caption{ Same as Figure \ref{fig:hist_excess_lya}, but with a different velocity range of $-700 < \Delta v / (\mathrm{km~s}^{-1}) \le -200$ and $\theta_\mathrm{tran} \le 2.5~\mathrm{arcsec}$ that focuses on the blueshifted component of \lya\ emission. No individual galaxy is an extreme outlier in terms of excess blueshifted \lya\ along the minor axis. }
\label{fig:hist_excess_lya_blue}
\end{figure}

We conducted an analysis on the \textit{blueshifted} emission similar to that done for the redshifted asymmetry, with results summarised in Figure \ref{fig:hist_excess_lya_blue}. There is no obvious outlier in the difference in integrated \lya\ flux between the major and minor axis azimuth bins except for Q0142-BX165, for which the excess again favors the {\it major} axis (i.e., it is in the direction opposite to the apparent blueshifted asymmetry identified in Figure~\ref{fig:2dspec_nooutlier2}). We also consecutively removed from the \lya\ stack galaxies with extreme excess emission along the minor axis, and found no sudden and significant changes in the composite spectra. Evidently, the blueshifted excess along the minor axis, while of about the same significance as the redshifted excess in the major axis direction, is a general property of the full sample rather than a result of a small number of outliers.

In summary, we found excess emission along the galaxy major axis for the redshifted component of \lya\ near $\Delta v \simeq 300~\mathrm{km~s}^{-1}$. However, this particular excess emission appears to be caused by galaxy outliers with extreme emission along the major axis. After removing them from the composite CP2D spectra we found excess emission along the galaxy minor axis for the blue peak within $-700 \lesssim \Delta v / (\mathrm{km~s}^{-1}) \lesssim -200$ that is not apparently affected by the extreme scenarios. Both detections are not particularly significant at the $\sim 2\sigma$ level. 

\subsection{A Closer Look at the Extreme Cases}
\label{sec:closer_look}

\begin{figure}
\centering
\includegraphics[width=8cm]{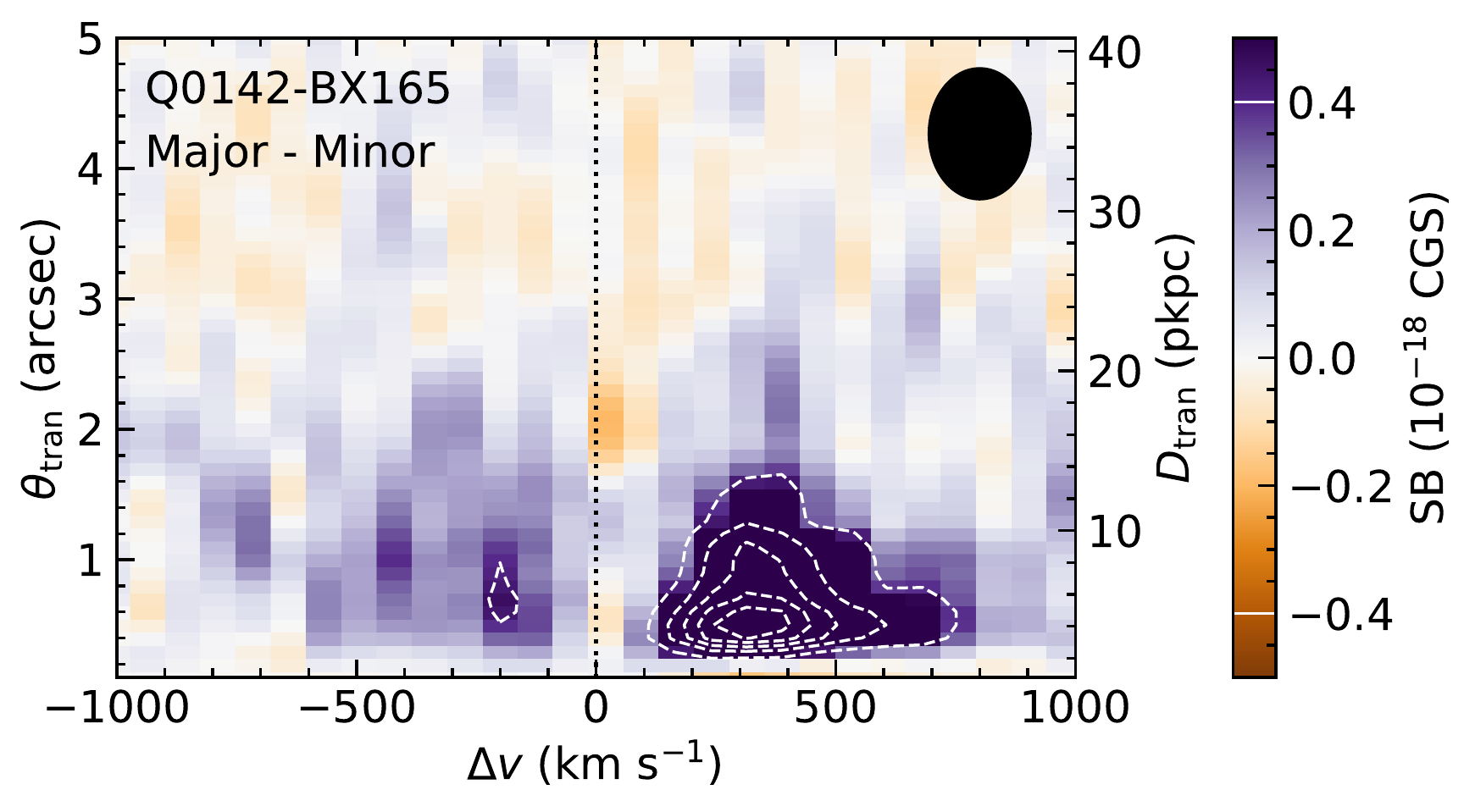}\\
\includegraphics[width=8cm]{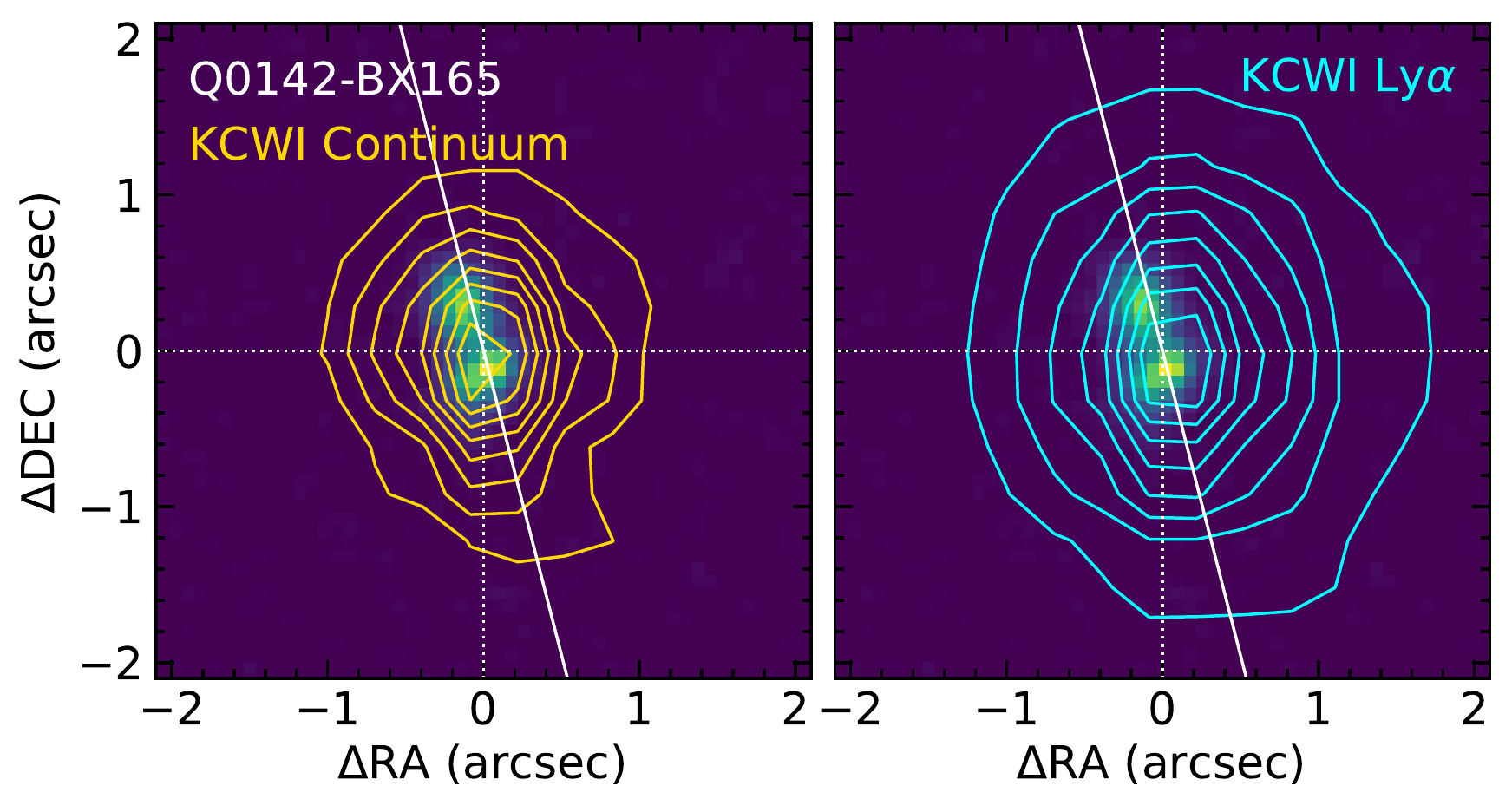}
\caption{ {\it Top}: Same as the top-right panel of Figure \ref{fig:2dspec}, but for a single galaxy, Q0142-BX165, which has the strongest excess \lya\ emission along the galaxy major axis. Note that the colour scale is 10 times that of Figure \ref{fig:2dspec}. {\it Bottom}: The HST F160w image of Q0142-BX165, overlaid with contours from the KCWI continuum image (left) and the narrow-band \lya\ image (right). \label{fig:q0142-BX165}}
\end{figure} 

Figure \ref{fig:q0142-BX165} shows the residual between the cylindrically projected 2D spectra extracted along the major and minor axes of Q0142-BX165, as well as continuum images from both HST and KCWI. Q0142-BX165 has two comparably bright components separated by $\simeq 3.5~\mathrm{pkpc}$ ($\simeq 0\secpoint4$). Careful inspection of Keck/LRIS and Keck/MOSFIRE spectra of this system revealed no sign of an object at a different redshift. There is no significant offset between the KCWI continuum and \lya\ centroids ($\le 0.5~{\rm pix}\sim 0\secpoint15$ separation), indicating that the apparent directional asymmetry in the CP2D spectra is not caused by a spatial shift between the continuum and \lya\ emission. Instead, the narrow-band \lya\ map shows that the \lya\ emission is elongated approximately along the N-S direction. However, after aligning the KCWI and HST astrometry with reference to a nearby compact galaxy, we found that both the KCWI stellar continuum near \lya\ and the narrow-band \lya\ emission are centered near the SW component (see Figure \ref{fig:q0142-BX165}). It is possible that the SW component alone is responsible for the \lya\ emission, in which case its $\pa_0$ would be $-59^\circ$,~ $\sim 70^\circ$ off from what was determined in \S\ref{sec:pa}. However, adopting $\pa_0 = -59^{\circ}$ would cause BX165 to become an outlier with excess \lya\ emission along the {\it minor} axis. Meanwhile, the elongation of the KCWI continuum aligns with the direction of the separation of the two components, and is roughly consistent with the direction of the \lya\ elongation as well. This seems to suggest that the \lya\ elongation simply reflects the asymmetry of the continuum source, albeit on a larger angular scale; however, as shown in Figure \ref{fig:hst_mmt_only}, many galaxies in the sample possess similar morphologies, but Q0142-BX165 is the only one that shows extraordinary asymmetry in \lya\ emission. In any case, Q0142-BX165 has a uniquely asymmetric \lya\ halo, possibly due to source confusion. Consequently, we exclude it from most of the analysis that follows. 

\begin{figure}
\centering
\includegraphics[width=8cm]{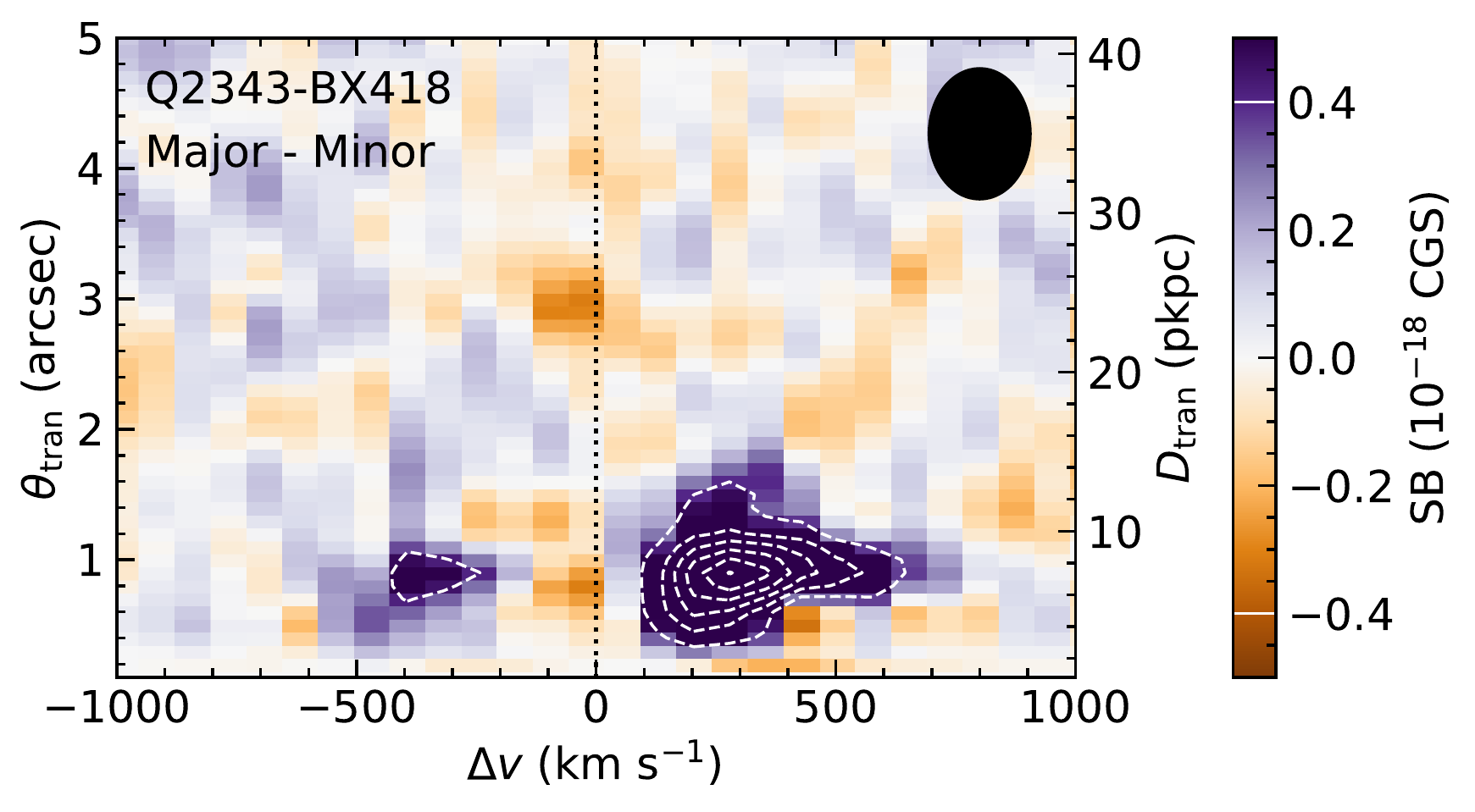}\\
\includegraphics[width=8cm]{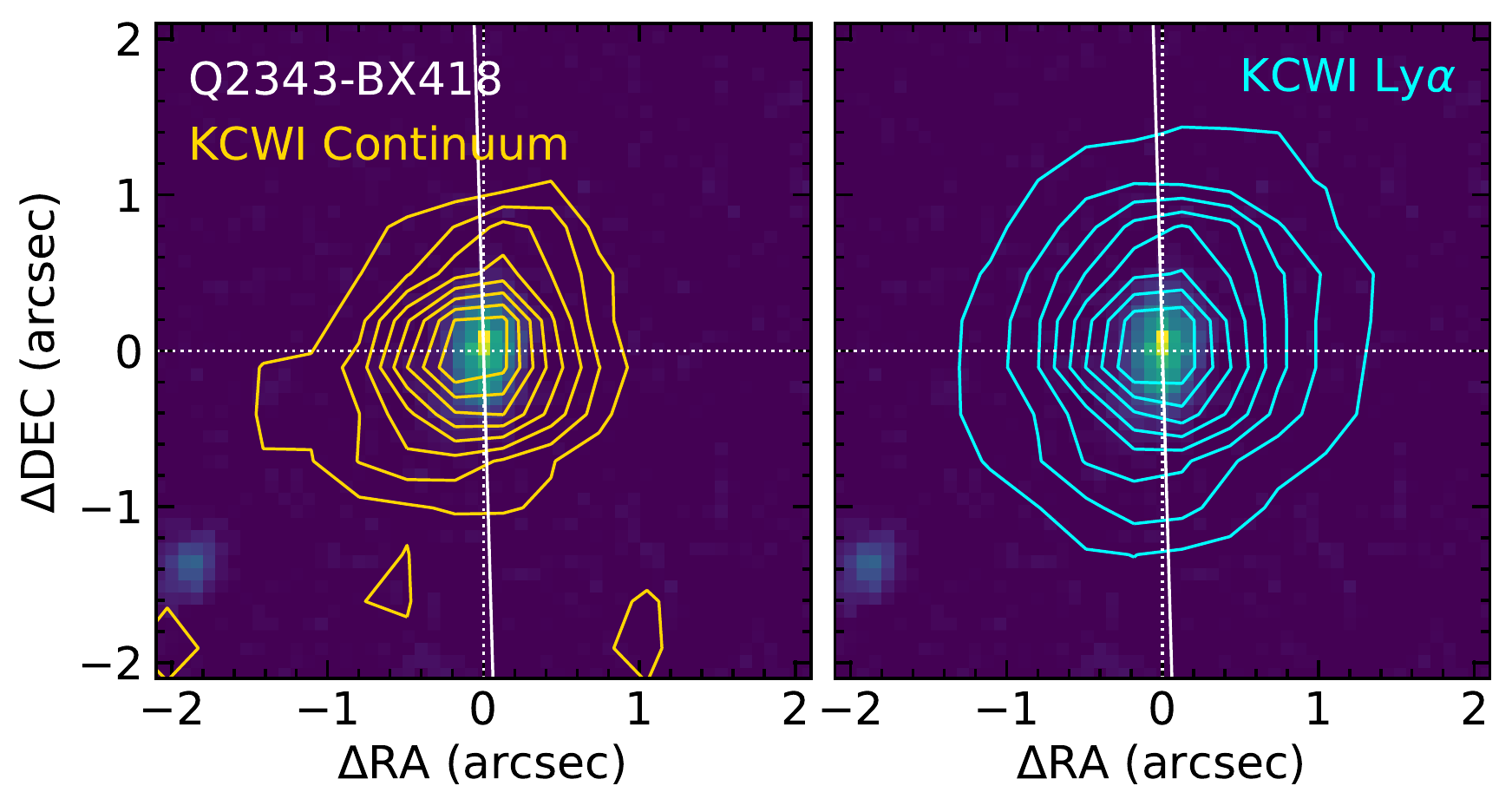}
\caption{  Same as Figure \ref{fig:q0142-BX165} but for Q2343-BX418, the object with the second strongest major axis \lya\ asymmetry. \label{fig:q2343-BX418}}
\end{figure} 

The KCWI data cube for Q2343-BX418 has been analysed previously by \citet{erb18}; here, we consider it in the context of the analysis of Q0142-BX165 above (see Figure~\ref{fig:q2343-BX418}). 
 The difference in peak SB between the major and minor axis CP2D spectra is nearly equal to that of Q0142-BX165 ($2.4 \times 10^{-18}~\mathrm{erg~s}^{-1}\mathrm{cm}^{-2}\mathrm{arcsec}^{-2}\textrm{\AA}^{-1}$ for both). However, the spatial extent of the excess emission is significantly smaller for Q2343-BX418.  The HST/WFC3, KCWI continuum, KCWI \lya\ images, and OSIRIS-H$\alpha$ images all show Q2343-BX418 comprise a single component whose centroids in the various images are consistent with one another. Despite its extreme SB asymmetry in \lya, Q2343-BX418 exhibits no other obviously peculiar property compared to the rest of the sample.

\section{  Integrated Line Flux and Azimuthal Asymmetry}
\label{sec:az_halo}

\begin{figure*}
\centering
\includegraphics[width=16cm]{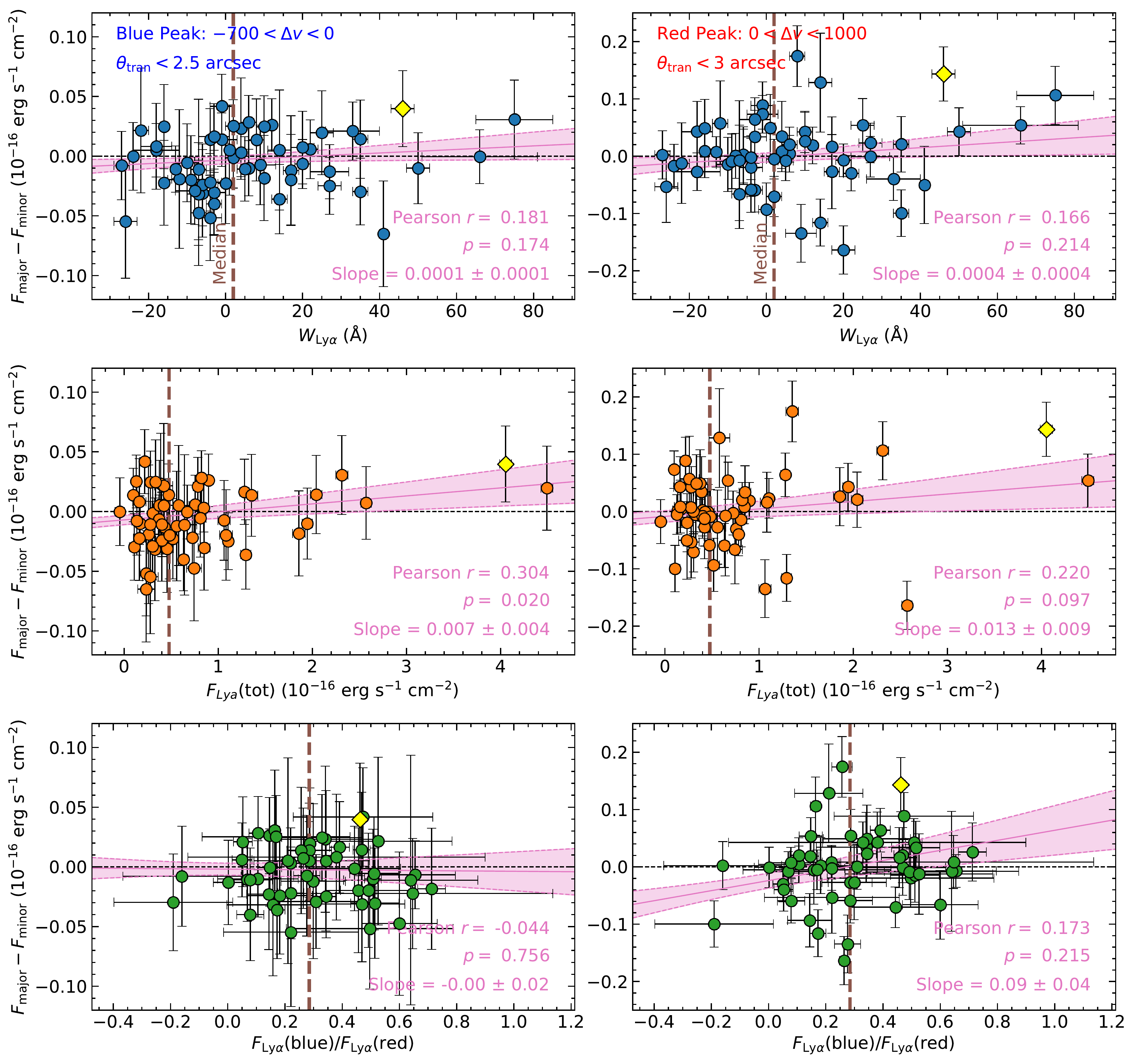}
\caption{ Relationship between the flux measurements of anisotropic (excess) \lya\ emission ($\Delta \flya = F_\mathrm{major} - F_\mathrm{minor}$) of the blueshifted component (left) and redshifted component (right) of \lya\ emission and properties of the integrated \lya\ halo [{\it Top}: central \lya\ equivalent width, \wlya; {\it Middle}: total \lya\ flux, $\flya(\mathrm{tot})$; {\it Bottom}: the ratio between the total blueshifted and redshifted components,  $\fblue / \fred$]. Galaxies without reliable $F_\mathrm{red}$ are omitted in the bottom panel since their $\fblue / \fred$ are dominated by noise. {The pink lines and shaded regions show the results and their 1$\sigma$ uncertainties of a linear regression accounting for the errors in both x- and y-directions.} The vertical dashed line in each panel marks the median value of the (x-axis) property for the full sample. The yellow diamond in each panel marks the location of Q2343-BX418, the outlier that caused the excess emission of the redshifted peak along the galaxy major axis as discussed in \S\ref{sec:closer_look}.}
\label{fig:df_halo}
\end{figure*}

As shown in the previous sections, the degree of \lya\ halo azimuthal asymmetry varies from case to case in our $z \simeq 2.3$ sample, but the correlation with the morphology of the central galaxy is sufficiently weak that, on average, \lya\ halos are remarkably symmetric and appear to be nearly independent  --  both kinematically and spatially -- of the apparent orientation of the galaxy at the center. 

Thus far we have treated the blueshifted and redshifted components of \lya\ emission separately. However, the overall \lya\ profile is expected to provide clues to the geometry and velocity field of  circumgalactic \ion{H}{I}. In this section, we compare the dependency between excess \lya\ emission and \wlya, total \lya\ flux $\flya(\mathrm{tot})$, and the ratio of the total flux of blueshifted and redshifted components of \lya\ emission  [$\fblue / \fred$]. The integration windows used to compute the values in Table~\ref{tab:sample} are $\flya(\mathrm{tot})$: $\theta_\mathrm{tran} \le 3~\mathrm{arcsec}$ and $-700 < \Delta v / (\mathrm{km~s}^{-1}) \le 1000$; \fblue: $\theta_\mathrm{tran} \le 2.5~\mathrm{arcsec}$ and $-700 < \Delta v / (\mathrm{km~s}^{-1}) \le 0$; \fred: $\theta_\mathrm{tran} \le 3~\mathrm{arcsec}$ and $0 < \Delta v / (\mathrm{km~s}^{-1}) \le 1000$. {Different integration windows were used for the two components in order to optimise the S/N of the integral; they were chosen based on the detected extent of each component in Figure~\ref{fig:2dspec_nooutlier1}, in both $\theta_{\rm tran}$ and $\Delta v$.}%

{Figure \ref{fig:df_halo} examines whether or not there is a connection between major axis/minor axis asymmetry in \lya\ flux and the overall \lya\ halo properties mentioned above.For each pair of variables in Figure~\ref{fig:df_halo}, we indicate the  value of the Pearson coefficient ($r$) and the corresponding probability $p$ that the observed data set could be drawn from an uncorrelated parent sample. 
We also performed a linear regression using Orthogonal Distance Regression (ODR) in \textit{SciPy}, which accounts for the estimated uncertainty in both x- and y-variables. As can be seen from the figure, most of the Pearson tests show no significant correlation and linear regression yields slopes consistent with zero. However, the Pearson test for the relation between $F_\mathrm{major}(\mathrm{blue}) - F_\mathrm{minor}(\mathrm{blue})$ and $\flya(\mathrm{tot})$ (middle left panel of Figure \ref{fig:df_halo}) yields $p = 0.02$, suggesting a marginally significant trend in which the asymmetry of the blueshifted component of \lya\ favors the minor axis when $F_{\lya}(\rm tot)$ is weak, and the major axis when $F_{\lya}({\rm tot})$ is strong. }

The second relationship that stands out is $F_\mathrm{major}(\mathrm{red}) - F_\mathrm{minor}(\mathrm{red})$ and $\fblue / \fred$ (bottom right panel of Figure \ref{fig:df_halo}), where a non-parametric test for correlation is not significant. The linear regression results in a marginally-significant positive slope, indicating that as the blueshifted component of \lya\ approaches the strength of the redshifted component, there is a tendency for excess emission along the major axis; for galaxies with $\fblue / \fred \simlt 0.3$ (i.e., smaller than the median value for the sample), the tendency is for excess \lya\ emission along the minor axis.

\begin{figure*}
\centering
\includegraphics[width=16cm]{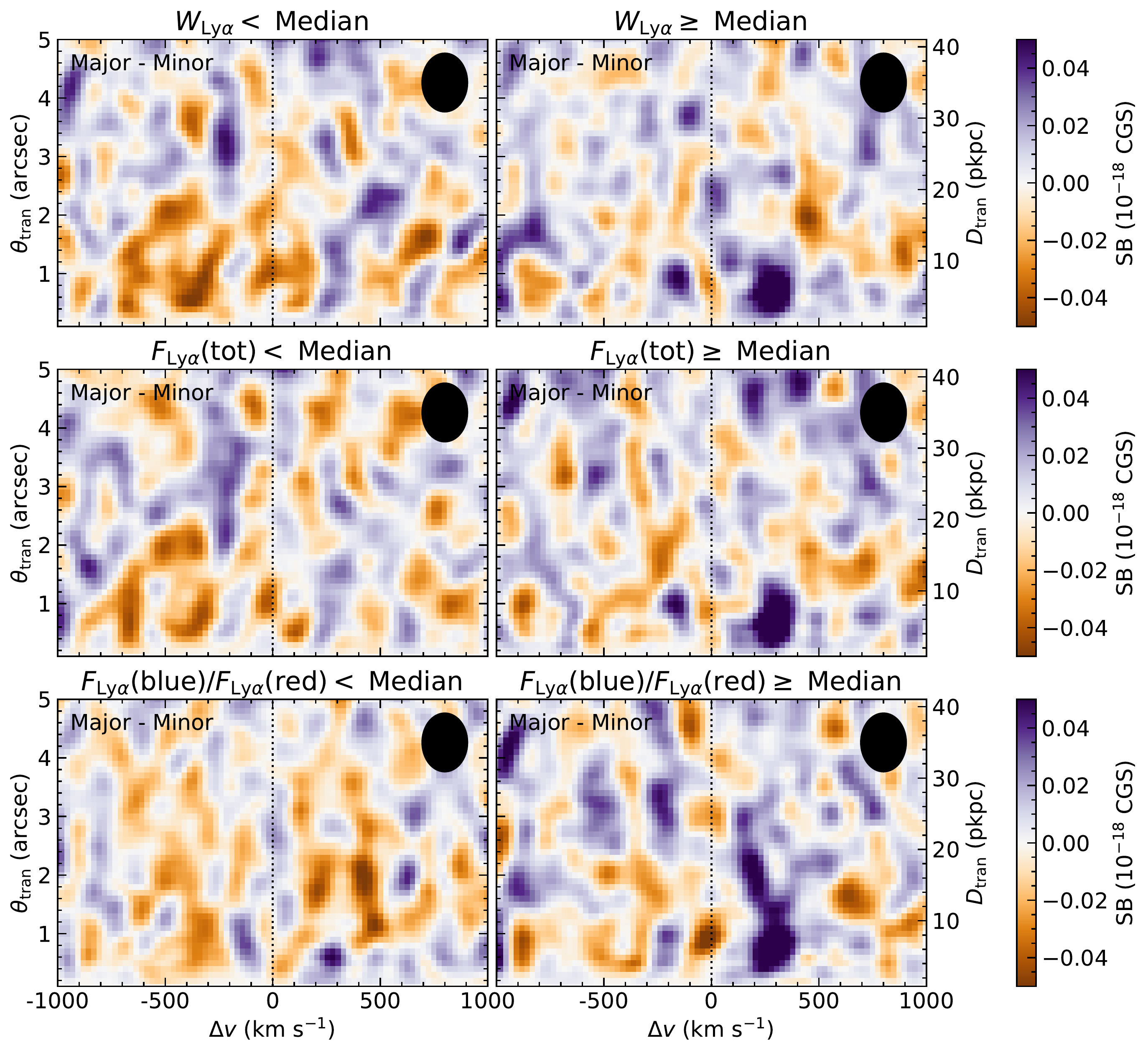}
\caption{  The difference between the CP2D spectra of \lya\ emission for the major and minor axes. The maps show the residual for CP2D stacks for two sub-samples representing those below (left) and above (right) the sample median, From top to bottom, the \lya\ halo properties are the central \lya\ equivalent width (\wlya), the integrated \lya\ flux ($\flya ({\rm tot})$), and the flux ratio between the blueshifted and redshifted components ($\fblue / \fred$). This figure confirms that the blueshifted excess \lya\ emission favours weak \lya\ emitting galaxies.}
\label{fig:sb2d_halo}
\end{figure*}

To further explore the reliability of the correlations, we split the sample in two halves according to the overall halo properties, and compared the subtracted CP2D spectra between the major and minor axes. Figure \ref{fig:sb2d_halo} shows the result. As discussed in \S\ref{sec:closer_look}, the excess \lya\ emission along the galaxy major axis for the red peak at $\Delta v \simeq 300~\mathrm{km~s}^{-1}$ within $\theta_\mathrm{tran} \lesssim 1~\mathrm{arcsec}$ can be attributed to a single outlier (Q2343-BX418), which happens to fall above the median in all 3 quantities considered in Figure~\ref{fig:sb2d_halo} (i.e., on the righthand panels of the figure). The subtracted CP2D spectra also indicate that essentially the entire excess \lya\ emission for the blue peak along the minor axis -- as identified earlier (\S\ref{sec:closer_look}) --  is contributed by galaxies below the median \wlya\ and  $F_{\lya}({\rm tot})$ (top and middle lefthand panels of Figure~\ref{fig:sb2d_halo}). In particular, the integrated significance within $-700 < \Delta v / (\kms) \le -200$ and $\theta_\mathrm{tran} \le 2\secpoint5$ exceeds $2.5\sigma$ for the $\wlya < \mathrm{Median}$ bin. Comparison of the top two panels also illustrate the same trend of $F_\mathrm{major}(\mathrm{blue}) - F_\mathrm{minor}(\mathrm{blue})$ vs. \wlya\ and \flya\ correlations suggested by Figure~\ref{fig:df_halo}. 

For the subsamples divided based on the value of $\fblue/\fred$, the differences between the major axis and minor axis range of azimuthal angle are less significant: there is a marginally significant excess of \lya\ emission, more noticeable in the redshifted component, where it appears to extend over the range $\theta_{\rm tran} \simeq 1-3$ arcsec, and the bin with higher $\fblue/\fred$ appears to have a major axis excess over approximately the same range of angular distances. If Q2343-BX418 is removed from the stack of larger $\fblue/\fred$ galaxies, 
the residual remains, showing that it is not attributable to a single outlier. However, none of the residuals in the bottom panels of Figure~\ref{fig:sb2d_halo} reaches a threshold of $2\sigma$ per resolution element.  

In summary, a small statistical azimuthal asymmetry of \lya\ halos persists when the galaxy sample is divided into two according to central \wlya, total $F_{\lya}$, and the flux ratio of blueshifted and redshifted emission. Perhaps most intriguing is that galaxies with small or negative central \wlya\ have a tendency to exhibit excess \lya\ emission along galaxy minor axes extending over a fairly large range of both $\theta_{\rm tran}$ and velocity ($-700 < (\Delta v/\kms) < -200$).  The apparent excess along the major axis of redshifted \lya\ for the subsample with stronger \lya\ emission and larger $\fblue/\fred$, on the other hand, is confined to a smaller range of (redshifted) velocities, again roughly coincident with the typical location of the``red peak''.

\section{Finer Division of Galaxy Azimuthal Angles}
\label{sec:three_bins}

At lower redshifts ($z < 1$), the covering fraction and column density of gas in various ionisation stages are commonly found to be related to the orientation of the gaseous disk relative to the line of sight. Many authors have used used background QSO or galaxy sightlines to detect strong \ion{Mg}{II} absorbers associated with galaxies at redshifts to allow the foreground galaxy orientations to be measured (e.g., \citealt{steidel02,bordoloi11,bouche12,schroetter19, lundgren21}). A common conclusion is that the sightlines giving rise to strong absorption tend to be those located at azimuthal angles corresponding to  both the galaxy major and minor axes, but fewer (strong) absorbers are found at intermediate angles,  $30^\circ < \phi < 60^\circ$. The kinematics of the absorbing gas also appear to be related to $\phi$ \citep[e.g., ][]{ho17,martin19}, with the broadest (and therefore strongest) systems found along the minor axis, presumably due to fast bi-conical outflows oriented perpendicular to the disk, followed by major axis sightlines sampling accreting or galactic fountain gas sharing the halo's angular momentum and thus exhibiting disk-like rotation.

\begin{figure*}
\centering
\includegraphics[width=16cm]{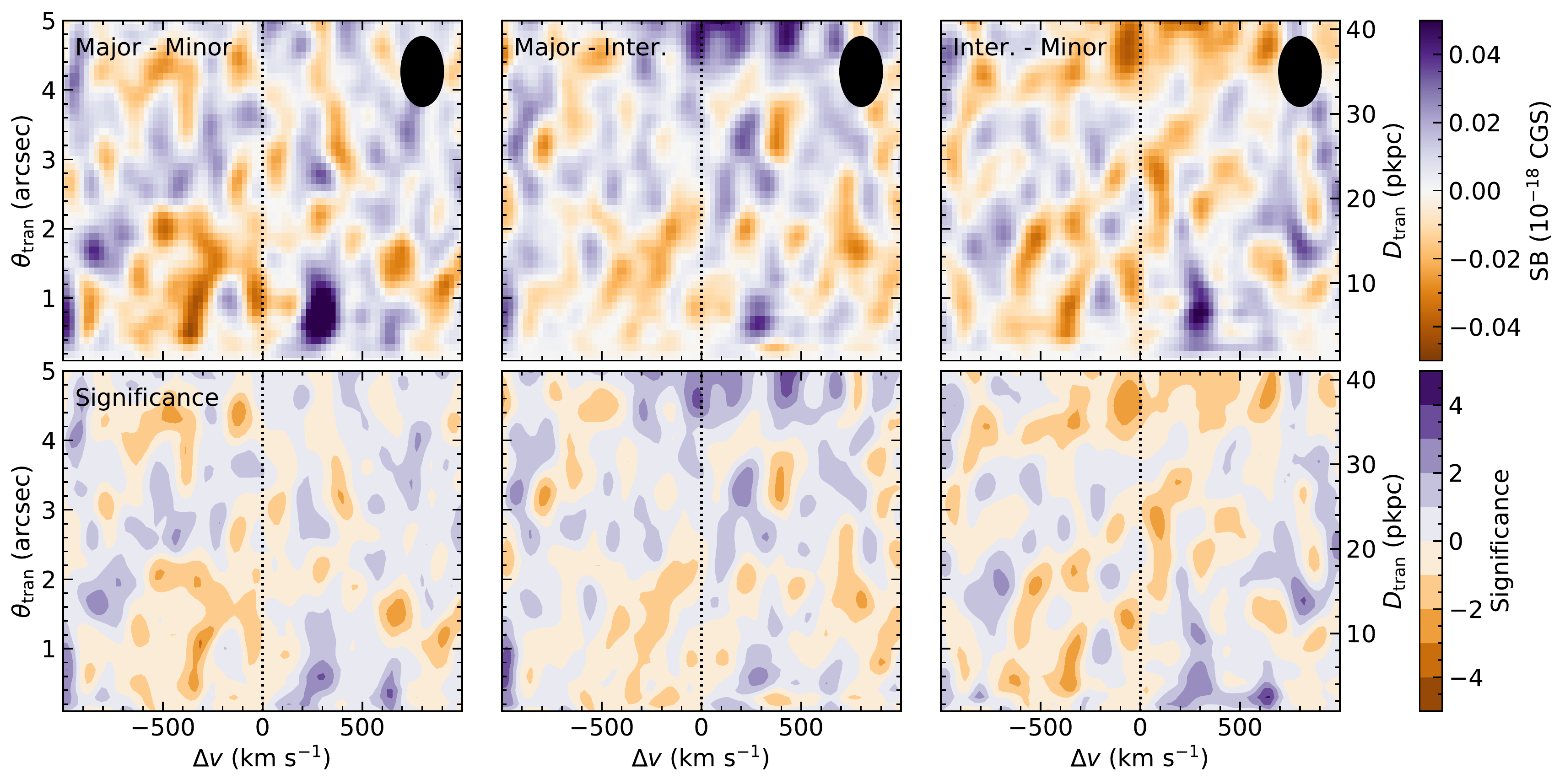}
\caption{  Similar to the right panel of Figure \ref{fig:2dspec} but residual maps are between the major and minor (left), major and intermediate (middle), and intermediate and minor (right) axes with each bin size of only $\Delta \phi = 30^\circ$. The strong residual beyond $\theta_\mathrm{tran} > 4~\mathrm{arcsec}$ is caused by a contaminating source near a single object. No sign of a bimodal distribution of \lya\ emission is present. The sample in this figure is the same as in Figure \ref{fig:2dspec_nooutlier1}. (Q0142-BX165: discarded; Q2343-BX418: included). }
\label{fig:sb30}
\end{figure*} 

If the \lya\ emission around $z = 2$--3 galaxies arises in CGM gas with properties similar to that of the low-ionization metallic absorbers at lower redshift, one might expect to see similar evidence for asymmetries along the two principal axes relative to intermediate azimuthal angles. We tested this possibility by expanding our analysis in \S\ref{sec:2dspec} by dividing $\phi$ into three azimuthal bins -- major axis ( $0^\circ \le \phi < 30^\circ$), minor axis ($60^\circ \le \phi \le 90^\circ$), and intermediate ($30^\circ \le \phi \le 60^\circ$). CP2D difference spectra among these 3 bins are shown in Figure~\ref{fig:sb30}.

Figure \ref{fig:sb30} shows no obvious sign of a bimodal distribution of \lya\ emission with respect to $\phi$, in which case the middle and right panels would show residuals of opposite sign. 
Instead, the residual maps of ``Major $-$Intermediate'' and ``Intermediate $-$ Minor'' suggest a gradual transition of the small asymmetries between major and minor axes in the larger bins of $\phi$ shown previously. 

In any case, the main conclusion to draw from Figure~\ref{fig:sb30} is once again that \lya\ emission halos are remarkably similar in all directions with respect to the projected principal axes of $z \sim 2-3$ galaxies.

\section{Discussion} 
\label{sec:discussions}
\subsection{Comparison to previous work}

The morphology of \lya\  emission from the CGM and its relation to the host galaxies has been analysed in various works between $z = 0$ and $z \lesssim 4$. However, analyses quantifying the \lya\ emission with respect to the host-galaxy orientation is limited. In this section, we attempt to compare the existing research on \lya\ halo morphology with our findings.

At very low redshifts ($z = 0.02 - 0.2$), \citet{guaita15} studied the \lya\ halos of 14 \lya-emitting galaxies in the Lyman Alpha Reference Sample (LARS; \citealt{ostlin14}) and their connection to the host-galaxy morphologies. They found that the \lya\ halos for the subset of the sample that would be considered LAEs are largely axisymmetric, and there is no single galaxy morphological property that can be easily connected to the overall shape of the \lya\ emitting regions.  Indeed, the \lya\ images of the individual galaxies in \citet{hayes14} appear to be independent of galaxy orientation beyond $D_{\rm tran} \sim 2$ pkpc. Meanwhile, the stacked \lya\ image of the LARS galaxies is elongated in the same direction as the far-UV continuum, i.e., along the major axis.
However, this stack could be significantly affected by sample variance from galaxies that are particularly bright in rest-UV and \lya, as we saw for the KBSS sample before removing outliers. Interestingly, \citet{duval16} studied a special galaxy in the LARS sample which is almost perfectly edge-on -- they found two small \lya\ emitting components (each of extent $< 1$ pkpc) near the disk consistent with \lya\ emission escaping from the ISM through ``holes'' in the galactic disk, akin to that expected in a classical Galactic fountain \citep{bregman80}. While this is likely to be driven by stellar feedback allowing \lya\ to escape in the direction perpendicular to the galaxy disk, the observed \lya\ features are far closer to the galaxy than could be measured in our $z \sim 2-3$ sample. Moreover, most of the galaxies in our sample do not have organised thin disks, and so likely have much lower dust column densities obscuring active star forming regions.

At $z > 3$, \citet{bacon17} (and subsequent papers from the same group) conducted a systematic survey of \lya\ emitting galaxies in the Hubble Ultra Deep Field using VLT/MUSE. For example, \citet{leclercq17} found  significant variation of \lya\ halo morphology among 145 galaxies, and identified correlations between the halo size and the size and brightness of the galaxy UV continuum. However, its connection with the galaxy morphological orientation remain to be investigated. 

Meanwhile, \citet{mchen20} examined a case of a strongly-lensed pair of galaxies with extended \lya\ emission at $z > 3$. 
The continuum image in the reconstructed source plane of their system A has at least three subcomponents extending over $> 1$ arcsec, which may be similar to the subsample of galaxies shown in our Figure \ref{fig:hst_mmt_only}, in terms of morphological complexity. In the context of the analysis we describe in the present work, this arrangement of \lya\ with respect to continuum emission would be classified as excess minor axis \lya\ emission. Unfortunately, the \lya\ halo of a second $z > 3$ system is truncated in the source plane reconstruction. 

In summary, we compared our result with galaxies and their \lya\ emission morphology at $z\sim 0$ and $z > 3$ in previous work, finding that although our results are qualitatively consistent with earlier results, the comparison is hampered by limited sample sizes, as well as by differences in redshift and intrinsic galaxy properties represented in each sample.

\subsection{Theoretical predictions}

Many existing studies of the distribution and kinematics of CGM gas have focused on simulated galaxies within cosmological hydrodynamic simulations.  \citet{peroux20} analysed how inflowing and outflowing gas might distribute differently as a function of the galactic azimuthal angle within the EAGLE and IllustrisTNG simulations, finding significant angular dependence of the flow rate and direction, as well as the CGM metallicity, with outflows of higher metallicity gas favoring the galaxy minor axis, and accretion of more metal-poor gas tending to occur along the major axis, at $z < 1$. Although \citet{peroux20} focused on $z \sim 0.5$ for their study, they made clear that the predicted trends would weaken significantly with increasing redshift.

At much higher redshifts ($z = 5-7$) around galaxies in the FIRE suite of simulations,  \citet{smith19} found that \lya\ escape is highly correlated  with the direction of the \ion{H}{I} outflow. Naively, one might expect that more \lya\ would be found along the galaxy minor axis, which is the direction along which gaseous outflows would encounter the least resistance to propagation to large galactocentric radii.  However, the galaxies experiencing the most active star formation at these redshifts tend to be altered on short timescales ($\sim 10^7$ yrs) by episodic accretion, star formation, and feedback events. Thus, the direction of outflows may change on similar timescales, while the CGM will evolve on a longer timescale, possibly erasing any clear signatures of alignment of outflows and \lya\ emission. For the same reason, rapidly star-forming galaxies at $z \sim 2-3$, most of which have not yet established stable stellar disks, are likely to be surrounded by gas that is similarly turbulent and disordered.

Meanwhile, analytic or semi-analytic models of \lya\ resonant scattering for idealised outflow geometries have focused primarily on the integrated \lya\ emission profile from the entire galaxy or \lya\ halo. Although many models account for the impact of the geometry and kinematics of gaseous outflows or accretion on the integrated \lya\ emission profiles, there have been fewer efforts to predict the two-dimensional spatial and spectral profiles for detailed comparison to IFU observations.

For example, \citet{carr18} constructed a model to predict the \lya\ spectral morphology assuming biconical \ion{H}{I} outflows with resonant scattering. The model predicts the integrated spectral profile without spatial information. In the context of the biconical outflow model, an integrated profile resembling our observation is predicted when the minor axis is perpendicular to the line of sight, with a large outflow having a small opening angle.  However, this particular configuration would likely give rise to a highly asymmetric {\it spatial} distribution of \lya\ emission, which we have shown is unlikely to be consistent with our observations.

Our results highlight the need for 3-D models of the cool gas in the CGM around rapidly-star-forming galaxies at high redshifts, prior to the development of stable disk configuration, for which the assumption of axisymmetry of outflows, at least on average, may be closer to reality. In any case, predicting the spatial and spectral properties of \lya\ {\it emission}  will require realistic treatment of the kinematics, small-scale structure, and radiative transfer of \lya\ photons from the sites of production to their escape last scattering from the CGM toward an observer. 

\citet{gronke16b} has devised a model that assumes a two-phase CGM, composed of optically-thick clumps embedded in a highly-ionised diffuse ``inter-clump''  medium. This method has been used to fit the \lya\ profiles at multiple locations within a spatially-resolved ``\lya\ blob'' at $z \simeq 3.1$ (LAB) \citep{li20}. More recently, Li et al., in prep, have shown that the clumpy outflow models can be applied successfully to fit multiple regions within a spatially resolved \lya\ halo simultaneously, i.e., using a central source producing \lya\ which then propagates through a clumpy medium with an axisymmetric outflow (see also \citealt{steidel11}, who showed that \lya\ emission halos similar in extent to those presented in this paper are predicted naturally given the observed velocity fields of outflows viewed ``down the barrel'' to the galaxy center and the same radial dependence of clump covering fraction inferred from absorption line studies of background objects).

\subsection{Implications for \ion{H}{I} kinematics and \lya\ Radiative Transfer}

A spectral profile with a dominant redshifted component of \lya\ emission line and a weaker blueshifted component -- with peaks shifted by similar $|\Delta v|$ relative to the systemic redshift -- is a typical signature of an expanding geometry, a central \lya\ source function, and resonant scattering.
It has also been shown that, for an ensemble of galaxies also drawn from the same KBSS redshift survey, the mean \lya\ {\it absorption} signature measured in the spectra of background objects within $D_{\rm tran} \simlt 50$ pkpc ($\theta_{\rm tran} \simlt 6\secpoint$) are outflow-dominated \citep{chen20}. However, although the observed \lya\ halos and their dependence on galaxy properties may be naturally explained by central \lya\ sources scattering through the CGM, many authors have emphasized that collisionally-excited \lya\ emission from accreting gas (i.e., graviational cooling -- see e.g., \citealt{fg10,goerdt10,lake15}) and {\it in situ} photoionization by the UV background and/or local sources of ionizing photons (e.g., \citealt{leclercq20})  may also contribute significantly to extended \lya\ halos. 

Due to the complex nature of \lya\ radiative transfer, our results cannot resolve this issue definitively. However, the fact that the stacked CP2D \lya\ spectra show asymmetry of $< 2$\% between the major and minor axis for the ensemble, combined with (1) the empirical correlation between the central \lya\ line strength ($\wlya$) and the \lya\ flux integrated within the entire halo,  and (2) the consistently red-peak-dominated kinematics of both the central and integrated \lya\ line, all favor scattering of \lya\ photons produced near the galaxy center through an outflowing, clumpy medium -- at least within $D_\mathrm{tran} \lesssim  30$ pkpc. The remarkable statistical symmetry of the full 2D profiles, both spatially and spectrally, suggests that most of the galaxies in our sample lack persistent disk-like configurations, an inference supported also by the ubiquity of blue-shifted absorption profiles in DTB spectra of similar galaxies and their lack of dependence on HST morphology (\citealt{law12b,law12c}). As a consequence, outflows do not behave in the manner expected for central starbursts in disk galaxies; in other words, $z \sim 2-3$ galaxies on average appear to be more axisymmetric than their lower-redshift counterparts. This may have important implications for the cycling of gas and metals into and out of forming galaxies.

On the other hand, we have detected marginal ($2\sigma$) excess \lya\ emission along the galaxy major axis for the red peak, and along the galaxy minor axis for the blue peak. While most of the excess emission along the major axis can be easily explained by the relatively small sample size and the presence of one or two extreme cases, the excess blueshifted emission along the galaxy minor axis cannot be. While it is possible that the observed asymmetries  indicate the prevalence of outflows along the major axis and inflows along the minor axis (i.e., the opposite of the behavior of galaxies at $z < 1$), we regard such an interpretation as unlikely. An important clue may be that most of the blueshifted, minor-axis excess is contributed by galaxies with weaker than the median \lya\ emission strength -- many in that subset have central $\wlya < 0$, meaning that \lya\ photons must scatter from higher-velocity material or in directions with a more porous distribution of optically thick gas  to have a high probability of escaping. Since we can only observe the photons that escape, galaxies with lower overall \lya\ escape fractions will mean that those that do escape must take more extreme paths on average.  As shown in \S\ref{sec:az_halo}, the galaxies with relatively weak emission also tend to have lower values of $\fblue / \fred$, which makes smaller absolute differences in emission strength vs. azimuthal angle more noticeable. 

{Finally, we would like to emphasise the fact that the lack of a strong statistical  correlation between the morphology of extended  \lya\ emission and the galaxy orientation does {\it not} imply that the individual \lya\ halos are symmetric. In fact, as shown in Figures \ref{fig:hist_excess_lya}, \ref{fig:hist_excess_lya_blue}, and \ref{fig:df_halo}, individual \lya\ halos are often asymmetric, particularly for objects with weak central \lya\ emission. Rather, our results indicate that morphological variations of \lya\ halos are uncorrelated with the apparent orientation of the host galaxy starlight. }

\section{Summary}
\label{sec:summary}

In this paper, we have presented the first statistical results of an IFU survey of star-forming galaxies at $\langle z \rangle =2.43$ drawn from the Keck Baryonic Structure Survey and observed with the Keck Cosmic Web Imager on the Keck 2 telescope. The 59 galaxies, with stellar mass and SFR typical of the full KBSS galaxy sample, comprise the subset of the KBSS-KCWI survey with both deep KCWI observations (typical exposure times of $\sim 5$ hours) and existing high-spatial-resolution images from Hubble Space Telescope and/or Keck/OSIRIS. The high resolution images were used to determine the direction of the projected major axis of the stellar continuum light of each galaxy; {the KCWI IFU data cubes were used to detect spatially- and spectrally-resolved \lya\ emission from the CGM around each galaxy to
a limiting surface brightness of $\lesssim 1\times 10^{-19}$ ergs s$^{-1}$ cm$^{-2}$ arcsec$^{-2}$ in the composite data (Figure \ref{fig:lya_profile}),} enabling detection of diffuse \lya\  emission halos to projected distances of $\theta_{\rm tran} \simeq 4$\arcs\ ($D_{\rm tran}\simeq 30$ pkpc).

Our major findings are summarised below: \\
\begin{enumerate}
\item We introduced ``cylindrically projected 2D spectra'' (CP2D) in order to visualise and quantify \lya\ spectra as a function of projected galactocentric distance $D_{\rm tran}$.  The CP2D spectra are averages of spaxels over a specified range of azimuthal angle ($\phi$) at a common galactocentric distance, enabling statistical analyses of \lya\ spectral profiles and their spatial variation simultaneously. The CP2D spectra clearly show distinct redshifted and blueshifted components of \lya\ emission that remain distinct out to projected distances of at least 25 pkpc, with rest-frame velocity extending from $-700 \le (\Delta v_{\rm sys}/\kms) \simlt 1000$ with respect to the galaxy systemic redshift, with blue and red peaks at $\simeq -300$ \kms\ and $\simeq +300$ \kms, respectively. (\S\ref{sec:2dspec})

\item We stacked the CP2D spectra of individual galaxies after aligning their continuum major axes, in bins of azimuthal angle $\phi$ measured with respect to the major axis. By creating difference images of the CP2D projections in independent ranges of $\phi$, we showed that residual differences between ``major axis'' and ``minor axis'' -- which would reflect asymmetries in either the spatial or spectral dimension along different ranges of $\phi$ --  are very small, with amplitude: $\simlt 2\times 10^{-20}~\mathrm{erg~s}^{-1} \mathrm{cm}^{-2} \mathrm{arcsec}^{-2} \textrm{\AA}^{-1}$, corresponding to asymmetries in \lya\ flux amounting to $\le 2\%$ of the total, between galaxy major and minor axis directions. (\S\ref{sec:2dspec_azimuthal})

\item We found little evidence of statitically significant assymetry of the \lya\ emission, except for an excess ($\simeq 2\sigma$) of \lya\ emission along galaxy major axes for the redshifted component of  \lya\ emission, with a peak near $\sim +300~\mathrm{km~s}^{-1}$. However, closer inspection revealed that most of the signal was caused by two galaxies with unusually asymmetric \lya\ halos. 
After discarding these outliers, another excess emission feature, with integrated significance $\simeq 2\sigma$, manifests as excess emission in the {\it blueshifted} component of \lya\ along galaxy {\it minor} axes. This feature extends over a large range of velocity, and appears to be contributed primarily by galaxies with weaker than the sample median \lya\ emission, and central \lya\ equivalent width $\wlya  < 0$.  The same weak-\lya\  subsample includes many of highest $M_{\ast}$ galaxies in the sample, as well as many of the galaxies with the smallest flux ratio between blueshifted and redshifted components ($\fblue/\fred$). We speculate that this asymmetry may indicate that significant azimuthal variation of \lya\ emission morphology exists only for galaxies with the smallest \lya\ escape fractions within the sample, i.e., weaker \lya\ emitting galaxies possess more developed rotational structure. Evidently, one sees only the photons managing to find rare low-\nhi\ holes or that scatter from the highest velocity material, both of which are more likely along the minor axis ( i.e., similar to expectations based on the standard picture of biconical starburst-driven outflows from disk-like systems). (\S\ref{sec:closer_look}, \S\ref{sec:az_halo})

\item  Taken together, the results show that, statistically, the \lya\ halo around galaxies in this sample (and, by extension, the population of relatively massive star-forming galaxies at $z \sim 2-3$) have remarkably little correlation -- either kinematically or spatially -- with the morphological  distribution of stellar continuum light of the host galaxy. The observations suggest that most of the galaxies do not conform to expectations in which outflows are bi-conical and oriented along the minor axis of disk-like configurations, with accretion occurring preferentially along the major axis, suggested by observations of CGM gas in star-forming galaxies with $z < 1$. Instead, the lack of systematic variation in the kinematics and spatial extent with azimuthal angle of \lya\ emission, together with the fact that $\fblue / \fred$  is universally smaller than unity,
suggests that the bulk of \lya\ at galactocentric distances $\simlt 30$ pkpc is scattered from the inside out. The vast majority of scattered photons propagate through a scattering medium whose kinematics are dominated by outflows that statistically symmetric with respect to the apparent morphology of a galaxy's starlight. (\S\ref{sec:discussions})
\end{enumerate}

This paper marks the first attempt to understand the relationship between Ly$\alpha$ emission in the CGM and its host-galaxy properties in the KBSS-KCWI sample. As the central DTB \lya\ emission was shown to be correlated with the host galaxy properties for the KBSS galaxies \citep[e.g.,][]{trainor15, trainor16, trainor19}, in forthcoming work, we will utilise the CP2D spectra to further investigate the connection between the \lya\ halo and observable properties of the host galaxies (e.g., stellar mass,  star-formation rate, star-formation rate surface density, etc.), to understand whether galaxies at $z = 2 - 3$ significantly impact the \ion{H}{I} distribution in the CGM and \textit{vice versa}, and to place additional constraints on the source functions and radiative transfer of \lya\ in forming galaxies. 
With increased sample size and improved data reduction processes, we are also pushing to higher sensitivity in the stacked spectral cubes that allow us to probe the \lya\ spectrum at larger galactocentric distances with high fidelity.

\section*{Acknowledgements}
This work has included data from Keck/KCWI \citep{morrissey18}, Keck/OSIRIS \citep{larkin06}, Keck/MOSFIRE \citep{mclean12}, Keck/LRIS-B \citep{steidel04}, HST/WFC3-IR and HST/ACS. We appreciate the contribution from the staff of the W. M. Keck Observatory and the Space Telescope Science Institute. 

The following software packages have been crucial to the results presented: Astropy \citep{astropy18}, the SciPy and NumPy system \citep{scipy20, numpy20}, QFitsView\footnote{https://www.mpe.mpg.de/~ott/QFitsView/}, CWITools \citep{osullivan20b}, Montage\footnote{http://montage.ipac.caltech.edu/}, GALFIT \citep{peng02,peng10}, and DrizzlePac\footnote{https://www.stsci.edu/scientific-community/software/drizzlepac.html}. 

This work has been supported in part by grant AST-2009278 from the US NSF, by NASA through grant HST-GO15287.001, and the JPL/Caltech President's and Director's Program (YC, CS). 

The authors wish to recognise and acknowledge the very significant cultural role and reverence that the summit of Maunakea has always had within the indigenous Hawaiian community.  We are most fortunate to have the opportunity to conduct observations from this mountain. {We would like to thank the anonymous referee for providing constructive feedback.} We would like to acknowledge Kurt Adelberger, Milan Bogosavljevi\'{c}, Max Pettini, and Rachel Theios for their contribution to the KBSS survey. It is a great pleasure for us to thank Don Neill, Mateusz Matuszewski, Luca Rizzi, Donal O'Sullivan, and Sebastiano Cantalupo for their help in handling the KCWI data, and Cameron Hummels and Max Gronke for insightful discussions. YC would like to acknowledge his grandfather, Chen Yizong, who passed away during the preparation of this manuscript. 

\section*{Data Availability}
The composite CP2D spectra and the Python program used to generate figures in this article are available upon reasonable request.




\bibliographystyle{mnras}
\bibliography{main.bib} 

\begin{thebibliography}{}
\makeatletter
\relax
\def\mn@urlcharsother{\let\do\@makeother \do\$\do\&\do\#\do\^\do\_\do\%\do\~}
\def\mn@doi{\begingroup\mn@urlcharsother \@ifnextchar [ {\mn@doi@}
  {\mn@doi@[]}}
\def\mn@doi@[#1]#2{\def\@tempa{#1}\ifx\@tempa\@empty \href
  {http://dx.doi.org/#2} {doi:#2}\else \href {http://dx.doi.org/#2} {#1}\fi
  \endgroup}
\def\mn@eprint#1#2{\mn@eprint@#1:#2::\@nil}
\def\mn@eprint@arXiv#1{\href {http://arxiv.org/abs/#1} {{\tt arXiv:#1}}}
\def\mn@eprint@dblp#1{\href {http://dblp.uni-trier.de/rec/bibtex/#1.xml}
  {dblp:#1}}
\def\mn@eprint@#1:#2:#3:#4\@nil{\def\@tempa {#1}\def\@tempb {#2}\def\@tempc
  {#3}\ifx \@tempc \@empty \let \@tempc \@tempb \let \@tempb \@tempa \fi \ifx
  \@tempb \@empty \def\@tempb {arXiv}\fi \@ifundefined
  {mn@eprint@\@tempb}{\@tempb:\@tempc}{\expandafter \expandafter \csname
  mn@eprint@\@tempb\endcsname \expandafter{\@tempc}}}

\bibitem[\protect\citeauthoryear{{Ao} et~al.,}{{Ao} et~al.}{2020}]{ao20}
{Ao} Y.,  et~al., 2020, \mn@doi [Nature Astronomy] {10.1038/s41550-020-1033-3},
  \href {https://ui.adsabs.harvard.edu/abs/2020NatAs...4..670A} {4, 670}

\bibitem[\protect\citeauthoryear{{Astropy Collaboration} et~al.,}{{Astropy
  Collaboration} et~al.}{2018}]{astropy18}
{Astropy Collaboration} et~al., 2018, \mn@doi [\aj] {10.3847/1538-3881/aabc4f},
  \href {https://ui.adsabs.harvard.edu/abs/2018AJ....156..123A} {156, 123}

\bibitem[\protect\citeauthoryear{{Bacon} et~al.,}{{Bacon}
  et~al.}{2010}]{bacon10}
{Bacon} R.,  et~al., 2010, in {McLean} I.~S.,  {Ramsay} S.~K.,   {Takami} H.,
  eds,  Society of Photo-Optical Instrumentation Engineers (SPIE) Conference
  Series Vol. 7735, Ground-based and Airborne Instrumentation for Astronomy
  III. p. 773508, \mn@doi{10.1117/12.856027}

\bibitem[\protect\citeauthoryear{{Bacon} et~al.,}{{Bacon}
  et~al.}{2017}]{bacon17}
{Bacon} R.,  et~al., 2017, \mn@doi [\aap] {10.1051/0004-6361/201730833}, \href
  {https://ui.adsabs.harvard.edu/abs/2017A&A...608A...1B} {608, A1}

\bibitem[\protect\citeauthoryear{{Bordoloi} et~al.,}{{Bordoloi}
  et~al.}{2011}]{bordoloi11}
{Bordoloi} R.,  et~al., 2011, \mn@doi [\apj] {10.1088/0004-637X/743/1/10},
  \href {https://ui.adsabs.harvard.edu/abs/2011ApJ...743...10B} {743, 10}

\bibitem[\protect\citeauthoryear{{Borisova} et~al.,}{{Borisova}
  et~al.}{2016}]{borisova16}
{Borisova} E.,  et~al., 2016, \mn@doi [\apj] {10.3847/0004-637X/831/1/39},
  \href {https://ui.adsabs.harvard.edu/abs/2016ApJ...831...39B} {831, 39}

\bibitem[\protect\citeauthoryear{{Bouch{\'e}}, {Hohensee}, {Vargas},
  {Kacprzak}, {Martin}, {Cooke}  \& {Churchill}}{{Bouch{\'e}}
  et~al.}{2012}]{bouche12}
{Bouch{\'e}} N.,  {Hohensee} W.,  {Vargas} R.,  {Kacprzak} G.~G.,  {Martin}
  C.~L.,  {Cooke} J.,   {Churchill} C.~W.,  2012, \mn@doi [\mnras]
  {10.1111/j.1365-2966.2012.21114.x}, \href
  {https://ui.adsabs.harvard.edu/abs/2012MNRAS.426..801B} {426, 801}

\bibitem[\protect\citeauthoryear{{Bregman}}{{Bregman}}{1980}]{bregman80}
{Bregman} J.~N.,  1980, \mn@doi [\apj] {10.1086/157776}, \href
  {https://ui.adsabs.harvard.edu/abs/1980ApJ...236..577B} {236, 577}

\bibitem[\protect\citeauthoryear{{Bruzual} \& {Charlot}}{{Bruzual} \&
  {Charlot}}{2003}]{bruzual03}
{Bruzual} G.,  {Charlot} S.,  2003, \mn@doi [\mnras]
  {10.1046/j.1365-8711.2003.06897.x}, \href
  {https://ui.adsabs.harvard.edu/abs/2003MNRAS.344.1000B} {344, 1000}

\bibitem[\protect\citeauthoryear{{Byrohl}, {Nelson}, {Behrens}, {Pillepich},
  {Hernquist}, {Marinacci}  \& {Vogelsberger}}{{Byrohl}
  et~al.}{2020}]{byrohl20}
{Byrohl} C.,  {Nelson} D.,  {Behrens} C.,  {Pillepich} A.,  {Hernquist} L.,
  {Marinacci} F.,   {Vogelsberger} M.,  2020, arXiv e-prints, \href
  {https://ui.adsabs.harvard.edu/abs/2020arXiv200907283B} {p. arXiv:2009.07283}

\bibitem[\protect\citeauthoryear{{Cai} et~al.,}{{Cai} et~al.}{2019}]{cai19}
{Cai} Z.,  et~al., 2019, \mn@doi [\apjs] {10.3847/1538-4365/ab4796}, \href
  {https://ui.adsabs.harvard.edu/abs/2019ApJS..245...23C} {245, 23}

\bibitem[\protect\citeauthoryear{{Calzetti}, {Armus}, {Bohlin}, {Kinney},
  {Koornneef}  \& {Storchi-Bergmann}}{{Calzetti} et~al.}{2000}]{calzetti00}
{Calzetti} D.,  {Armus} L.,  {Bohlin} R.~C.,  {Kinney} A.~L.,  {Koornneef} J.,
   {Storchi-Bergmann} T.,  2000, \mn@doi [\apj] {10.1086/308692}, \href
  {https://ui.adsabs.harvard.edu/abs/2000ApJ...533..682C} {533, 682}

\bibitem[\protect\citeauthoryear{{Cantalupo}, {Arrigoni-Battaia}, {Prochaska},
  {Hennawi}  \& {Madau}}{{Cantalupo} et~al.}{2014}]{cantalupo14}
{Cantalupo} S.,  {Arrigoni-Battaia} F.,  {Prochaska} J.~X.,  {Hennawi} J.~F.,
  {Madau} P.,  2014, \mn@doi [\nat] {10.1038/nature12898}, \href
  {https://ui.adsabs.harvard.edu/abs/2014Natur.506...63C} {506, 63}

\bibitem[\protect\citeauthoryear{{Carr}, {Scarlata}, {Panagia}  \&
  {Henry}}{{Carr} et~al.}{2018}]{carr18}
{Carr} C.,  {Scarlata} C.,  {Panagia} N.,   {Henry} A.,  2018, \mn@doi [\apj]
  {10.3847/1538-4357/aac48e}, \href
  {https://ui.adsabs.harvard.edu/abs/2018ApJ...860..143C} {860, 143}

\bibitem[\protect\citeauthoryear{{Chabrier}}{{Chabrier}}{2003}]{chabrier03}
{Chabrier} G.,  2003, \mn@doi [\apjl] {10.1086/374879}, \href
  {https://ui.adsabs.harvard.edu/abs/2003ApJ...586L.133C} {586, L133}

\bibitem[\protect\citeauthoryear{{Chen}, {Chen}, {Gronke}, {Rauch}  \&
  {Broadhurst}}{{Chen} et~al.}{2020a}]{mchen20}
{Chen} M.~C.,  {Chen} H.-W.,  {Gronke} M.,  {Rauch} M.,   {Broadhurst} T.,
  2020a, arXiv e-prints, \href
  {https://ui.adsabs.harvard.edu/abs/2020arXiv201203959C} {p. arXiv:2012.03959}

\bibitem[\protect\citeauthoryear{{Chen} et~al.,}{{Chen} et~al.}{2020b}]{chen20}
{Chen} Y.,  et~al., 2020b, \mn@doi [\mnras] {10.1093/mnras/staa2808}, \href
  {https://ui.adsabs.harvard.edu/abs/2020MNRAS.499.1721C} {499, 1721}

\bibitem[\protect\citeauthoryear{{Claeyssens} et~al.,}{{Claeyssens}
  et~al.}{2019}]{claeyssens19}
{Claeyssens} A.,  et~al., 2019, \mn@doi [\mnras] {10.1093/mnras/stz2492}, \href
  {https://ui.adsabs.harvard.edu/abs/2019MNRAS.489.5022C} {489, 5022}

\bibitem[\protect\citeauthoryear{{Dijkstra}}{{Dijkstra}}{2014}]{dijkstra14}
{Dijkstra} M.,  2014, \mn@doi [\pasa] {10.1017/pasa.2014.33}, \href
  {https://ui.adsabs.harvard.edu/abs/2014PASA...31...40D} {31, e040}

\bibitem[\protect\citeauthoryear{{Duval} et~al.,}{{Duval}
  et~al.}{2016}]{duval16}
{Duval} F.,  et~al., 2016, \mn@doi [\aap] {10.1051/0004-6361/201526876}, \href
  {https://ui.adsabs.harvard.edu/abs/2016A&A...587A..77D} {587, A77}

\bibitem[\protect\citeauthoryear{{Erb}, {Steidel}, {Shapley}, {Pettini}  \&
  {Adelberger}}{{Erb} et~al.}{2004}]{erb04}
{Erb} D.~K.,  {Steidel} C.~C.,  {Shapley} A.~E.,  {Pettini} M.,   {Adelberger}
  K.~L.,  2004, \mn@doi [\apj] {10.1086/422464}, \href
  {https://ui.adsabs.harvard.edu/abs/2004ApJ...612..122E} {612, 122}

\bibitem[\protect\citeauthoryear{{Erb}, {Steidel}  \& {Chen}}{{Erb}
  et~al.}{2018}]{erb18}
{Erb} D.~K.,  {Steidel} C.~C.,   {Chen} Y.,  2018, \mn@doi [\apjl]
  {10.3847/2041-8213/aacff6}, \href
  {https://ui.adsabs.harvard.edu/abs/2018ApJ...862L..10E} {862, L10}

\bibitem[\protect\citeauthoryear{{Faucher-Gigu{\`e}re}, {Kere{\v{s}}},
  {Dijkstra}, {Hernquist}  \& {Zaldarriaga}}{{Faucher-Gigu{\`e}re}
  et~al.}{2010}]{fg10}
{Faucher-Gigu{\`e}re} C.-A.,  {Kere{\v{s}}} D.,  {Dijkstra} M.,  {Hernquist}
  L.,   {Zaldarriaga} M.,  2010, \mn@doi [\apj] {10.1088/0004-637X/725/1/633},
  \href {https://ui.adsabs.harvard.edu/abs/2010ApJ...725..633F} {725, 633}

\bibitem[\protect\citeauthoryear{{F{\"o}rster Schreiber} et~al.,}{{F{\"o}rster
  Schreiber} et~al.}{2009}]{forster09}
{F{\"o}rster Schreiber} N.~M.,  et~al., 2009, \mn@doi [\apj]
  {10.1088/0004-637X/706/2/1364}, \href
  {https://ui.adsabs.harvard.edu/abs/2009ApJ...706.1364F} {706, 1364}

\bibitem[\protect\citeauthoryear{{F{\"o}rster Schreiber} et~al.,}{{F{\"o}rster
  Schreiber} et~al.}{2018}]{fs18}
{F{\"o}rster Schreiber} N.~M.,  et~al., 2018, \mn@doi [\apjs]
  {10.3847/1538-4365/aadd49}, \href
  {https://ui.adsabs.harvard.edu/abs/2018ApJS..238...21F} {238, 21}

\bibitem[\protect\citeauthoryear{{Goerdt}, {Dekel}, {Sternberg}, {Ceverino},
  {Teyssier}  \& {Primack}}{{Goerdt} et~al.}{2010}]{goerdt10}
{Goerdt} T.,  {Dekel} A.,  {Sternberg} A.,  {Ceverino} D.,  {Teyssier} R.,
  {Primack} J.~R.,  2010, \mn@doi [\mnras] {10.1111/j.1365-2966.2010.16941.x},
  \href {http://adsabs.harvard.edu/abs/2010MNRAS.407..613G} {407, 613}

\bibitem[\protect\citeauthoryear{{Gronke}}{{Gronke}}{2017}]{gronke17}
{Gronke} M.,  2017, \mn@doi [\aap] {10.1051/0004-6361/201731791}, \href
  {https://ui.adsabs.harvard.edu/abs/2017A&A...608A.139G} {608, A139}

\bibitem[\protect\citeauthoryear{{Gronke} \& {Dijkstra}}{{Gronke} \&
  {Dijkstra}}{2016}]{gronke16a}
{Gronke} M.,  {Dijkstra} M.,  2016, \mn@doi [\apj]
  {10.3847/0004-637X/826/1/14}, \href
  {https://ui.adsabs.harvard.edu/abs/2016ApJ...826...14G} {826, 14}

\bibitem[\protect\citeauthoryear{{Gronke}, {Dijkstra}, {McCourt}  \&
  {Oh}}{{Gronke} et~al.}{2016}]{gronke16b}
{Gronke} M.,  {Dijkstra} M.,  {McCourt} M.,   {Oh} S.~P.,  2016, \mn@doi
  [\apjl] {10.3847/2041-8213/833/2/L26}, \href
  {https://ui.adsabs.harvard.edu/abs/2016ApJ...833L..26G} {833, L26}

\bibitem[\protect\citeauthoryear{{Guaita} et~al.,}{{Guaita}
  et~al.}{2015}]{guaita15}
{Guaita} L.,  et~al., 2015, \mn@doi [\aap] {10.1051/0004-6361/201425053}, \href
  {https://ui.adsabs.harvard.edu/abs/2015A&A...576A..51G} {576, A51}

\bibitem[\protect\citeauthoryear{{Harris} et~al.,}{{Harris}
  et~al.}{2020}]{numpy20}
{Harris} C.~R.,  et~al., 2020, \mn@doi [\nat] {10.1038/s41586-020-2649-2},
  \href {https://ui.adsabs.harvard.edu/abs/2020Natur.585..357H} {585, 357}

\bibitem[\protect\citeauthoryear{{Hayes} et~al.,}{{Hayes}
  et~al.}{2014}]{hayes14}
{Hayes} M.,  et~al., 2014, \mn@doi [\apj] {10.1088/0004-637X/782/1/6}, \href
  {https://ui.adsabs.harvard.edu/abs/2014ApJ...782....6H} {782, 6}

\bibitem[\protect\citeauthoryear{{Ho}, {Martin}, {Kacprzak}  \&
  {Churchill}}{{Ho} et~al.}{2017}]{ho17}
{Ho} S.~H.,  {Martin} C.~L.,  {Kacprzak} G.~G.,   {Churchill} C.~W.,  2017,
  \mn@doi [\apj] {10.3847/1538-4357/835/2/267}, \href
  {https://ui.adsabs.harvard.edu/abs/2017ApJ...835..267H} {835, 267}

\bibitem[\protect\citeauthoryear{{Jones}, {Stark}  \& {Ellis}}{{Jones}
  et~al.}{2012}]{jones12}
{Jones} T.,  {Stark} D.~P.,   {Ellis} R.~S.,  2012, \mn@doi [\apj]
  {10.1088/0004-637X/751/1/51}, \href
  {https://ui.adsabs.harvard.edu/abs/2012ApJ...751...51J} {751, 51}

\bibitem[\protect\citeauthoryear{{Kacprzak}, {Churchill}, {Evans}, {Murphy}  \&
  {Steidel}}{{Kacprzak} et~al.}{2011}]{kacprzak11}
{Kacprzak} G.~G.,  {Churchill} C.~W.,  {Evans} J.~L.,  {Murphy} M.~T.,
  {Steidel} C.~C.,  2011, \mn@doi [\mnras] {10.1111/j.1365-2966.2011.19261.x},
  \href {https://ui.adsabs.harvard.edu/abs/2011MNRAS.416.3118K} {416, 3118}

\bibitem[\protect\citeauthoryear{{Kacprzak}, {Churchill}  \&
  {Nielsen}}{{Kacprzak} et~al.}{2012}]{kacprzak12}
{Kacprzak} G.~G.,  {Churchill} C.~W.,   {Nielsen} N.~M.,  2012, \mn@doi [\apjl]
  {10.1088/2041-8205/760/1/L7}, \href
  {https://ui.adsabs.harvard.edu/abs/2012ApJ...760L...7K} {760, L7}

\bibitem[\protect\citeauthoryear{{Kacprzak}, {Muzahid}, {Churchill}, {Nielsen}
  \& {Charlton}}{{Kacprzak} et~al.}{2015}]{kacprzak15}
{Kacprzak} G.~G.,  {Muzahid} S.,  {Churchill} C.~W.,  {Nielsen} N.~M.,
  {Charlton} J.~C.,  2015, \mn@doi [\apj] {10.1088/0004-637X/815/1/22}, \href
  {https://ui.adsabs.harvard.edu/abs/2015ApJ...815...22K} {815, 22}

\bibitem[\protect\citeauthoryear{{Kakiichi} \& {Dijkstra}}{{Kakiichi} \&
  {Dijkstra}}{2018}]{kakiichi18}
{Kakiichi} K.,  {Dijkstra} M.,  2018, \mn@doi [\mnras] {10.1093/mnras/sty2214},
  \href {https://ui.adsabs.harvard.edu/abs/2018MNRAS.480.5140K} {480, 5140}

\bibitem[\protect\citeauthoryear{{Kollmeier}, {Zheng}, {Dav{\'e}}, {Gould},
  {Katz}, {Miralda-Escud{\'e}}  \& {Weinberg}}{{Kollmeier}
  et~al.}{2010}]{kollmeier10}
{Kollmeier} J.~A.,  {Zheng} Z.,  {Dav{\'e}} R.,  {Gould} A.,  {Katz} N.,
  {Miralda-Escud{\'e}} J.,   {Weinberg} D.~H.,  2010, \mn@doi [\apj]
  {10.1088/0004-637X/708/2/1048}, \href
  {http://adsabs.harvard.edu/abs/2010ApJ...708.1048K} {708, 1048}

\bibitem[\protect\citeauthoryear{{Kornei}, {Shapley}, {Erb}, {Steidel},
  {Reddy}, {Pettini}  \& {Bogosavljevi{\'c}}}{{Kornei} et~al.}{2010}]{kornei10}
{Kornei} K.~A.,  {Shapley} A.~E.,  {Erb} D.~K.,  {Steidel} C.~C.,  {Reddy}
  N.~A.,  {Pettini} M.,   {Bogosavljevi{\'c}} M.,  2010, \mn@doi [\apj]
  {10.1088/0004-637X/711/2/693}, \href
  {http://adsabs.harvard.edu/abs/2010ApJ...711..693K} {711, 693}

\bibitem[\protect\citeauthoryear{{Kulas}, {Shapley}, {Kollmeier}, {Zheng},
  {Steidel}  \& {Hainline}}{{Kulas} et~al.}{2012}]{kulas12}
{Kulas} K.~R.,  {Shapley} A.~E.,  {Kollmeier} J.~A.,  {Zheng} Z.,  {Steidel}
  C.~C.,   {Hainline} K.~N.,  2012, \mn@doi [\apj]
  {10.1088/0004-637X/745/1/33}, \href
  {https://ui.adsabs.harvard.edu/abs/2012ApJ...745...33K} {745, 33}

\bibitem[\protect\citeauthoryear{{Lake}, {Zheng}, {Cen}, {Sadoun}, {Momose}  \&
  {Ouchi}}{{Lake} et~al.}{2015}]{lake15}
{Lake} E.,  {Zheng} Z.,  {Cen} R.,  {Sadoun} R.,  {Momose} R.,   {Ouchi} M.,
  2015, \mn@doi [\apj] {10.1088/0004-637X/806/1/46}, \href
  {https://ui.adsabs.harvard.edu/abs/2015ApJ...806...46L} {806, 46}

\bibitem[\protect\citeauthoryear{{Lan} \& {Mo}}{{Lan} \& {Mo}}{2018}]{lan18}
{Lan} T.-W.,  {Mo} H.,  2018, \mn@doi [\apj] {10.3847/1538-4357/aadc08}, \href
  {https://ui.adsabs.harvard.edu/abs/2018ApJ...866...36L} {866, 36}

\bibitem[\protect\citeauthoryear{{Larkin} et~al.,}{{Larkin}
  et~al.}{2006}]{larkin06}
{Larkin} J.,  et~al., 2006, in \procspie. p. 62691A, \mn@doi{10.1117/12.672061}

\bibitem[\protect\citeauthoryear{{Law}, {Steidel}, {Erb}, {Larkin}, {Pettini},
  {Shapley}  \& {Wright}}{{Law} et~al.}{2007}]{law07}
{Law} D.~R.,  {Steidel} C.~C.,  {Erb} D.~K.,  {Larkin} J.~E.,  {Pettini} M.,
  {Shapley} A.~E.,   {Wright} S.~A.,  2007, \mn@doi [\apj] {10.1086/521786},
  \href {https://ui.adsabs.harvard.edu/abs/2007ApJ...669..929L} {669, 929}

\bibitem[\protect\citeauthoryear{{Law}, {Steidel}, {Erb}, {Larkin}, {Pettini},
  {Shapley}  \& {Wright}}{{Law} et~al.}{2009}]{law09}
{Law} D.~R.,  {Steidel} C.~C.,  {Erb} D.~K.,  {Larkin} J.~E.,  {Pettini} M.,
  {Shapley} A.~E.,   {Wright} S.~A.,  2009, \mn@doi [\apj]
  {10.1088/0004-637X/697/2/2057}, \href
  {https://ui.adsabs.harvard.edu/abs/2009ApJ...697.2057L} {697, 2057}

\bibitem[\protect\citeauthoryear{{Law}, {Shapley}, {Steidel}, {Reddy},
  {Christensen}  \& {Erb}}{{Law} et~al.}{2012a}]{law12b}
{Law} D.~R.,  {Shapley} A.~E.,  {Steidel} C.~C.,  {Reddy} N.~A.,  {Christensen}
  C.~R.,   {Erb} D.~K.,  2012a, \mn@doi [\nat] {10.1038/nature11256}, \href
  {https://ui.adsabs.harvard.edu/abs/2012Natur.487..338L} {487, 338}

\bibitem[\protect\citeauthoryear{{Law}, {Steidel}, {Shapley}, {Nagy}, {Reddy}
  \& {Erb}}{{Law} et~al.}{2012b}]{law12a}
{Law} D.~R.,  {Steidel} C.~C.,  {Shapley} A.~E.,  {Nagy} S.~R.,  {Reddy} N.~A.,
    {Erb} D.~K.,  2012b, \mn@doi [\apj] {10.1088/0004-637X/745/1/85}, \href
  {https://ui.adsabs.harvard.edu/abs/2012ApJ...745...85L} {745, 85}

\bibitem[\protect\citeauthoryear{{Law}, {Steidel}, {Shapley}, {Nagy}, {Reddy}
  \& {Erb}}{{Law} et~al.}{2012c}]{law12c}
{Law} D.~R.,  {Steidel} C.~C.,  {Shapley} A.~E.,  {Nagy} S.~R.,  {Reddy} N.~A.,
    {Erb} D.~K.,  2012c, \mn@doi [\apj] {10.1088/0004-637X/759/1/29}, \href
  {https://ui.adsabs.harvard.edu/abs/2012ApJ...759...29L} {759, 29}

\bibitem[\protect\citeauthoryear{{Law}, {Steidel}, {Chen}, {Strom}, {Rudie}  \&
  {Trainor}}{{Law} et~al.}{2018}]{law18}
{Law} D.~R.,  {Steidel} C.~C.,  {Chen} Y.,  {Strom} A.~L.,  {Rudie} G.~C.,
  {Trainor} R.~F.,  2018, \mn@doi [\apj] {10.3847/1538-4357/aae156}, \href
  {https://ui.adsabs.harvard.edu/abs/2018ApJ...866..119L} {866, 119}

\bibitem[\protect\citeauthoryear{{Leclercq} et~al.,}{{Leclercq}
  et~al.}{2017}]{leclercq17}
{Leclercq} F.,  et~al., 2017, \mn@doi [\aap] {10.1051/0004-6361/201731480},
  \href {https://ui.adsabs.harvard.edu/abs/2017A&A...608A...8L} {608, A8}

\bibitem[\protect\citeauthoryear{{Leclercq} et~al.,}{{Leclercq}
  et~al.}{2020}]{leclercq20}
{Leclercq} F.,  et~al., 2020, \mn@doi [\aap] {10.1051/0004-6361/201937339},
  \href {https://ui.adsabs.harvard.edu/abs/2020A&A...635A..82L} {635, A82}

\bibitem[\protect\citeauthoryear{{Li}, {Steidel}, {Gronke}  \& {Chen}}{{Li}
  et~al.}{2020}]{li20}
{Li} Z.,  {Steidel} C.~C.,  {Gronke} M.,   {Chen} Y.,  2020, arXiv e-prints,
  \href {https://ui.adsabs.harvard.edu/abs/2020arXiv200809130L} {p.
  arXiv:2008.09130}

\bibitem[\protect\citeauthoryear{{Lundgren} et~al.,}{{Lundgren}
  et~al.}{2021}]{lundgren21}
{Lundgren} B.~F.,  et~al., 2021, arXiv e-prints, \href
  {https://ui.adsabs.harvard.edu/abs/2021arXiv210210117L} {p. arXiv:2102.10117}

\bibitem[\protect\citeauthoryear{{Martin}, {Chang}, {Matuszewski}, {Morrissey},
  {Rahman}, {Moore}  \& {Steidel}}{{Martin} et~al.}{2014}]{martin14}
{Martin} D.~C.,  {Chang} D.,  {Matuszewski} M.,  {Morrissey} P.,  {Rahman} S.,
  {Moore} A.,   {Steidel} C.~C.,  2014, \mn@doi [\apj]
  {10.1088/0004-637X/786/2/106}, \href
  {http://adsabs.harvard.edu/abs/2014ApJ...786..106M} {786, 106}

\bibitem[\protect\citeauthoryear{{Martin}, {Matuszewski}, {Morrissey}, {Neill},
  {Moore}, {Steidel}  \& {Trainor}}{{Martin} et~al.}{2016}]{martin16}
{Martin} D.~C.,  {Matuszewski} M.,  {Morrissey} P.,  {Neill} J.~D.,  {Moore}
  A.,  {Steidel} C.~C.,   {Trainor} R.,  2016, \mn@doi [\apjl]
  {10.3847/2041-8205/824/1/L5}, \href
  {http://adsabs.harvard.edu/abs/2016ApJ...824L...5M} {824, L5}

\bibitem[\protect\citeauthoryear{{Martin}, {Ho}, {Kacprzak}  \&
  {Churchill}}{{Martin} et~al.}{2019}]{martin19}
{Martin} C.~L.,  {Ho} S.~H.,  {Kacprzak} G.~G.,   {Churchill} C.~W.,  2019,
  \mn@doi [\apj] {10.3847/1538-4357/ab18ac}, \href
  {https://ui.adsabs.harvard.edu/abs/2019ApJ...878...84M} {878, 84}

\bibitem[\protect\citeauthoryear{{Matsuda} et~al.,}{{Matsuda}
  et~al.}{2012}]{matsuda12}
{Matsuda} Y.,  et~al., 2012, \mn@doi [\mnras]
  {10.1111/j.1365-2966.2012.21143.x}, \href
  {https://ui.adsabs.harvard.edu/abs/2012MNRAS.425..878M} {425, 878}

\bibitem[\protect\citeauthoryear{{Matthee} et~al.,}{{Matthee}
  et~al.}{2021}]{matthee21}
{Matthee} J.,  et~al., 2021, arXiv e-prints, \href
  {https://ui.adsabs.harvard.edu/abs/2021arXiv210207779M} {p. arXiv:2102.07779}

\bibitem[\protect\citeauthoryear{{McLean} et~al.,}{{McLean}
  et~al.}{2012}]{mclean12}
{McLean} I.~S.,  et~al., 2012, in {McLean} I.~S.,  {Ramsay} S.~K.,   {Takami}
  H.,  eds,  Society of Photo-Optical Instrumentation Engineers (SPIE)
  Conference Series Vol. 8446, Ground-based and Airborne Instrumentation for
  Astronomy IV. p. 84460J, \mn@doi{10.1117/12.924794}

\bibitem[\protect\citeauthoryear{{Momose} et~al.,}{{Momose}
  et~al.}{2014}]{momose14}
{Momose} R.,  et~al., 2014, \mn@doi [\mnras] {10.1093/mnras/stu825}, \href
  {https://ui.adsabs.harvard.edu/abs/2014MNRAS.442..110M} {442, 110}

\bibitem[\protect\citeauthoryear{{Morrissey} et~al.,}{{Morrissey}
  et~al.}{2018}]{morrissey18}
{Morrissey} P.,  et~al., 2018, \mn@doi [\apj] {10.3847/1538-4357/aad597}, \href
  {https://ui.adsabs.harvard.edu/abs/2018ApJ...864...93M} {864, 93}

\bibitem[\protect\citeauthoryear{{Mostardi}, {Shapley}, {Steidel}, {Trainor},
  {Reddy}  \& {Siana}}{{Mostardi} et~al.}{2015}]{mostardi15}
{Mostardi} R.~E.,  {Shapley} A.~E.,  {Steidel} C.~C.,  {Trainor} R.~F.,
  {Reddy} N.~A.,   {Siana} B.,  2015, \mn@doi [\apj]
  {10.1088/0004-637X/810/2/107}, \href
  {https://ui.adsabs.harvard.edu/abs/2015ApJ...810..107M} {810, 107}

\bibitem[\protect\citeauthoryear{{Nelson} et~al.,}{{Nelson}
  et~al.}{2019}]{nelson19}
{Nelson} D.,  et~al., 2019, \mn@doi [\mnras] {10.1093/mnras/stz2306}, \href
  {https://ui.adsabs.harvard.edu/abs/2019MNRAS.490.3234N} {490, 3234}

\bibitem[\protect\citeauthoryear{{Nielsen}, {Churchill}, {Kacprzak}, {Murphy}
  \& {Evans}}{{Nielsen} et~al.}{2015}]{nielsen15}
{Nielsen} N.~M.,  {Churchill} C.~W.,  {Kacprzak} G.~G.,  {Murphy} M.~T.,
  {Evans} J.~L.,  2015, \mn@doi [\apj] {10.1088/0004-637X/812/1/83}, \href
  {https://ui.adsabs.harvard.edu/abs/2015ApJ...812...83N} {812, 83}

\bibitem[\protect\citeauthoryear{{O'Sullivan} \& {Chen}}{{O'Sullivan} \&
  {Chen}}{2020}]{osullivan20b}
{O'Sullivan} D.,  {Chen} Y.,  2020, arXiv e-prints, \href
  {https://ui.adsabs.harvard.edu/abs/2020arXiv201105444O} {p. arXiv:2011.05444}

\bibitem[\protect\citeauthoryear{{O'Sullivan}, {Martin}, {Matuszewski},
  {Hoadley}, {Hamden}, {Neill}, {Lin}  \& {Parihar}}{{O'Sullivan}
  et~al.}{2020}]{osullivan20}
{O'Sullivan} D.~B.,  {Martin} C.,  {Matuszewski} M.,  {Hoadley} K.,  {Hamden}
  E.,  {Neill} J.~D.,  {Lin} Z.,   {Parihar} P.,  2020, \mn@doi [\apj]
  {10.3847/1538-4357/ab838c}, \href
  {https://ui.adsabs.harvard.edu/abs/2020ApJ...894....3O} {894, 3}

\bibitem[\protect\citeauthoryear{{Oke} et~al.,}{{Oke} et~al.}{1995}]{oke95}
{Oke} J.~B.,  et~al., 1995, \pasp, 107, 375

\bibitem[\protect\citeauthoryear{{{\"O}stlin} et~al.,}{{{\"O}stlin}
  et~al.}{2014}]{ostlin14}
{{\"O}stlin} G.,  et~al., 2014, \mn@doi [\apj] {10.1088/0004-637X/797/1/11},
  \href {https://ui.adsabs.harvard.edu/abs/2014ApJ...797...11O} {797, 11}

\bibitem[\protect\citeauthoryear{{Ouchi}, {Ono}  \& {Shibuya}}{{Ouchi}
  et~al.}{2020}]{ouchi20}
{Ouchi} M.,  {Ono} Y.,   {Shibuya} T.,  2020, \mn@doi [\araa]
  {10.1146/annurev-astro-032620-021859}, \href
  {https://ui.adsabs.harvard.edu/abs/2020ARA&A..58..617O} {58, 617}

\bibitem[\protect\citeauthoryear{{Pahl}, {Shapley}, {Steidel}, {Chen}  \&
  {Reddy}}{{Pahl} et~al.}{2021}]{pahl21}
{Pahl} A.~J.,  {Shapley} A.,  {Steidel} C.~C.,  {Chen} Y.,   {Reddy} N.~A.,
  2021, arXiv e-prints, \href
  {https://ui.adsabs.harvard.edu/abs/2021arXiv210402081P} {p. arXiv:2104.02081}

\bibitem[\protect\citeauthoryear{{Partridge} \& {Peebles}}{{Partridge} \&
  {Peebles}}{1967}]{partridge67}
{Partridge} R.~B.,  {Peebles} P.~J.~E.,  1967, \mn@doi [\apj] {10.1086/149079},
  \href {http://adsabs.harvard.edu/abs/1967ApJ...147..868P} {147, 868}

\bibitem[\protect\citeauthoryear{{Peng}, {Ho}, {Impey}  \& {Rix}}{{Peng}
  et~al.}{2002}]{peng02}
{Peng} C.~Y.,  {Ho} L.~C.,  {Impey} C.~D.,   {Rix} H.-W.,  2002, \mn@doi [\aj]
  {10.1086/340952}, \href
  {https://ui.adsabs.harvard.edu/abs/2002AJ....124..266P} {124, 266}

\bibitem[\protect\citeauthoryear{{Peng}, {Ho}, {Impey}  \& {Rix}}{{Peng}
  et~al.}{2010}]{peng10}
{Peng} C.~Y.,  {Ho} L.~C.,  {Impey} C.~D.,   {Rix} H.-W.,  2010, \mn@doi [\aj]
  {10.1088/0004-6256/139/6/2097}, \href
  {https://ui.adsabs.harvard.edu/abs/2010AJ....139.2097P} {139, 2097}

\bibitem[\protect\citeauthoryear{{P{\'e}roux}, {Nelson}, {van de Voort},
  {Pillepich}, {Marinacci}, {Vogelsberger}  \& {Hernquist}}{{P{\'e}roux}
  et~al.}{2020}]{peroux20}
{P{\'e}roux} C.,  {Nelson} D.,  {van de Voort} F.,  {Pillepich} A.,
  {Marinacci} F.,  {Vogelsberger} M.,   {Hernquist} L.,  2020, \mn@doi [\mnras]
  {10.1093/mnras/staa2888}, \href
  {https://ui.adsabs.harvard.edu/abs/2020MNRAS.499.2462P} {499, 2462}

\bibitem[\protect\citeauthoryear{{Peter}, {Shapley}, {Law}, {Steidel}, {Erb},
  {Reddy}  \& {Pettini}}{{Peter} et~al.}{2007}]{peter07}
{Peter} A.~H.~G.,  {Shapley} A.~E.,  {Law} D.~R.,  {Steidel} C.~C.,  {Erb}
  D.~K.,  {Reddy} N.~A.,   {Pettini} M.,  2007, \mn@doi [\apj]
  {10.1086/521184}, \href {http://adsabs.harvard.edu/abs/2007ApJ...668...23P}
  {668, 23}

\bibitem[\protect\citeauthoryear{{Pettini}, {Shapley}, {Steidel}, {Cuby},
  {Dickinson}, {Moorwood}, {Adelberger}  \& {Giavalisco}}{{Pettini}
  et~al.}{2001}]{pettini01}
{Pettini} M.,  {Shapley} A.~E.,  {Steidel} C.~C.,  {Cuby} J.-G.,  {Dickinson}
  M.,  {Moorwood} A. F.~M.,  {Adelberger} K.~L.,   {Giavalisco} M.,  2001,
  \mn@doi [\apj] {10.1086/321403}, \href
  {https://ui.adsabs.harvard.edu/abs/2001ApJ...554..981P} {554, 981}

\bibitem[\protect\citeauthoryear{{Reddy} \& {Steidel}}{{Reddy} \&
  {Steidel}}{2009}]{reddy09}
{Reddy} N.~A.,  {Steidel} C.~C.,  2009, \mn@doi [\apj]
  {10.1088/0004-637X/692/1/778}, \href
  {http://adsabs.harvard.edu/abs/2009ApJ...692..778R} {692, 778}

\bibitem[\protect\citeauthoryear{{Rudie} et~al.,}{{Rudie}
  et~al.}{2012}]{rudie12a}
{Rudie} G.~C.,  et~al., 2012, \mn@doi [\apj] {10.1088/0004-637X/750/1/67},
  \href {http://adsabs.harvard.edu/abs/2012ApJ...750...67R} {750, 67}

\bibitem[\protect\citeauthoryear{{Schroetter} et~al.,}{{Schroetter}
  et~al.}{2019}]{schroetter19}
{Schroetter} I.,  et~al., 2019, \mn@doi [\mnras] {10.1093/mnras/stz2822}, \href
  {https://ui.adsabs.harvard.edu/abs/2019MNRAS.490.4368S} {490, 4368}

\bibitem[\protect\citeauthoryear{{S{\'e}rsic}}{{S{\'e}rsic}}{1963}]{sersic63}
{S{\'e}rsic} J.~L.,  1963, Boletin de la Asociacion Argentina de Astronomia La
  Plata Argentina, \href
  {https://ui.adsabs.harvard.edu/abs/1963BAAA....6...41S} {6, 41}

\bibitem[\protect\citeauthoryear{{Shapley}, {Steidel}, {Pettini}  \&
  {Adelberger}}{{Shapley} et~al.}{2003}]{shapley03}
{Shapley} A.~E.,  {Steidel} C.~C.,  {Pettini} M.,   {Adelberger} K.~L.,  2003,
  \mn@doi [\apj] {10.1086/373922}, \href
  {https://ui.adsabs.harvard.edu/abs/2003ApJ...588...65S} {588, 65}

\bibitem[\protect\citeauthoryear{{Smith}, {Ma}, {Bromm}, {Finkelstein},
  {Hopkins}, {Faucher-Gigu{\`e}re}  \& {Kere{\v{s}}}}{{Smith}
  et~al.}{2019}]{smith19}
{Smith} A.,  {Ma} X.,  {Bromm} V.,  {Finkelstein} S.~L.,  {Hopkins} P.~F.,
  {Faucher-Gigu{\`e}re} C.-A.,   {Kere{\v{s}}} D.,  2019, \mn@doi [\mnras]
  {10.1093/mnras/sty3483}, \href
  {https://ui.adsabs.harvard.edu/abs/2019MNRAS.484...39S} {484, 39}

\bibitem[\protect\citeauthoryear{{Song}, {Seon}  \& {Hwang}}{{Song}
  et~al.}{2020}]{song20}
{Song} H.,  {Seon} K.-I.,   {Hwang} H.~S.,  2020, \mn@doi [\apj]
  {10.3847/1538-4357/abac02}, \href
  {https://ui.adsabs.harvard.edu/abs/2020ApJ...901...41S} {901, 41}

\bibitem[\protect\citeauthoryear{{Sravan} et~al.,}{{Sravan}
  et~al.}{2016}]{sravan16}
{Sravan} N.,  et~al., 2016, \mn@doi [\mnras] {10.1093/mnras/stw1962}, \href
  {https://ui.adsabs.harvard.edu/abs/2016MNRAS.463..120S} {463, 120}

\bibitem[\protect\citeauthoryear{{Stanway} \& {Eldridge}}{{Stanway} \&
  {Eldridge}}{2018}]{stanway18}
{Stanway} E.~R.,  {Eldridge} J.~J.,  2018, \mn@doi [\mnras]
  {10.1093/mnras/sty1353}, \href
  {https://ui.adsabs.harvard.edu/abs/2018MNRAS.479...75S} {479, 75}

\bibitem[\protect\citeauthoryear{{Steidel}, {Kollmeier}, {Shapley},
  {Churchill}, {Dickinson}  \& {Pettini}}{{Steidel} et~al.}{2002}]{steidel02}
{Steidel} C.~C.,  {Kollmeier} J.~A.,  {Shapley} A.~E.,  {Churchill} C.~W.,
  {Dickinson} M.,   {Pettini} M.,  2002, \mn@doi [\apj] {10.1086/339792}, \href
  {http://adsabs.harvard.edu/abs/2002ApJ...570..526S} {570, 526}

\bibitem[\protect\citeauthoryear{{Steidel}, {Shapley}, {Pettini}, {Adelberger},
  {Erb}, {Reddy}  \& {Hunt}}{{Steidel} et~al.}{2004}]{steidel04}
{Steidel} C.~C.,  {Shapley} A.~E.,  {Pettini} M.,  {Adelberger} K.~L.,  {Erb}
  D.~K.,  {Reddy} N.~A.,   {Hunt} M.~P.,  2004, \apj, \href
  {http://adsabs.harvard.edu/cgi-bin/nph-bib_query?bibcode=2004ApJ...604..534S&amp;db_key=AST}
  {604, 534}

\bibitem[\protect\citeauthoryear{{Steidel}, {Erb}, {Shapley}, {Pettini},
  {Reddy}, {Bogosavljevi{\'c}}, {Rudie}  \& {Rakic}}{{Steidel}
  et~al.}{2010}]{steidel10}
{Steidel} C.~C.,  {Erb} D.~K.,  {Shapley} A.~E.,  {Pettini} M.,  {Reddy} N.,
  {Bogosavljevi{\'c}} M.,  {Rudie} G.~C.,   {Rakic} O.,  2010, \mn@doi [\apj]
  {10.1088/0004-637X/717/1/289}, \href
  {https://ui.adsabs.harvard.edu/abs/2010ApJ...717..289S} {717, 289}

\bibitem[\protect\citeauthoryear{{Steidel}, {Bogosavljevi{\'c}}, {Shapley},
  {Kollmeier}, {Reddy}, {Erb}  \& {Pettini}}{{Steidel}
  et~al.}{2011}]{steidel11}
{Steidel} C.~C.,  {Bogosavljevi{\'c}} M.,  {Shapley} A.~E.,  {Kollmeier} J.~A.,
   {Reddy} N.~A.,  {Erb} D.~K.,   {Pettini} M.,  2011, \mn@doi [\apj]
  {10.1088/0004-637X/736/2/160}, \href
  {https://ui.adsabs.harvard.edu/abs/2011ApJ...736..160S} {736, 160}

\bibitem[\protect\citeauthoryear{{Steidel} et~al.,}{{Steidel}
  et~al.}{2014}]{steidel14}
{Steidel} C.~C.,  et~al., 2014, \mn@doi [\apj] {10.1088/0004-637X/795/2/165},
  \href {http://adsabs.harvard.edu/abs/2014ApJ...795..165S} {795, 165}

\bibitem[\protect\citeauthoryear{{Steidel}, {Strom}, {Pettini}, {Rudie},
  {Reddy}  \& {Trainor}}{{Steidel} et~al.}{2016}]{steidel16}
{Steidel} C.~C.,  {Strom} A.~L.,  {Pettini} M.,  {Rudie} G.~C.,  {Reddy} N.~A.,
    {Trainor} R.~F.,  2016, \mn@doi [\apj] {10.3847/0004-637X/826/2/159}, \href
  {https://ui.adsabs.harvard.edu/abs/2016ApJ...826..159S} {826, 159}

\bibitem[\protect\citeauthoryear{{Steidel}, {Bogosavljevi{\'c}}, {Shapley},
  {Reddy}, {Rudie}, {Pettini}, {Trainor}  \& {Strom}}{{Steidel}
  et~al.}{2018}]{steidel18}
{Steidel} C.~C.,  {Bogosavljevi{\'c}} M.,  {Shapley} A.~E.,  {Reddy} N.~A.,
  {Rudie} G.~C.,  {Pettini} M.,  {Trainor} R.~F.,   {Strom} A.~L.,  2018,
  \mn@doi [\apj] {10.3847/1538-4357/aaed28}, \href
  {http://adsabs.harvard.edu/abs/2018ApJ...869..123S} {869, 123}

\bibitem[\protect\citeauthoryear{{Strom}, {Steidel}, {Rudie}, {Trainor},
  {Pettini}  \& {Reddy}}{{Strom} et~al.}{2017}]{strom17}
{Strom} A.~L.,  {Steidel} C.~C.,  {Rudie} G.~C.,  {Trainor} R.~F.,  {Pettini}
  M.,   {Reddy} N.~A.,  2017, \mn@doi [\apj] {10.3847/1538-4357/836/2/164},
  \href {http://adsabs.harvard.edu/abs/2017ApJ...836..164S} {836, 164}

\bibitem[\protect\citeauthoryear{{Theios}, {Steidel}, {Strom}, {Rudie},
  {Trainor}  \& {Reddy}}{{Theios} et~al.}{2019}]{theios19}
{Theios} R.~L.,  {Steidel} C.~C.,  {Strom} A.~L.,  {Rudie} G.~C.,  {Trainor}
  R.~F.,   {Reddy} N.~A.,  2019, \mn@doi [\apj] {10.3847/1538-4357/aaf386},
  \href {http://adsabs.harvard.edu/abs/2019ApJ...871..128T} {871, 128}

\bibitem[\protect\citeauthoryear{{Trainor}, {Steidel}, {Strom}  \&
  {Rudie}}{{Trainor} et~al.}{2015}]{trainor15}
{Trainor} R.~F.,  {Steidel} C.~C.,  {Strom} A.~L.,   {Rudie} G.~C.,  2015,
  \mn@doi [\apj] {10.1088/0004-637X/809/1/89}, \href
  {https://ui.adsabs.harvard.edu/abs/2015ApJ...809...89T} {809, 89}

\bibitem[\protect\citeauthoryear{{Trainor}, {Strom}, {Steidel}  \&
  {Rudie}}{{Trainor} et~al.}{2016}]{trainor16}
{Trainor} R.~F.,  {Strom} A.~L.,  {Steidel} C.~C.,   {Rudie} G.~C.,  2016,
  \mn@doi [\apj] {10.3847/0004-637X/832/2/171}, \href
  {https://ui.adsabs.harvard.edu/abs/2016ApJ...832..171T} {832, 171}

\bibitem[\protect\citeauthoryear{{Trainor}, {Strom}, {Steidel}, {Rudie}, {Chen}
   \& {Theios}}{{Trainor} et~al.}{2019}]{trainor19}
{Trainor} R.~F.,  {Strom} A.~L.,  {Steidel} C.~C.,  {Rudie} G.~C.,  {Chen} Y.,
   {Theios} R.~L.,  2019, \mn@doi [\apj] {10.3847/1538-4357/ab4993}, \href
  {https://ui.adsabs.harvard.edu/abs/2019ApJ...887...85T} {887, 85}

\bibitem[\protect\citeauthoryear{{Tumlinson}, {Peeples}  \& {Werk}}{{Tumlinson}
  et~al.}{2017}]{tumlinson17}
{Tumlinson} J.,  {Peeples} M.~S.,   {Werk} J.~K.,  2017, \mn@doi [\araa]
  {10.1146/annurev-astro-091916-055240}, \href
  {https://ui.adsabs.harvard.edu/abs/2017ARA&A..55..389T} {55, 389}

\bibitem[\protect\citeauthoryear{{Vanzella} et~al.,}{{Vanzella}
  et~al.}{2017}]{vanzella17}
{Vanzella} E.,  et~al., 2017, \mn@doi [\mnras] {10.1093/mnras/stw2442}, \href
  {https://ui.adsabs.harvard.edu/abs/2017MNRAS.465.3803V} {465, 3803}

\bibitem[\protect\citeauthoryear{{Veilleux}, {Cecil}  \&
  {Bland-Hawthorn}}{{Veilleux} et~al.}{2005}]{veilleux05}
{Veilleux} S.,  {Cecil} G.,   {Bland-Hawthorn} J.,  2005, \mn@doi [\araa]
  {10.1146/annurev.astro.43.072103.150610}, \href
  {https://ui.adsabs.harvard.edu/abs/2005ARA&A..43..769V} {43, 769}

\bibitem[\protect\citeauthoryear{{Verhamme}, {Schaerer}  \&
  {Maselli}}{{Verhamme} et~al.}{2006}]{verhamme06}
{Verhamme} A.,  {Schaerer} D.,   {Maselli} A.,  2006, \mn@doi [\aap]
  {10.1051/0004-6361:20065554}, \href
  {https://ui.adsabs.harvard.edu/abs/2006A&A...460..397V} {460, 397}

\bibitem[\protect\citeauthoryear{{Verhamme}, {Dubois}, {Blaizot}, {Garel},
  {Bacon}, {Devriendt}, {Guiderdoni}  \& {Slyz}}{{Verhamme}
  et~al.}{2012}]{verhamme12}
{Verhamme} A.,  {Dubois} Y.,  {Blaizot} J.,  {Garel} T.,  {Bacon} R.,
  {Devriendt} J.,  {Guiderdoni} B.,   {Slyz} A.,  2012, \mn@doi [\aap]
  {10.1051/0004-6361/201218783}, \href
  {https://ui.adsabs.harvard.edu/abs/2012A&A...546A.111V} {546, A111}

\bibitem[\protect\citeauthoryear{{Verhamme} et~al.,}{{Verhamme}
  et~al.}{2018}]{verhamme18}
{Verhamme} A.,  et~al., 2018, \mn@doi [\mnras] {10.1093/mnrasl/sly058}, \href
  {https://ui.adsabs.harvard.edu/abs/2018MNRAS.478L..60V} {478, L60}

\bibitem[\protect\citeauthoryear{{Virtanen} et~al.,}{{Virtanen}
  et~al.}{2020}]{scipy20}
{Virtanen} P.,  et~al., 2020, \mn@doi [Nature Methods]
  {10.1038/s41592-019-0686-2}, \href
  {https://ui.adsabs.harvard.edu/abs/2020NatMe..17..261V} {17, 261}

\bibitem[\protect\citeauthoryear{{Whitaker} et~al.,}{{Whitaker}
  et~al.}{2014}]{whitaker14}
{Whitaker} K.~E.,  et~al., 2014, \mn@doi [\apj] {10.1088/0004-637X/795/2/104},
  \href {https://ui.adsabs.harvard.edu/abs/2014ApJ...795..104W} {795, 104}

\bibitem[\protect\citeauthoryear{{Wisotzki} et~al.,}{{Wisotzki}
  et~al.}{2016}]{wisotzki16}
{Wisotzki} L.,  et~al., 2016, \mn@doi [\aap] {10.1051/0004-6361/201527384},
  \href {https://ui.adsabs.harvard.edu/abs/2016A&A...587A..98W} {587, A98}

\bibitem[\protect\citeauthoryear{{Wisotzki} et~al.,}{{Wisotzki}
  et~al.}{2018}]{wisotzki18}
{Wisotzki} L.,  et~al., 2018, \mn@doi [\nat] {10.1038/s41586-018-0564-6}, \href
  {https://ui.adsabs.harvard.edu/abs/2018Natur.562..229W} {562, 229}

\bibitem[\protect\citeauthoryear{{Zheng}, {Cen}, {Trac}  \&
  {Miralda-Escud{\'e}}}{{Zheng} et~al.}{2011}]{zheng11}
{Zheng} Z.,  {Cen} R.,  {Trac} H.,   {Miralda-Escud{\'e}} J.,  2011, \mn@doi
  [\apj] {10.1088/0004-637X/726/1/38}, \href
  {https://ui.adsabs.harvard.edu/abs/2011ApJ...726...38Z} {726, 38}

\makeatother
\end{thebibliography}






\bsp	
\label{lastpage}

\end{CJK*}
\end{document}